\documentclass{aa}

\usepackage{graphicx}
\usepackage{subcaption}
\usepackage{float}
\usepackage{wrapfig}
\usepackage{pdflscape}
\usepackage{amsmath}
\usepackage{ulem}
\usepackage{txfonts}
\DeclareUnicodeCharacter{2212}{\textendash}
\begin{document} 
   
   \title{CONCERTO: Simulating the CO, [CII], and [CI]  line emission of galaxies in a $\rm 117 \, deg^2$ field and the impact of field-to-field variance}

%   \subtitle{}

   \author{A. Gkogkou\inst{1}, 
   M. Béthermin\inst{1,2}, 
   G. Lagache \inst{1}, 
   M. Van Cuyck\inst{1}, 
   E. Jullo\inst{1}, 
   M. Aravena\inst{3}, 
   A. Beelen\inst{1}, 
   A. Benoit\inst{4,5}, 
   J. Bounmy\inst{6,5}, 
   M. Calvo\inst{4,5}, 
   A. Catalano\inst{6,5},
   S. Cora\inst{8,9}, 
   D. Croton\inst{11, 12},
   S. de la Torre\inst{1}, 
   A. Fasano\inst{1}, 
   A. Ferrara\inst{13}, 
   J. Goupy\inst{4,5}, 
   C. Hoarau\inst{6,5}, 
   W. Hu\inst{1}, 
   T. Ishiyama\inst{15},
   K. K. Knudsen\inst{7}, 
   J.-C. Lambert\inst{1}, 
   J. F. Macías-Pérez\inst{6,5}, 
   J. Marpaud\inst{6,5},
   G. Mellema\inst{16},
   A. Monfardini\inst{4,5}, 
   A. Pallottini\inst{13}, 
   N. Ponthieu\inst{4,5}, 
   F. Prada\inst{10}, 
   Y. Roehlly\inst{1}, 
   L. Vallini\inst{13}, 
   F. Walter\inst{14}
   }

   \institute{Aix Marseille Univ, CNRS, CNES, LAM, Marseille, France, \email{athanasia.gkogkou@lam.fr}
     \and
         Université de Strasbourg, CNRS, Observatoire astronomique de Strasbourg, UMR 7550, 67000 Strasbourg, France
     \and
        Núcleo de Astronomía, Facultad de Ingeniería y Ciencias, Universidad Diego Portales, Av. Ejército 441, Santiago, Chile
    \and
        Univ. Grenoble Alpes, CNRS, Grenoble INP, Institut Néel, 38000 Grenoble, France
    \and
        Groupement d’Interet Scientifique KID, 38000 Grenoble and 38400 Saint Martin d’Héres, France
    \and
        Univ. Grenoble Alpes, CNRS, LPSC/IN2P3, 38000 Grenoble, France
    \and
        Department of Space, Earth and Environment, Chalmers University of Technology Onsala Space Observatory, SE-439 92 Onsala, Sweden
    \and
        Instituto de Astrof\'isica de La Plata (CCT La Plata, CONICET, UNLP),
        Observatorio Astron\'omico, Paseo del Bosque S/N, B1900FWA, La Plata, Argentina
    \and
        Facultad de Ciencias Astron\'omicas y Geof\'isicas, Universidad Nacional de La Plata,
        Observatorio Astron\'omico, Paseo del Bosque S/N, B1900FWA, La Plata, Argentina
    \and
        Instituto de Astrofísica de Andalucía (CSIC), Glorieta de la Astronomía, E-18080 Granada, Spain
    \and
        Centre for Astrophysics \& Supercomputing, Swinburne University of Technology, Hawthorn, VIC 3122, Australia
    \and
        ARC Centre of Excellence for All Sky Astrophysics in 3 Dimensions (ASTRO 3D)"
    \and
        Scuola Normale Superiore, Piazza dei Cavalieri 7, 56126 Pisa, Italy
    \and
        Max-Planck-Institut für Astronomie, Königstuhl 17, D-69117 Heidelberg, Germany
    \and 
        Institute of Management and Information Technologies, Chiba University, 1-33, Yayoi-cho, Inage-ku, Chiba, 263-8522, Japan
    \and
        The Oskar Klein Centre, Department of Astronomy, Stockholm University, AlbaNova, SE-10691 Stockholm, Sweden
         }
        
   \date{Received 05/10/2022; accepted 29/11/2022}

\abstract{In the submillimeter regime, spectral line scans and line intensity mapping (LIM) are new promising probes for the cold gas content and star formation rate of galaxies across cosmic time. However, both of these two measurements suffer from field-to-field variance. We study the effect of field-to-field variance on the predicted CO and [CII] power spectra from future LIM experiments such as CONCERTO, as well as on the line luminosity functions (LFs) and the cosmic molecular gas mass density that are currently derived from spectral line scans. We combined a $\rm 117 \, deg^2$ dark matter lightcone from the Uchuu cosmological simulation with the simulated infrared dusty extragalactic sky (SIDES) approach. The clustering of the dusty galaxies in the SIDES-Uchuu product is validated by reproducing the cosmic infrared background anisotropies measured by \textit{Herschel} and \textit{Planck}. We find that in order to constrain the CO LF with an uncertainty below 20\%, we need survey sizes of at least $\rm 0.1 \, deg^2$. Furthermore, accounting for the field-to-field variance using only the Poisson variance can underestimate the total variance by up to 80\%. The lower the luminosity is and the larger the survey size is, the higher the level of underestimate. At $z<3$, the impact of field-to-field variance on the cosmic molecular gas density can be as high as 40\% for the 4.6 arcmin$^2$ field, but drops below 10\% for areas larger than 0.2 deg$^2$. However, at $z>3$ the variance decreases more slowly with survey size and for example drops below 10\% for 1 deg$^2$ fields. Finally, we find that the CO and [CII] LIM power spectra can vary by up to 50\% in $\rm 1 \, deg^2$ fields. This limits the accuracy of the constraints provided by the first 1\,deg$^2$ surveys. In addition the level of the shot noise power is always dominated by the sources that are just below the detection thresholds, which limits its potential for deriving number densities of faint [CII] emitters. We provide an analytical formula to estimate the field-to-field variance of current or future LIM experiments given their observed frequency and survey size. The underlying code to derive the field-to-field variance and the full SIDES-Uchuu products (catalogs, cubes, and maps) are publicly available.}


  \titlerunning{CONCERTO: The impact of field-to-field variance}\space
  \authorrunning{Gkogkou et al.}\space
  \maketitle
%
%-------------------------------------------------------------------

\section{Introduction}
\label{sec:intro}
    The cosmic star formation rate density (SFRD), which is an integral constraint on star formation averaged over the volume of the observable universe at a given redshift, is a critical measure for theoretical models. Several ultraviolet (UV), optical, and infrared (IR) surveys have aimed to constrain the SFRD up to $\rm z \sim 10$ \citep{madau2014, bouwens2015, ishigaki2018, gruppioni2020, khusanova2021,fudamoto2021,zavala2021,wang2021} revealing that the SFRD rises at early times, peaks at $\rm z \sim 2$ and drops again toward present day.
    
    The cold molecular gas is tightly linked to the star formation through cosmic time, since it constitutes the fuel of star formation in galaxies \citep[see][for a review]{carilli2013, tacconi2020}. It is therefore essential to probe its abundance during different epochs in order to understand the cosmic star formation history (SFH). The molecular hydrogen lines are either too faint to be observed or the UV absorption is very strong. Additionally since they are quadrupole transitions, they require a high temperature to be excited. They thus trace only warm/dense molecular gas (e.g., shock-heated molecular gas), which is a minor fraction of the total amount of the molecular gas. Therefore, the most suitable tracer of the cold molecular gas is the carbon monoxide molecule CO (with $^{12}$C$^{16}$O being the most common CO isotope), which is the second most abundant molecule in the Universe and its rotational lines are bright enough to observe even at high redshifts \citep[e.g.,][]{venemans2017, decarli2022}.
    
   The method of spectral line scans is a powerful tool to study the evolution of the molecular gas throughout cosmic time. With this method we can search for targets in a given volume without preselecting them, avoiding biasing the sample toward given galaxy types. Several such surveys have aimed to constrain the CO luminosity function (\citealt{walter2014, decarli2019, decarli2020, riechers2019}), which is the comoving volume density of sources per luminosity, and infer the cosmic molecular gas density at different redshifts. They found that there is an evident evolution of the luminosity functions (LFs) with redshift and, as a consequence, of molecular gas abundance ($\rho_{\rm H_2}$). The cosmic molecular gas density increases at early times, peaks at $\rm z \sim 1-3$, and finally decreases down to the present day, suggesting that there is indeed a coevolution with the SFRD \citep[see also][]{walter2020}.
    
    Several observational studies have investigated the SFRD at different redshifts, but there is limited knowledge of its spatial distribution at large scales. The fluctuations of the integrated emission of dust in galaxies across cosmic times, which is the cosmic infrared background (CIB), can fill this gap of information, especially at high redshifts \citep{lagache2007, viero2009, planckcollaboration2011, amblard2011, planckcollaboration2014}. However, it is difficult to break the degeneracies among the different redshift slices especially at $z$ > 4 \citep{maniyar2018}.
    
    The line intensity mapping (LIM) technique can break these degeneracies. LIM measures the spatial fluctuations of the emission of a given spectral line in multiple frequencies, obtaining a 3D map \citep[see][for a review]{kovetz2017}. Therefore, using different lines, it is possible to trace the star formation and gas content in a specific redshift slice. The [CII] line at 158 $\rm \mu m$ is among the brightest far-infrared emission lines and it has been found that there is an empirical correlation between the [CII] emission and the star formation rate (SFR) \citep{delooze2014, lagache2018, schaerer2020}. Moreover, \cite{zanella2018} and \cite{vizgan2022} argue that [CII] is a convenient molecular gas tracer, especially at $z$ > 5. Theoretical studies also support this picture as shown in \cite{pallottini2017} and \cite{ferrara2019} via a numerical simulation and analytical model, respectively. This therefore constitutes [CII] as a great candidate for LIM.
    
    Multiple on-going experiments aim to measure the [CII] fluctuations at high redshifts, such as the Carbon [CII] line in post-reionization and reionization epoch project \citep[CONCERTO,][]{theconcertocollaboration2020}, the instrumentation for the tomographic ionized-carbon intensity mapping experiment \citep[TIME,][]{crites2014}, and the Prime-Cam spectro-imager \citep[mounted on
    the Fred Young Submillimeter Telescope telescope - FYST,][]{stacey2018}. A strong advantage of LIM experiments  is that they can observe much larger areas compared to traditional deep spectral scans. For example, compared to GOODS-North/CO Luminosity Density at High Redshift (COLDz, 0.014 deg$^2$ field), CONCERTO will observe a 1.4 deg$^2$ field, that is a 100 times larger area.

    Nevertheless, the field-to-field variance effect (also referred as cosmic variance in the literature) can still be an obstacle. Rare and randomly distributed bright sources can significantly alter the level of shot noise, while the density fluctuations of faint sources at large scales can impact the clustering component of the power spectra. Therefore the selection of the observed region could introduce significant uncertainties on the power spectra of the matter density or the line intensity fluctuations. It can also affect several other observables, such as the number density of galaxies, the LFs, and all the inferred quantities.
    
    There are studies that investigate the effect of the field-to-field variance on the galaxy number density and other related quantities like the LFs. \cite{moster2011} used N-body simulations and a recipe for the computation of cosmic variance to quantify its significant excess with respect to the Poisson variance. Similarly, \cite{driver2010} used Sloan Digital Sky Survey (SDSS) 0.03 < $z$ < 0.1 data to measure the cosmic variance as a function of survey volume and field aspect ratio. These studies reached a similar conclusion, that any density measurement of normal galaxies is subject to uncertainties derived from the cosmic variance. More specifically, they show that the cosmic variance in the number density of galaxies is $\sim 70\%$ for $\rm 1 \, deg^2$ fields and drops to $\sim 25\%$ for $\rm 100 \, deg^2$ fields. At such low redshift, big areas in the sky correspond to small comoving volumes. The variance effect is thus expected to be weaker at high redshift.
    
    \cite{keenan2020} used simulated data to investigate the effect of the cosmic variance on the shape of the CO LF and the evolution of the molecular gas mass density with redshift. They found that for a survey size of $\rm \sim 50 \, arcmin^2$ and apparent luminosity of $\sim \, 10^{10} \, \rm K \, km \, s^{-1} pc^2$ the Poisson and cosmic variance uncertainties are equal. Hence, not accounting for cosmic variance leads to underestimating of the total uncertainty by a factor of $\sim \sqrt{2}$. Regarding the molecular gas mass density, they point out that the volume required to detect the evolution around the peak of cosmic SFRD should be larger than 100\,arcmin$^2$.

    Although intensity mapping is gaining more and more attention as a technique, there are not enough studies on the effect of the field-to-field variance on the power spectrum. The clustering part of the power spectrum dominates at large scales ($k \lesssim 1$ arcmin$^{-1}$) and is linked to the distribution of galaxies in the large-scale structure. The shot noise appears due to the randomly distributed bright galaxies and it is dominant at small scales ($k \gtrsim 4$ arcmin$^{-1}$). Estimating the expected variance of these two components is key for scientific interpretation of intensity mapping data.
    
    In this paper, we combined the Simulated Infrared Dusty Extragalactic Sky (SIDES), which is a simulation of the far-infrared and submillimeter sky based on observed empirical relations, with the Uchuu\footnote{\url{http://www.skiesanduniverses.org/Simulations/Uchuu/}} N-body cosmological dark matter -only simulation \citep{ishiyama2021}. Thanks to the large volume that Uchuu provides alongside its high mass resolution, we can study the field-to-field variance of the power spectra obtained by line intensity maps, as well as its effect on other observables, as luminosity functions.
    
    The paper is organized as follows. In Sect.\,\ref{sec:simulations} we briefly describe the Simulated Infrared Dusty Extragalactic Sky (SIDES) approach as well as the building and the properties of the 117\,deg$^2$ simulated lightcone. In Sect.\,\ref{sec:model_validation} we compare the simulated power spectra with observations of the CIB fluctuations in order to validate our simulation. In Sect.\,\ref{sec:LFs} we also compare the simulated LFs with CO and [CII] observations and we investigate the field-to-field variance of the LFs and the cosmic molecular gas density. In Sect. \ref{sec:pk_variance} we investigate the contribution to the shot noise. We present a model that can estimate the field-to-field variance of the power spectra given the characteristics of any LIM experiment. Additionally, using this model we make forecasts for CONCERTO and other current/future LIM experiments. Finally, in Sect. \ref{sec:conclusion} we summarize our conclusions.
    
    For the SIDES simulation we assume a \cite{planckcollaboration2016c} cosmology, while for the Uchuu cosmological simulation we assume a \cite{planckcollaboration2020} cosmology \footnote{The discrepancy of the luminosity distance as computed using both cosmologies is less than 0.1\%. We thus skipped any recalibration of the used quantities.}. Throughout the paper we use a \cite{chabrier2003} initial mass function (IMF).

\section{Simulations}
\label{sec:simulations}

\subsection{Simulated Infrared Dusty Extragalactic Sky (SIDES)}
\label{subsec:sides}

SIDES is a simulation of the far-infrared (FIR) and submillimeter sky based on observed empirical relations. The initial frameworks \citep{sargent2012, bethermin2012} connect the stellar mass with several properties of galaxies (e.g., SFR $L_{\rm IR}$, $L_{\rm FIR}$) using a population of normal galaxies \citep[e.g.,][]{daddi2007, schreiber2015} and starburst galaxies. These are the galaxies above the main sequence of star forming galaxies, which is the tight relation between galaxy SFR and stellar mass. This formalism is extended in \cite{bethermin2013, bethermin2017} to perform the connection with dark matter halos using abundance matching.  Finally, the far-infrared and submillimeter lines are included in \cite{bethermin2022} (hereafter B22). The total size of the simulation presented in B22 is 2\,$\rm deg^2$ and the maximum redshift is 10, although the SIDES model is reliable only up to redshift 7 as also explained in B22. Moreover, in B22 for dark matter, SIDES uses the Bolshoi-Planck cosmological simulation \citep{rodriguez-puebla2016}, which has a volume of $(\rm 250 \, h^{-1} Mpc)^3$ and dark matter particles of $\rm 1.5 \times 10^8 \, h^{-1} M_{\odot}$ mass (halos resolved at $\gtrsim \, 100$ particles).

A dark matter lightcone, which encapsulates the information of the position and abundance of the dark matter halos, serves as the starting point for the SIDES simulation. The stellar masses of the galaxies populating the halos is determined using an abundance matching technique \citep[e.g.,][]{kravtsov2004,vale2004}. Subsequently, the generated galaxies are split into passive and star forming with a probability determined by observations \citep{davidzon2018}. It is assumed that only the star forming galaxies emit in the FIR and millimeter, so only these type of galaxies need to be assigned with a SFR value. This assumption is supported by the results presented in \cite{whitaker2021}, where it was found that passive galaxies are extremely faint in the submillimeter regime, so essentially their fluxes do not contribute to the FIR/submm sky.

The stellar mass of a galaxy is highly correlated with its SFR. Therefore, in our model the SFR of the galaxies is defined by their stellar mass. The SFR values of the main sequence and starburst galaxies are drawn accordingly based on the parameterized fit of the SFR-$M_{\rm stellar}$ relation described in \cite{schreiber2015}. We add a scatter of 0.3\,dex to this relation. The drawn SFR value defines the $L_{\rm IR}$ of each galaxy \citep{kennicutt1998} and consequently the normalization of its spectral energy distribution (SED). SEDs are selected from the library given in \cite{magdis2012} and they are updated at $z > 2$ in \cite{bethermin2015} and \cite{bethermin2017}. The shape of the SED depends on the galaxy type (main sequence or starburst) and the mean intensity radiation field ($\langle U \rangle$), which is correlated to the temperature of the dust. It is modeled such as $\langle U \rangle = 1$ corresponds to the local interstellar UV radiation field. Finally, a magnification value ($\mu$) due to the lensing effect is randomly drawn from the distributions presented in \cite{hezaveh2011} and \cite{hilbert2007} for strong and weak lensing, respectively. Therefore, all the sources are either magnified or demagnified following the distributions. This is a simplification since the magnification is not consistent with the underlying dark matter simulation.

SIDES can also simulate the emission of the strongest high-redshift submillimeter lines as [CII], CO and [CI]. The [CII] emission of the galaxies is generated using the $\rm L_{[CII]}-$SFR relation either from \cite{delooze2014} (DL14) or \cite{lagache2018} (L18). The L18 relation predicts a lower [CII] emission at high redshifts. The DL14 relation was derived in the local Universe; however, its validity at high redshift was tested using the follow-up of 118 optically selected galaxies at $4.4<z<5.9$ recently observed with ALMA \citep{capak2015,lefevre2020,bethermin2020,faisst2020,schaerer2020}. It was found that the DL14 relation still holds at these redshifts. \cite{carniani2020} reached the same result for galaxies at $z \,$=6-7. A caveat of this recipe is that environmental dependencies are not taken into account.

For CO, the fundamental transition is modeled from the $L_{\rm IR}-L_{\rm CO(1-0)}^{'}$ correlation \citep{sargent2014} for the main sequence galaxies, while for the starbursting systems there is an offset of -0.46\,dex for $L_{\rm IR}$ at a given $ L_{\rm CO(1-0)}^{'}$. $ L_{\rm CO(1-0)}^{'}$ is the observed luminosity and $L_{\rm IR}$ is the integrated continuum emission over 8-1000$\mu$m. The flux of the other transitions is computed using a clumpy and diffuse spectral line energy distribution (SLED) template from \cite{bournaud2015} and following the empirical relation presented in \cite{daddi2015} which connects the flux ratio of CO(5-4) and CO(2-1) transitions with $\langle U \rangle$.

Finally, the two transitions of [CI] also contribute to the millimeter regime. The modeling of the first transition is achieved by taking advantage of the tight correlation between $L_{[\rm CI](1-0)}$ and $ L_{\rm CO(4-3)}$. The flux of the second transition is obtained using the correlation between the ratio of the two [CI] transitions and the ratio of CO(7-6) and CO(4-3) fluxes as calibrated and presented in B22.

\subsection{The Uchuu dark matter simulation}
\label{subsec:UCHUU}

Nowadays cosmological simulations often achieve either high mass resolution or large volumes \citep{skillman2014, ishiyama2012, rodriguez-puebla2016, potter2017, cheng2020}. The ultimate goal, however, is to satisfy both requirements simultaneously. Significant progress has been made by zoom-in cosmological simulations \citep[e.g.,][]{hopkins2018, pallottini2022}. The Uchuu collaboration has created a suite of N-body cosmological simulations with various comoving volumes and mass resolutions \citep{ishiyama2021}. The Uchuu suite consists of four simulations: Uchuu, mini-Uchuu, micro-Uchuu, and shin-Uchuu with comoving volumes of 2000 $\rm h^{-1}Mpc$, 400 $\rm h^{-1}Mpc$, 100 $\rm h^{-1}Mpc$, 140 $\rm h^{−1}Mpc$, respectively and mass resolutions of $ \rm 3.27 \times 10^8 \, h^{-1}M_{\odot}$ (halos resolved at $\gtrsim \, 40$ particles) for all except for shin-Uchuu which has a resolution of $ \rm 8.97 \times 10^5 \, h^{-1} M_{\odot}$. For our study, we chose Uchuu, the biggest simulation in size, because it satisfies both requirements of large comoving volume and sufficient mass resolution. With this new simulation in hand, we can thus extensively validate the SIDES simulation and study the field-to-field variance introduced in the intensity mapping observables.

\begin{figure*}[h]
\begin{center}
\includegraphics[width=\linewidth]{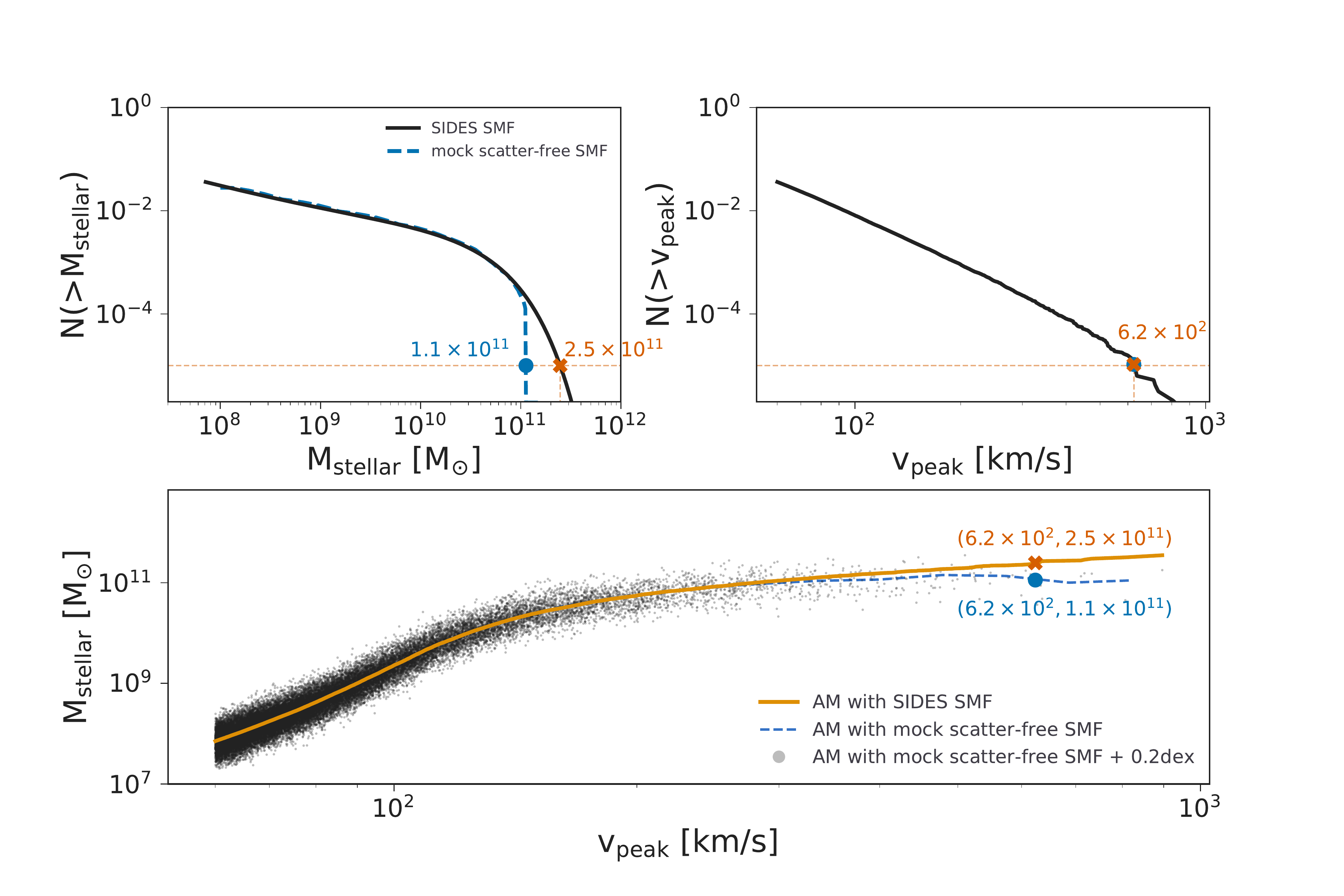}
\caption{Explanation of our abundance matching (AM) method between dark matter halos and galaxies. 
In the top left panel we show the cumulative SIDES stellar mass function (SMF, black solid line) and the mock scatter-free SMF (dashed blue line, Sect. \ref{subsec:AM}) which is obtained after the deconvolution of the SIDES SMF with the 0.2\,dex scatter. In the top right panel we show the cumulative halo velocity function. In the bottom panel we show the resulting relation between the stellar mass and the halo peak velocity when performing direct abundance matching using the SIDES SMF (orange line) and the mock scatter-free SMF (dashed blue line). The black points are the stellar masses assigned to the halos of our simulated catalog using the scatter-free relation (blue dashed line) and adding a 0.2 dex log-normal scatter. All plots are made for the 0.05 < $z$ < 0.1 redshift range.}
\label{fig:AM_scheme}
\end{center}
\end{figure*}

\begin{table*}[h]
\begin{center}
    {\centering
    \begin{tabular}{c | c c | c c | c c | c c }
    \hline\hline 
     $z$ & \multicolumn{2}{c|}{$\rm [CII]_{correlated}$ } & \multicolumn{2}{|c|}{$\rm CO(5-4)_{correlated}$} & \multicolumn{2}{c|}{$\rm [CII]_{SN}$} & \multicolumn{2}{|c}{$\rm CO(5-4)_{SN}$}\\
     & B22 & this work & B22 & this work & B22 & this work & B22 & this work \\
     & (\%) & (\%) & (\%) & (\%) & (\%) & (\%) & (\%) & (\%) \\
    \hline 
    1 & 1.53 & 0.68 & 3.87 & 2.94 & 0.41 & 0.05 & 0.48 & 0.13 \\
    2 & 0.38 & 0.18 & 1.16 & 1.11 & 0.04 & 0.0 & 0.1 & 0.01 \\
    3 & 2.0 & 1.33 & 8.58 & 7.73 & 0.02 & 0.0 &  0.1 & 0.03 \\
    4 & 1.67 & 0.96 & 6.82 & 5.66 & 0.01 & 0.0 &  0.06 & 0.01 \\
    5 & 2.09 & 1.04 & 8.53 & 6.36 &0.01 & 0.0 & 0.07 & 0.01 \\
    6 & 1.71 & 1.05 & 7.04 & 6.2 & 0.01 & 0.0 & 0.07 & 0.01 \\
    \hline
    \end{tabular}}
       \caption{        \label{table:mh_limit_effect_values} Percentage of correlated and shot noise (SN) power we miss due to the halo mass limit of the cosmological simulation we use. We give the values for this work (SIDES-Uchuu) and B22 (SIDES-Bolshoi) for both [CII] and CO(5-4) at different redshift.}
\end{center}
\end{table*}

The halo masses of the lightcone we obtained span almost 6 orders of magnitude at z=0, while the range gradually decreases to 3-4 orders of magnitude up to z=7. After comparing the halo mass functions of Uchuu with the theoretical ones from \cite{despali2016} at various redshifts, we found that the Uchuu lightcone is complete down to $\rm \sim 1.3 \times 10^{10} \, h^{-1}M_{\odot}$ at all redshifts. To investigate the effect of this halo mass limit, we computed the missing fraction of the correlated and shot noise power as a function of the galaxies host halo mass. We followed the same steps as in B22 (Sect. 4.4), where they estimate the contribution of galaxies to the shot noise and clustering of the [CII] power spectrum, as a function of their host halo mass. They do it by computing the sum of the luminosity squared and the product of the luminosity by the linear bias, respectively. Our results are summarized in Table \ref{table:mh_limit_effect_values}. The effect of the halo mass limit is small for both SIDES-Bolshoi and SIDES-Uchuu but it is even smaller in SIDES-Uchuu. This is because at $z \gtrsim 0.5$ the Bolshoi simulation becomes gradually less complete than Uchuu, even though at $z=0$ both simulations share a similar level of completeness.

The cosmological parameters in the Uchuu simulation are consistent with the results from \cite{planckcollaboration2020} and the initial redshift is set to 127. The rest of the initial conditions are generated using the parallel 2LPTic code\footnote{\url{http://cosmo.nyu.edu/roman/2LPT/}} \citep{crocce2006}. The GreeM code\footnote{\url{http://hpc.imit.chiba-u.jp/~ishiymtm/greem/}} \citep{ishiyama2009, ishiyama2012} has been used to track the gravitational evolution while the ROCKSTAR\footnote{\url{https://bitbucket.org/gfcstanford/rockstar/}} \citep{behroozi2013b} (sub)halo finder has been used for the identification of the position and velocity of each halo and subhalo. Then merger trees were constructed using the CONSISTENT TREES code\footnote{\url{https://bitbucket.org/pbehroozi/consistent-trees/src/main/}} \citep{behroozi2013a}.

The outputs of the cosmological simulation are saved as discrete snapshots which represent the evolution stage at different time steps and, hence, different redshifts. We used the different snapshots of the simulation box to create a 9 deg $\times$ 13.6 deg lightcone within the redshift range $\rm 0 < z < 7$, which corresponds to a comoving volume of 7.9 $\rm Gpc^3$. The total simulation area we have is 122\,deg$^2$ (9 deg $\times$ 13.6 deg) but the final exploitable simulation area is 117\,deg$^2$, because we cut the total area into smaller square subfields of 1\,deg$^2$. For the construction of the lightcone, we remapped the $\rm (2 \;  Gpc \: h^{-1} )^3$ cubical volume of the simulation, following the prescription described in \cite{carlson2010}. Taking advantage of the periodicity of the simulated box we broke it into cells, which were then translated by integer offsets to form cuboids. This remapping procedure keeps local structures intact, meaning that the structures inside each cuboid are not cut in half but they rather keep their continuity. We adjusted the remapping parameters in order to get a lightcone long enough ($\rm z\sim7$). Finally, for each snapshot, we remapped the positions of the halos and subhalos converting from the coordinates of the simulation box to the ones of the newly created lightcone.

\subsection{From dark matter halos to galaxies: abundance matching}
\label{subsec:AM}

Galaxies form in the gravitational potential wells of the dark matter halos. We thus need to connect the galaxies with their dark matter halos, which can be succeeded by abundance matching. This empirical technique assumes that there is a monotonical relation between some property of the galaxies and some property of the dark matter halos \citep{kravtsov2004, vale2004, conroy2006, shankar2006, behroozi2013, moster2013}. 

There are several proxies that can be used to apply the abundance matching method. In the case of galaxies it can be either the stellar mass or the luminosity. In the case of the dark matter halos it could be either the halo mass or the circular velocity. The evolutionary stage of the halo at which we compute these quantities can strongly affect the results. For instance, during a merging episode between two or more dark matter halos, their dark matter content gets stripped faster than their stars. For instance, \cite{niemiec2019} investigated the stripping timescale of the subhalos and satellite galaxies confirming the theoretically expected trend. Therefore, the usual technique is to place the N-th most massive galaxy in the halo with the N-th highest velocity (or highest halo mass).

The galaxy properties are strongly correlated with the halo properties before any stripping event. This is why the most suitable proxies for matching the abundances of the dark matter halos with those of the galaxies are either the $M_{\rm peak}$ or the $v_{\rm peak}$ \citep{behroozi2019, behroozi2020}. $M_{\rm peak}$ is the maximum halo mass throughout the entire past merging history of the halo and $v_{\rm peak}$ is the maximum circular velocity throughout the entire merging history of the halo.

In \cite{reddick2013} it is shown that $v_{\rm peak}$ gives better results at low redshifts. They reach to this conclusion after comparing the projected two-point correlation function of the SDSS catalog and the two-point correlation function of mock catalogs. The latter are constructed after populating the Bolshoi cosmological simulation \citep{klypin2011} with abundance-matched galaxies using several different proxies. It turns out that only the $v_{\rm peak}$-based abundance-matched galaxies can properly recreate the SDSS two-point correlation function. 

In order to investigate the effect of the selection of the proxy quantity on our results we used both quantities ($M_{\rm peak}$, $v_{\rm peak}$). While the abundance matching quantity can have an impact on the two-point correlation function, as described above, we show in Sect. \ref{subsec:vpeak_vs_mpeak} that this choice has no impact on the CIB anisotropies. We also checked and found that the impact of this choice on CO and [CII] intensity mapping power spectra is lower than 10\,\%, which is much less than the typical uncertainty between different line emission models.

While the observed SMF, which is used as a starting point in SIDES, is already corrected for the scatter between the true and the measured SMF, we still have to take into account that not all the galaxies are exactly on the relation between the proxy and the stellar mass. \citet{reddick2013} estimates that the scatter around this relation is $\sim$0.2\,dex. In the presence of scatter, the abundance matching is thus more complex. We used the approach of \citet{behroozi2010}, implemented in the \texttt{abundancematching} \footnote{\url{https://github.com/yymao/abundancematching}} python module developed by Y. Mao, to deal with the scatter. The code deconvolves the SIDES SMF by the 0.2\,dex log-normal scatter to obtain a mock SMF. Then it performs abundance matching using this mock SMF and obtains a proxy-$M_{\rm stellar}$ relation which is scatter-free. Finally, the stellar mass assigned to each dark matter (sub)halo is the value defined by the scatter-free proxy-$M_{\rm stellar}$ relation plus the 0.2\,dex log-normal scatter around this value.

The different outcomes of abundance matching with or without considering scatter are shown in Fig. \ref{fig:AM_scheme}. The mean of the black points coincides with the resulting $M_{\rm stellar} - v_{\rm peak}$ relation when using the virtual scatter-free SMF. One can notice that at high $v_{\rm peak}$ values the mean of the black points (blue dashed line) is below the $M_{\rm stellar} - v_{\rm peak}$ relation resulting after the direct abundance matching between the SIDES SMF and halo peak velocity function, which ignores the scatter. Compared to the virtual scatter-free SMF, the SIDES "true" SMF has a shallower massive end and implies a steeper relation.

We applied the SIDES method to derive the galaxy properties determined above, from the stellar masses and redshifts. The resulting simulated SIDES-Uchuu catalog is publicly available \footnote{\url{https://cesamsi.lam.fr/instance/sides/home}}.

\section{Validation of the model at large scales using CIB anisotropies}
\label{sec:model_validation}

The original SIDES-Bolshoi simulation \citep[][]{bethermin2017} is validated by comparing several observables with real data, like the observed continuum number counts from the mid-infrared to the millimeter and the \textit{Herschel} data of CIB anisotropies. The latter aims to validate the ability of SIDES to reproduce the clustering of galaxies at intermediate scales ($\lesssim$1\,deg). However, the much larger simulation area of SIDES-Uchuu ($\rm 117\, deg^2$ compared to $\rm 2\, deg^2$ for SIDES-Bolshoi) offers a more robust validation of the simulation recipes at large scales.

The clustering of SIDES is validated by comparing it with data of individually detected dusty galaxies \citep{cooray2010, maddox2010}. However, measuring the clustering of individual galaxies can be problematic due to the confusion impacting the source extraction at these wavelengths. Because of this difficulty, the measurements disagree with each other depending on the source extraction technique. It is thus impossible to validate a model using the measured two-point correlation function. The CIB power spectrum, though, does not suffer from such limitations. Thus, comparing our model with the CIB power spectra anisotropies offers a more accurate model validation.

\subsection{Simulated maps}
\label{subsec:simulated_maps}

The simulated catalogs created by SIDES contain a comprehensive set of information for each galaxy, like its position on the sky, its redshift and its flux in the bandpasses of given experiments. Using the information of the simulated catalogs we can generate maps at different wavelengths, for example, \textit{Spitzer}, \textit{Herschel}, NIKA2, \textit{Planck}-HFI. For more details on the map generator we refer to B22. The generated maps include the emission of the sources at low angular resolution (in the confusion-limited regime). They thus simulate the CIB, which is the cumulative emission of all dusty galaxies along every line of sight.

\begin{figure}[h]
\begin{center}
\includegraphics[width=\linewidth]{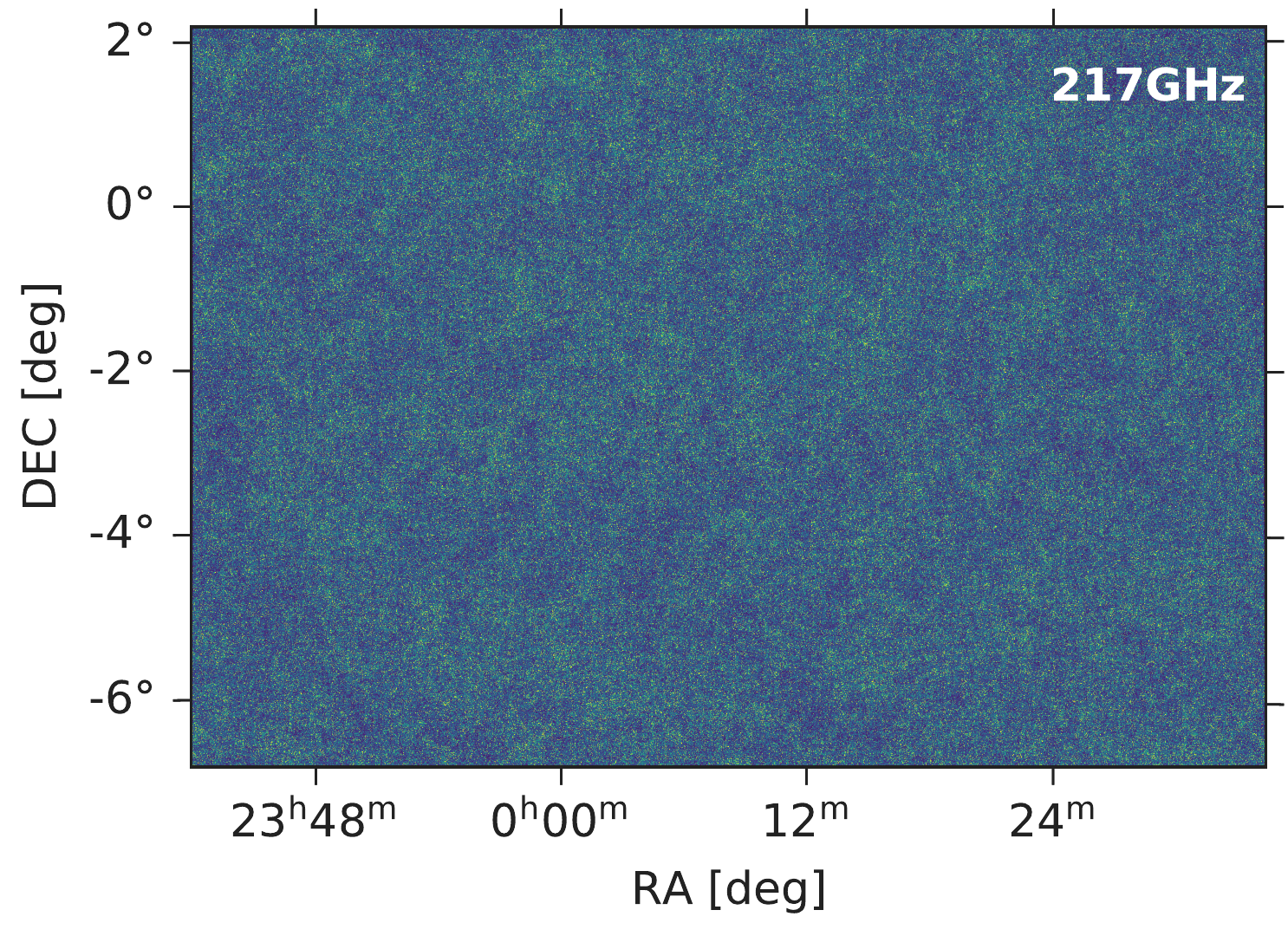}
\caption{Simulated $\rm 13.6^o \times 9^o$ CIB \textit{Planck} map at 217\,GHz. Sources with flux higher than 225\,mJy have been removed to mimic the analysis of CIB anisotropies of \cite{planckcollaboration2014}.}
\label{fig:Planck217_map}
\end{center}
\end{figure}

We generated maps for the \textit{Herschel}/SPIRE three frequency bands, 600, 857, and 1200\,GHz \citep{amblard2011, viero2013, thacker2013} as well as for the \textit{Planck} high frequency bands, at 217, 353, 545, and 857\,GHz \citep{planckcollaboration2011}. In Fig.\,\ref{fig:Planck217_map} we show an exemplary simulated map for \textit{Planck} at 217\,GHz. 

\subsection{Power spectrum estimate from the model and comparison with observations}
\label{subsec:maps_pk_estimation_color_corrections}

We computed the auto power spectrum of each of the maps mentioned in Sect.\,\ref{subsec:simulated_maps} as well as the cross power spectra for all the combinations of bands that have observational constraints. We compared the model with CIB observational data from \textit{Planck} and \textit{Herschel} in order to assess how well SIDES can reproduce the clustering of the galaxies. For the power spectra computation we used the \texttt{powspec}\footnote{Public code by Alexandre Beelen hosted at \url{https://zenodo.org/record/4507624#.YTiIfyZR3mE}} python package, which computes the square of the Fourier transform of the map and averages it in k-scale.

The randomly distributed bright sources cause a significant enhancement of the Poisson component (shot noise) in the observed power spectrum. Hence each experiment masks the bright sources by applying certain flux cuts, which depend on the frequency. The CIB power spectrum data we use were observed in various filters using different flux cuts. Thus in order to consistently compare our model with the observations we masked the bright sources in the same way. For simplicity, we excluded from the simulated catalogs all the sources with fluxes above the given flux cut and then generated the maps, instead of applying a mask on the already generated map. The discrepancy between the two methods in the case of a diffuse field like the CIB emission, is negligible (more details about the effect of masking in Van Cuyck et al. in prep.). For \textit{Planck} the flux cuts are 225, 315, 350, and 710\,mJy at 217, 353, 545, and 857\,GHz, respectively \citep{planckcollaboration2014} while for \textit{Herschel}/SPIRE we used two sets of simulated data, that follow the two different masking techniques used by the corresponding observational data sets. \citet{viero2013} mask all the sources above 300\,mJy in the analyzed band (method m1), while \citet{viero2019} mask sources brighter than 300\,mJy at 1200\,GHz whatever the analyzed band (method m2).

Subsequently, in order to compare the model with the data, we had to apply the appropriate color corrections. We multiplied our model with a factor to convert from the theoretical SED flux to the measured flux that follows the $\rm \nu I_{\nu} = $ constant convention \citep[][Appendix A]{lagache2020}. The color correction factors for all the bands are given in Table\,\ref{table:cc}.

\begin{table}
\caption{Color corrections factors (cc) used to multiply the SIDES model before comparing with observations: $\rm C_{\ell,\nu,\nu'}^{data} = cc_{\nu} \times cc_{\nu'} \times C_{\ell,\nu,\nu'}^{model}$. }
\label{table:cc}
\centering
\begin{tabular}{c c c}
\hline\hline 
band & $\rm cc_{\nu}$ & ref. \\
\hline 
\textit{Planck} 217 GHz & 1.119 & a \\
\textit{Planck} 353 GHz & 1.097 & a \\
\textit{Planck} 545 GHz & 1.068 & a \\
\textit{Planck} 857 GHz & 0.995 & a \\
SPIRE 600 GHz & 0.9739 & b \\
SPIRE 857 GHz & 0.9887 & b \\
SPIRE 1200 GHz & 0.9880 & b \\
SPT 150 GHz & 1.1411 & b \\ %Table A.1.
SPT 220 GHz & 1.0059 & b \\ %Table A.1.
\hline
\end{tabular}
\footnotesize{\\a: \cite{planckcollaboration2014} b: \cite{lagache2020}}
\end{table}

\textit{Planck} used a planet model to calibrate the 545\,GHz and 857\,GHz channels. However, for 545 GHz the planet calibrations agreed with the CMB dipole calibrations within 1.5\%. Given the relative calibration errors between the two filters, the absolute calibration for 857 GHz is given to be within 2.5\% of the CMB calibration \citep{planckcollaboration2016}. The absolute calibration uncertainty of \textit{Herschel} for 600 and 857\,GHz is 8\% \citep{viero2013}. Therefore we consider \textit{Planck}'s highest calibration accuracy as the reference. We thus correct the \textit{Herschel}/SPIRE data by dividing with the 1.047 and 1.003 recalibration factors for 600 and 857\,GHz, respectively, as given in \cite{bertincourt2016}. In the end, when comparing the model with the \textit{Herschel} data, we added on top of the SIDES model the \textit{Planck} absolute calibration uncertainty mentioned above, for 545 and 857\,GHz. There is no recalibration factor for the 1200\,GHz band in the literature to correct the data. We have thus added to the SIDES model the 8\% given uncertainty of \textit{Herschel}.

\subsection{The impact of the abundance matching proxy selection}
\label{subsec:vpeak_vs_mpeak}

We followed the abundance matching procedure as explained in Sect.\,\ref{subsec:AM}, using both $v_{\rm peak}$ and $M_{\rm peak}$ as matching quantities. Subsequently, we created one \textit{Planck} map for each quantity and computed their power spectra, respectively.

In Fig.\,\ref{fig:Planck_pk_maps} we show the SIDES models for $v_{\rm peak}$ and $M_{\rm peak}$. In all the different filters the two models almost coincide with each other. They agree at better than 5\%. This proves that the choice between these two quantities has a negligible effect, at least in the context of this work. Following the argument presented in \cite{reddick2013} and taking into account the low impact of this choice, we chose the $v_{\rm peak}$ as proxy. All the results presented hereafter are obtained using this quantity.

\subsection{Comparison with the data}
\label{subsec:cib_pk_model_vs_data}

\begin{figure*}
  \centering
  \includegraphics[width=\linewidth]{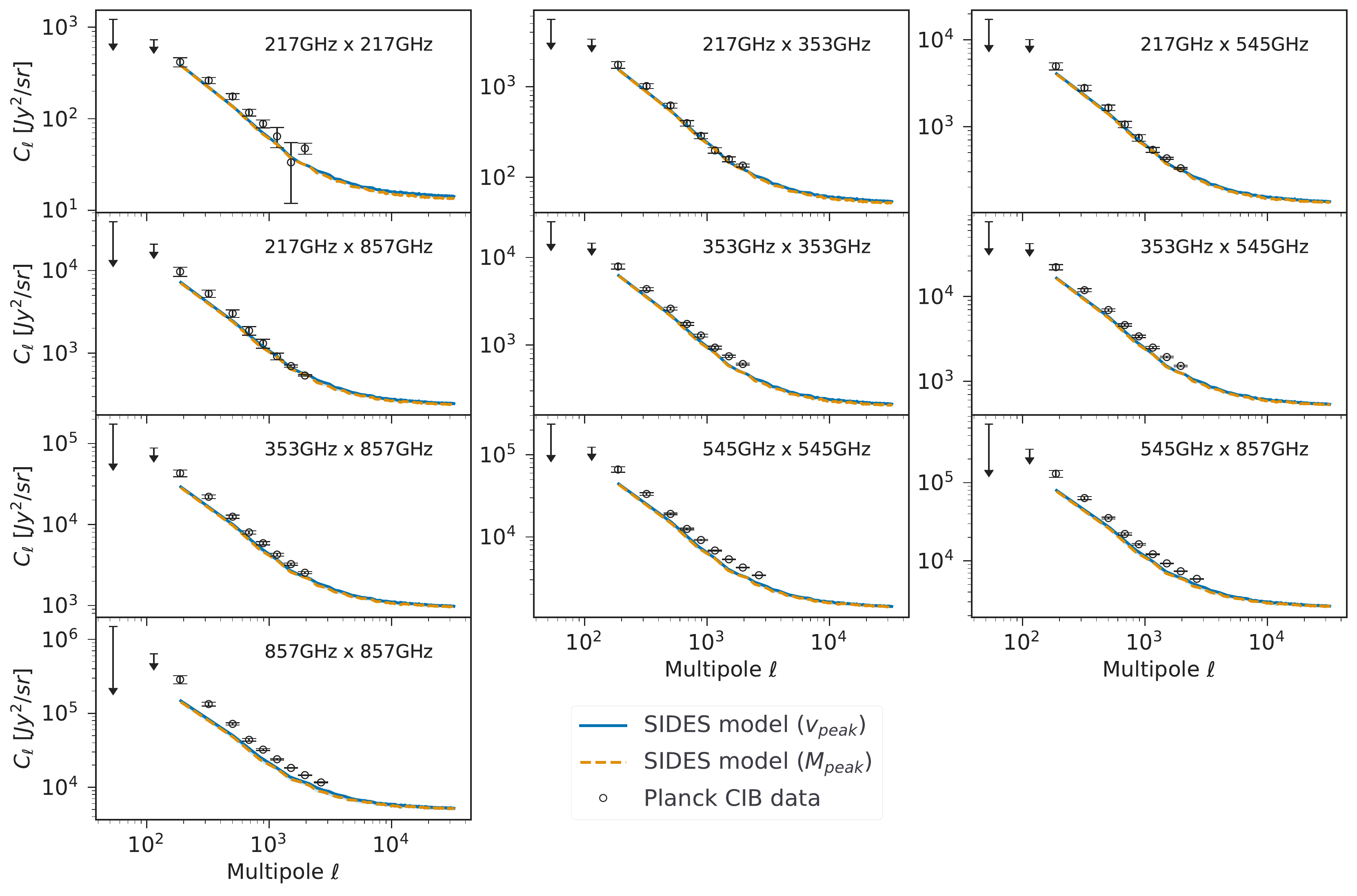}
  \caption{Comparison of SIDES and measured CIB power spectra for \textit{Planck}. The solid blue lines represent the SIDES power spectra for the \textit{Planck} bandpass, using $v_{\rm peak}$ as proxy for the abundance matching. The dashed orange line is the power spectrum obtained using the $M_{\rm peak}$ as proxy. The black points are the observational data from \textit{Planck} \citep{planckcollaboration2014}.}
  \label{fig:Planck_pk_maps}
\end{figure*}
\begin{figure*}
  \centering
  \includegraphics[width=\linewidth]{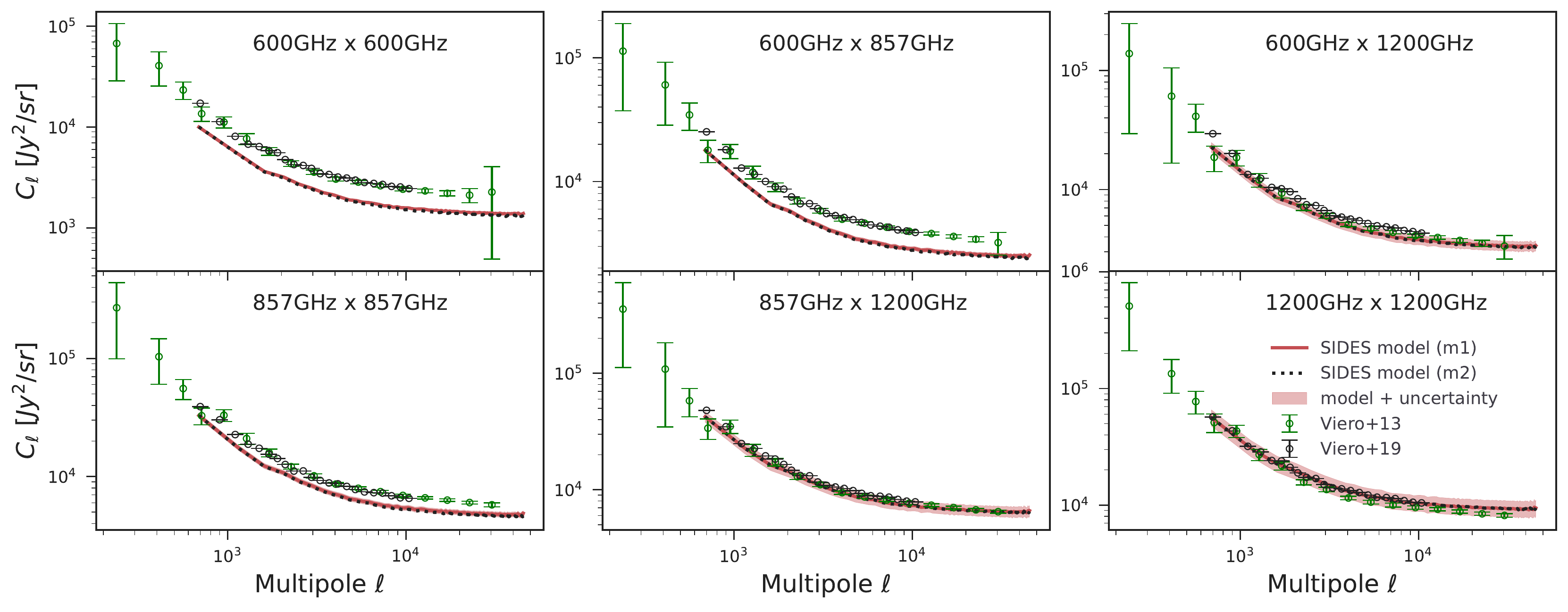}
  \caption{Comparison of the SIDES power spectrum model with \textit{Herschel}/SPIRE data. The solid red line corresponds to the SIDES model where we have followed the masking technique of \cite{viero2013}, noted as m1. On top of the SIDES models we illustrate the corresponding systematic uncertainty (i.e., absolute calibration) as described in Sect\,\ref{subsec:maps_pk_estimation_color_corrections} (red shaded area). The dotted black line is the SIDES model following the masking technique of \cite{viero2019}, noted as m2. The green points are the corresponding observational data presented in \cite{viero2013} while in black the data from \cite{viero2019}. }
  \label{fig:spire_pk_maps}
\end{figure*}

The comparison between the \textit{Planck} data and the SIDES model is shown in Fig.\,\ref{fig:Planck_pk_maps}. The black dots represent the measurements of the CIB anisotropy power spectrum \citep{planckcollaboration2014} and the blue lines the model of the power spectrum resulting from the SIDES simulation for the corresponding frequency bands. We also included the absolute calibration uncertainty as a shaded area on top of the model line, but it is not visible since it is almost negligible (1.5\% and 2.5\% for 545 and 857\,GHz, respectively). In each panel we show the data and model of the cross power spectra between different bands where we can see that there is an overall agreement. In the auto and cross power spectra that involve the 217 GHz band, the disagreement between the model and the CIB data is $\le$20\%. This is also the case for the rest of the auto and cross power spectra, except for the lowest $\rm \ell$ data points and the 857 $\times$ 857\,GHz case where the discrepancy is up to a factor of 2.

The comparison between the \textit{Herschel}/SPIRE data \citep{viero2013, viero2019} and the SIDES model is shown in Fig.\,\ref{fig:spire_pk_maps}. The model is able to adequately predict the power spectrum, except at 600\,GHz where there is a systematic offset. The level of disagreement in the cases of 600$\times$1200\,GHz, 857$\times$1200\,GHz, and 1200$\times$1200\,GHz is lower than 10\% in both the large and small scales. However, the discrepancy between the model and the data for 600$\times$600\,GHz, 600$\times$857\,GHz, and 857$\times$857\,GHz is 35\%, 25\%, and 16\%, respectively. From that we can conclude that the model at 1200\,GHz can reproduce the data to high accuracy while the lower the frequency the larger the discrepancy.

Previous studies in the literature that measured and modeled the CIB anisotropies have spotted inconsistencies between the data and the fitted models as well as between the different experiments. \cite{lagache2020} compare the shot noise power of the CIB anisotropies and the model-based predicted values both for \textit{Planck} and \textit{Herschel}. The expected behavior is to obtain a lower shot noise level when masking deeper. However, the \textit{Herschel}/SPIRE measurements at the 600, 857, and 1200\,GHz bands are incompatible with this expected behavior. The shot noise derived with a higher flux limit is lower than that derived with a lower flux limit, and the model used for the prediction is not systematically higher or lower than the measurements, preventing any robust conclusion. Moreover, \cite{maniyar2021} compare the measurements of the CIB power spectra taken from \textit{Planck} and \textit{Herschel} as well as cross power spectra measurements between bands of the same experiment. The power spectra of \textit{Planck} and \textit{Herschel} agree well on the large angular scales but they are clearly different on small angular scales. Since SIDES agrees with \textit{Planck} but not with Herschel at the same frequencies, it is possible that the discrepancy between SIDES and \textit{Herschel} comes from measurement systematics.

We also investigated the agreement between our model and South Pole Telescope (SPT) power spectra as well as SPT $\times$ SPIRE cross power spectra. The results are discussed in detail in Appendix \ref{ap:spt_and_spt_x_spire}.

The overall agreement between the SIDES model and the observational data taken from both the \textit{Planck} and \textit{Herschel} instruments, especially on the large scales, shows that the SIDES simulation can realistically reproduce the clustering of galaxies as revealed from their continuum emission. Hence, it is a reliable tool to use in order to simulate the CONCERTO \citep{theconcertocollaboration2020} observations and develop the needed tools to post process and interpret the observational data.

\section{Line luminosity functions}
\label{sec:LFs}

The observed LF provides important constraints to test our simulation. In Sect.\,\ref{subsec:validating_LFs}, we compare the LFs from our model with observational data. This allows us to validate the empirical relations embedded in SIDES-Uchuu to generate the emission of the different spectral lines. In Sect.\,\ref{subsec:cv_LFs} we investigate the field-to-field variance introduced in the LF and its dependence on survey size, while in Sect.\, \ref{subsec:cv_gas_mass} we study how this variance propagates through to the molecular gas density. Finally, in Sect.\, \ref{subsec:script_description_lf_rho_mol} we present a public tool for the computation of the LFs and the molecular gas variance based on our simulation.

\subsection{Validation of the spectral line recipes introduced in SIDES}
\label{subsec:validating_LFs}

In order to validate the newly added recipes for the spectral lines in SIDES (B22) we compared the CO and [CII] LFs with observational data provided by recent surveys. This test has already been carried out in B22 for a single simulated field of 2\,$\rm deg^2$. In this work we extended the validation to our simulated field of 117\,$\rm deg^2$ total size, thanks to the Uchuu cosmological simulation. This allowed us to test the validity of our simulation at a higher precision and test if the field-to-field variance could explain some discrepancies.
 
\begin{figure*}[h]
\begin{center}
\includegraphics[width=0.85\textwidth]{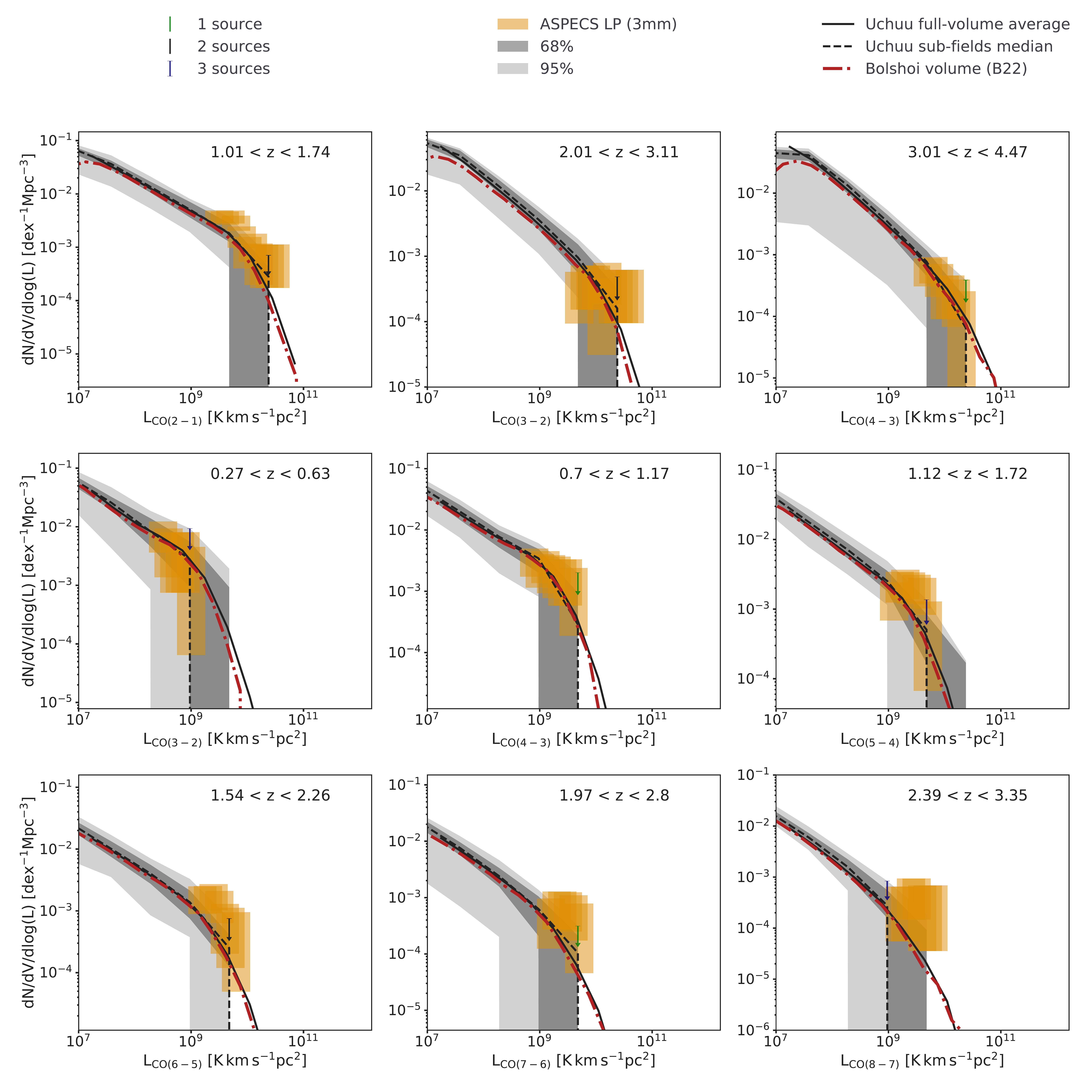}
\caption{Comparison of the CO LF resulting from the SIDES simulation with ASPECS observational data. Each LF is created using sources from 117 different $\rm 4.6 \, arcmin^2$-sized simulated subfields. This field size is chosen to match the size of ASPECS. The gray shaded areas with different transparencies correspond to the 16th-84th and 5th-95th percentile confidence intervals. The black solid line is the LF of the entire volume of the Uchuu simulation, and the dashed line is the median of the multiple LFs computed from all the subfields. The red line shows the resulting LF presented in B22 where the Bolshoi-Planck cosmological simulation was used, while the arrows show the last luminosity bin of the SIDES-Uchuu LFs that contains at least one source. The different colors stand for a different number of sources.}
\label{fig:CO_LF}
\end{center}
\end{figure*}

\subsubsection{CO luminosity functions}
\label{subsubsec:CO_LFs_results}

Millimeter interferometers provide rich constraints on various CO transitions. We used data from the ALMA spectroscopic survey in the Hubble Ultra Deep Field (ASPECS), which spans a $\rm 4.6 \, arcmin^2$ region and covers band 3 (84–115\,GHz) and 6 (212–272\,GHz) \citep{decarli2019, decarli2020}. The band 3 and band 6 windows correspond to different redshift ranges for each CO line. This offers a wide variety of constraints on the various CO transitions and redshifts.

We cut the simulated catalogs into subfields of 4.6\,arcmin$^2$ which is the size of the ASPECS survey. We then computed individually for each catalog the number of galaxies per comoving volume using the same redshift ranges as ASPECS \citep{decarli2019, decarli2020}. We compared all the resulting CO LFs with the ASPECS data and the LF of the total simulated field.

We show in Fig.\,\ref{fig:CO_LF} the LF constructed from the entire Uchuu field and the median of the multiple LFs from the 4.6\,arcmin$^{-1}$ Uchuu subfields together with the ASPECS data. The median LF sharply drops to zero faster than the LF of the entire Uchuu in all redshift cases. The bright sources are less common and thus the majority of the subfields have zero sources above a certain luminosity bin. The last nonzero luminosity bin is denoted with an arrow in Fig.\,\ref{fig:CO_LF}. However, this is not the case for the entire Uchuu field where there are more bright sources.

The LF from entire Uchuu field and the SIDES-Bolshoi LF agree very well at the faint-end and the knee. There are only some discrepancies ($\lesssim 20$\%) at the bright-end of the LF at some redshift slices (e.g., $2.01<z<3.11$, $0.27<z<0.63$, $0.7<z<1.17$). We computed the field-to-field variance, constructing multiple LFs from 2 $\rm deg^2$ Uchuu subfields and found that the SIDES-Bolshoi LF (red line in Fig.\,\ref{fig:CO_LF}) still lies inside the 2\,$\sigma$ range. We could thus conclude that the offset between the two LFs is just a statistical fluctuation.

Despite the significant uncertainty introduced in the LF by the small ASPECS survey size, the SIDES-Uchuu LFs are still within the 1\,$\rm \sigma$ confidence level of the observations. However, at the higher CO transitions ($\rm J_{upper} \geqslant 6$) the model appears to be systematically lower as also found in B22 for SIDES-Bolshoi. Considering the large Uchuu area, we can rule out that this trend is caused by an underdensity of the SIDES-Bolshoi simulation. This shift at high-J could be explained by the contamination of the ASPECS measurements by interlopers. This might be also related to the physics of the high-J CO emission. Toward high-z the interstellar medium of galaxies evolves, and in general it gets more turbulent and overall warmer. There are hints of a shift of the peak of the CO SLED toward higher J \citep{vallini2018}. We take this effect into account by linking the CO(5-4)/CO(2-1) ratio to $\langle U \rangle$, as mentioned in Sect.\,\ref{subsec:sides}. But this correlation is probably too weak to boost the high-J CO emission. The shift could also be due to the correlation between the luminosity bins, since ASPECS uses a sliding binning. Therefore, in the case of an over or underdensity the LF value at all the bins would be higher or lower, respectively. We also find that this effect also stands for nonoverlapping bins (see Sect.\,\ref{subsec:LF_bins_correlation}). However, the fact that the systematic shift of the model is always toward lower values is intriguing and makes it difficult to conclude.

\subsubsection{[CII] luminosity functions}
\label{subsubsec:CII_LFs_results}

Measuring the [CII] LF is more challenging compared to CO. The [CII] sources observed in the submm/mm atmospheric window reside at higher redshifts, therefore their counterparts can be very faint and difficult to observe. It is thus difficult to identify [CII], when a single line is detected. On top of that the number density of [CII] sources is rather small and larger fields are required. However, the ALPINE survey \citep{lefevre2020, bethermin2020, faisst2020} managed to target with ALMA the [CII] lines in 118 $4.4<z<5.9$ normal star forming galaxies and put further constraints on the [CII] LF \citep{loiacono2021, yan2020}.

For the computation of the [CII] LFs we followed the same steps as in CO. However, in this case we could not use the field size of the survey because ALPINE targeted selected galaxies from both the cosmic evolution survey (COSMOS) field and the Chandra deep field south (CDFS). This ALPINE sample is biased toward UV selected galaxies with well-defined spectroscopic redshifts \citep[see][for a discussion on the biases]{faisst2020}. Most of the ALPINE galaxies (89\,\%) are in the COSMOS field and they dominate the statistics. We thus chose to use the COSMOS size in our estimate. We computed the [CII] LFs at $\rm 4.4 < z < 4.6$ and $\rm 5.3 < z < 5.9$, which are the redshift ranges observed by ALPINE.

\begin{figure}[h]
\begin{center}
\includegraphics[width=\linewidth]{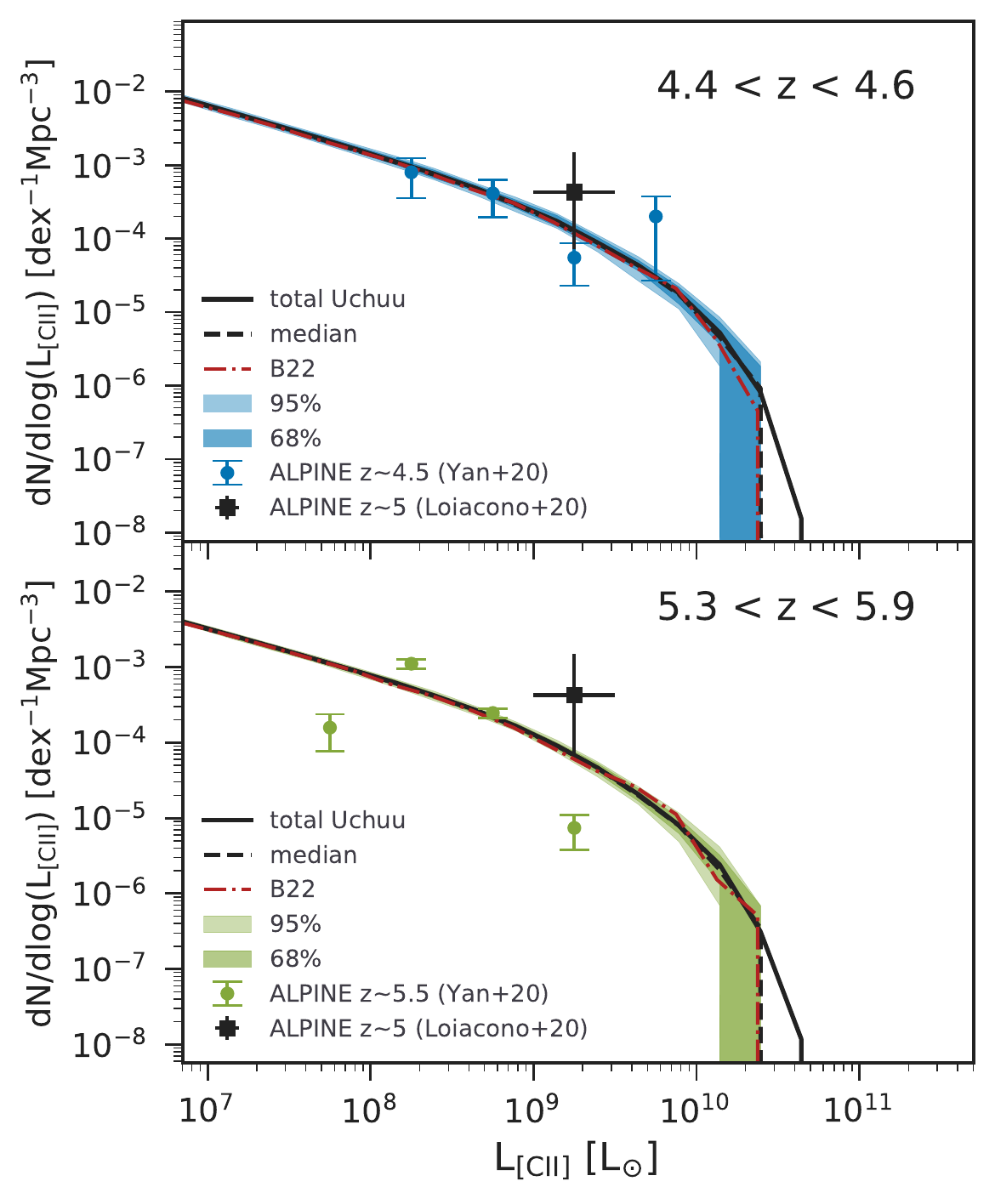}
\caption{Comparison of the SIDES [CII] LFs with ALPINE observations. Top panel: ALPINE data at $\rm z \sim 4.5$, bottom panel: ALPINE data at $\rm z \sim 5.5$. In both panels, the black solid line corresponds to the total Uchuu field LF and the dashed line corresponds to the median LF of 54 2 $\rm deg^2$ SIDES-Uchuu subfields. The shaded areas are the 16th-84th and 5th-95th percentile intervals while the red line is the SIDES-Bolshoi LF (B22). The blue and green points are the data from \cite{yan2020}, while the black square is the data point from \cite{loiacono2021}.}
\label{fig:CII_LFs}
\end{center}
\end{figure}

We show in Fig.\,\ref{fig:CII_LFs} a comparison of the SIDES-derived [CII] LFs with the ALPINE data at $\rm z \sim 4.5$ and $\rm z \sim 5.5$. In both cases we have also included the resulting LF of the SIDES-Bolshoi (B22) for comparison. Similarly to the CO LFs, the median LF drops to zero due to the lower number of bright sources included in the 2\,$\rm deg^2$-sized subfields compared to the total Uchuu field. The total Uchuu LF and Bolshoi LF perfectly agree with each other up to $\sim 10^{10} \, L_{\rm \odot}$ for both redshift slices, while at the bright end the agreement remains better than 2\,$\sigma$ of the field-to-field variance (shaded area). Similarly to the CO LFs and because of the much larger volume, the Uchuu LF goes to higher luminosities while Bolshoi is too small to contain such bright sources.

Similarly to the SIDES-Bolhoi LFs, the SIDES-Uchuu LFs agree at $\rm \sim 1 \, \sigma$ with the ALPINE measurements at $\rm z \sim 4.5$, except for the highest luminosity point. It is most probable that this is a statistical fluctuation since this point is derived from just two ALPINE objects \citep{yan2020}. At $\rm z \sim 5.5$ there is an overall agreement between $10^8$ and $10^9$ $L_{\odot}$, while the highest and the lowest luminosity points are lower than the simulation by $\rm 2 \, \sigma$. In the case of the faintest point, it could be explained by the incompleteness of detection of the ALPINE survey at lower luminosities, as discussed in \cite{yan2020}. The highest luminosity point may be affected by small number statistics at the bright end or a bias of the ALPINE sample against the most dusty and potentially [CII]-luminous galaxies. Finally, our simulation agrees at $\rm \sim 1 \, \sigma$ with the \cite{loiacono2021} measurement, which was taken from blind ALPINE detections and even though it is less precise (a few objects), it is less sensitive to assumptions or systematic effects.

The \cite{yan2020} LF measurements serve as a lower limit because they are obtained using a UV-selected sample. The survey targets only sources bright enough in the UV, possibly excluding bright [CII] sources that are faint in the UV. On the other hand, the LF measurements of \cite{loiacono2021} are obtained using the serendipitous sources of the ALPINE survey, that are strongly affected by the clustering. Their data point is therefore considered an upper limit. Our simulation is between the lower and upper limits offered by these measurements.

\subsection{Variance of line luminosity functions}
\label{subsec:cv_LFs}

Comparing the LFs of Fig.\,\ref{fig:CO_LF} (gray areas) and Fig.\,\ref{fig:CII_LFs} (blue and green area) we see, as expected, that smaller survey sizes lead to more uncertain LFs. It is common practice to estimate the errors of the LF using only Poisson statistics. However, clustering can lead to much larger uncertainties. Thanks to the large volume offered by the Uchuu cosmological simulation we can properly study the contribution of the clustering in the field-to-field variance.
 
 We cut the total exploitable Uchuu field  (117\,$\rm deg^2$) in multiple smaller subfields of the requested size. We constructed 12 sets of LFs from 12 different subfield sizes, respectively. The selected sizes are 0.0003, 0.0013, 0.0069, 0.0312, 0.125, 0.25, 0.5, 1, 2, 4, 8, and 16 $\rm deg^2$. For all the fields of size below $\rm 1 \, deg^2$ we could create a large number of subfields given the total Uchuu field. However, in order to save computational time we only used 117 out of the total subfields, by randomly selecting a smaller region out of each 1 $\rm deg^2$ subfield. For the fields with sizes of 1, 2, 4, 8, and 16\,$\rm deg^2$ we created 117, 54, 24, 12, and 6 subfields. Above 8 $\rm deg^2$ we had a low number of fields and thus poorer statistics. Furthermore, the subfields with size $\ge \rm 1 \, deg^2$ were contiguous, and thus not fully independent which could lead to an underestimation of the variance because of the large-scale modes. Finally, for each subfield size we computed the mean value ($\rm \mu_{LF}$) and standard deviation ($\rm \sigma_{LF}$) of each luminosity bin. We used the mean and not the median value because, as discussed above, the mean value does not drop to zero as fast as the median. We could thus define the relative variance ($\rm \sigma_{LF} / \mu_{LF}$) for a wider luminosity range.

The observed field-to-field variance in the LFs can be modeled as the combination of two components. The first one is the Poisson component, which is caused by the fluctuations of the number of sources in a given volume. The LF is the number of sources (N) per luminosity bin per unit of comoving volume. The mean value and standard deviation of the LFs from various subfields will thus be proportional to the Poisson mean and standard deviation, respectively, that is, $\rm \mu_{LF} \propto N$ and $\rm \sigma_{LF} \propto \sqrt{N}$. By dividing these two quantities, the luminosity bin size and comoving volume cancel out, and so $\rm \sigma_{LF} / \mu_{LF} = 1 / \sqrt{N}$. The larger the volume the higher the number of sources leading to lower relative variance. The second component is caused by the clustering, which practically describes the excess of probability to find a source next to another compared to the Poisson distribution. The clustering is linked to the distribution of the dark matter halos since galaxies are formed inside the halos. The total relative LF variance, which is the quadratic sum of the two, and can be modeled by

\begin{equation}
    \rm \frac{\sigma_{LF}}{\mu_{LF}} = \sqrt{y + \frac{1}{N}},
    \label{eq:s_mu_poisson_clustering}
\end{equation}
where $y$ is a term depending on the geometry of the field and the angular correlation function. It can be computed using
\citep{blake2002, bethermin2010}:
\begin{equation}
    \rm y = \frac{\int_{field} \int_{field} w(\theta) \,d \Omega_1 \,d \Omega_2}{\Omega ^2},
    \label{eq:y_factor}
\end{equation}
where $\rm w(\theta)$ is the angular two-point auto correlation function of the sources and $\Omega$ the angular size of the field. The $y$ term and thus the contribution of clustering to the relative uncertainties is proportional to the amplitude of the correlation. Also, it depends on the size of the field for a given shape. Indeed, $y$ is the average value of the correlation function for two points selected randomly in the field. Since usually $w(\theta)$ decreases with increasing $\theta$, $y$ will be smaller for wider fields. Both the relative Poisson (1/N) and clustering ($y$) uncertainties decrease with increasing field size but in a different way. There is thus a competition for the most dominant component at each field size. In order to investigate and compare these different trends, we modeled the variation of the relative uncertainties only as a function of the field size $\Omega$. The number of sources is linked to the survey size through the number density of galaxies $\rm \rho$:
\begin{equation}
\rm N = \rho \Omega.
\label{eq:n_rho_omega}
\end{equation}
The dependence of $y$ on the survey size is complex but in the case of a power law $\rm w(\theta) = A \,  \theta^{1-\gamma}$, \cite{blake2002} demonstrate that:
\begin{equation}
y \propto \Omega ^ {(1 - \gamma)/2}.
\label{eq:y_factor_size}
\end{equation}

The factor A depends on the flux and wavelength \citep{bethermin2010}. It is hence different for each luminosity and CO transition at a given redshift for simplicity. We thus considered it a constant for each luminosity and CO transition. Therefore, combining Eq. \ref{eq:s_mu_poisson_clustering} and Eq. \ref{eq:y_factor}, we get:

\begin{equation}
    \rm \frac{\sigma_{LF}}{\mu_{LF}} = \sqrt{A^{\prime} \Omega^{(1 - \gamma)/2} + (\rho \Omega)^{-1}},
    \label{eq:lf_variance_final}
\end{equation}
where $A^{\prime}$ is another constant derived from the integral in Eq. \ref{eq:y_factor} and proportional to $A$. The resulting best-fit $A^{\prime}$ and $\gamma$ values are summarized in Table\,\ref{table:Aprime_and_gamma_values}.

\begin{table}
\caption{Best-fit parameters of Eq.\,\ref{eq:lf_variance_final}.}
\label{table:Aprime_and_gamma_values}
\centering
\begin{tabular}{c c c}
    \hline\hline 
    $L_{\rm bin}$ & $A^{\prime}$ & $\gamma$ \\
    ($\rm K \, km \, s^{-1} pc^2$) & & \\
    \hline 
    $10^7$ & 0.003 & 1.67 \\
    $2.8 \times 10^7$ & 0.004 & 1.71 \\
    $7.7 \times 10^7$ & 0.005 & 1.67 \\
    $2.2 \times 10^8$ & 0.005 & 1.77 \\
    $6 \times 10^8$ & 0.006 & 1.71 \\
    $1.7 \times 10^9$ & 0.006 & 1.82 \\
    $4.6 \times 10^9$ & 0.01 & 1.68 \\
    $1.3 \times 10^{10}$ & 0.009 & 2.01 \\
    $3.6 \times 10^{10}$ & 0.01 & 1.28 \\
    \hline
\end{tabular}
\end{table}

The Poisson term $\rm (\rho \, \Omega)^{-1}$ was directly computed from the number of sources in the simulated catalog (Eq. \ref{eq:n_rho_omega}), where $\rm \rho$ was directly computed from our catalog. For the clustering term, $\rm A' \Omega^{(1 - \gamma)/2}$, we fit to our simulated data the parameterized function of Eq. \ref{eq:lf_variance_final} to get the parameters $A^{\prime}$ and $\rm \gamma$ that define the contribution of the clustering.

\begin{figure}[h]
\begin{center}
\includegraphics[width=\linewidth]{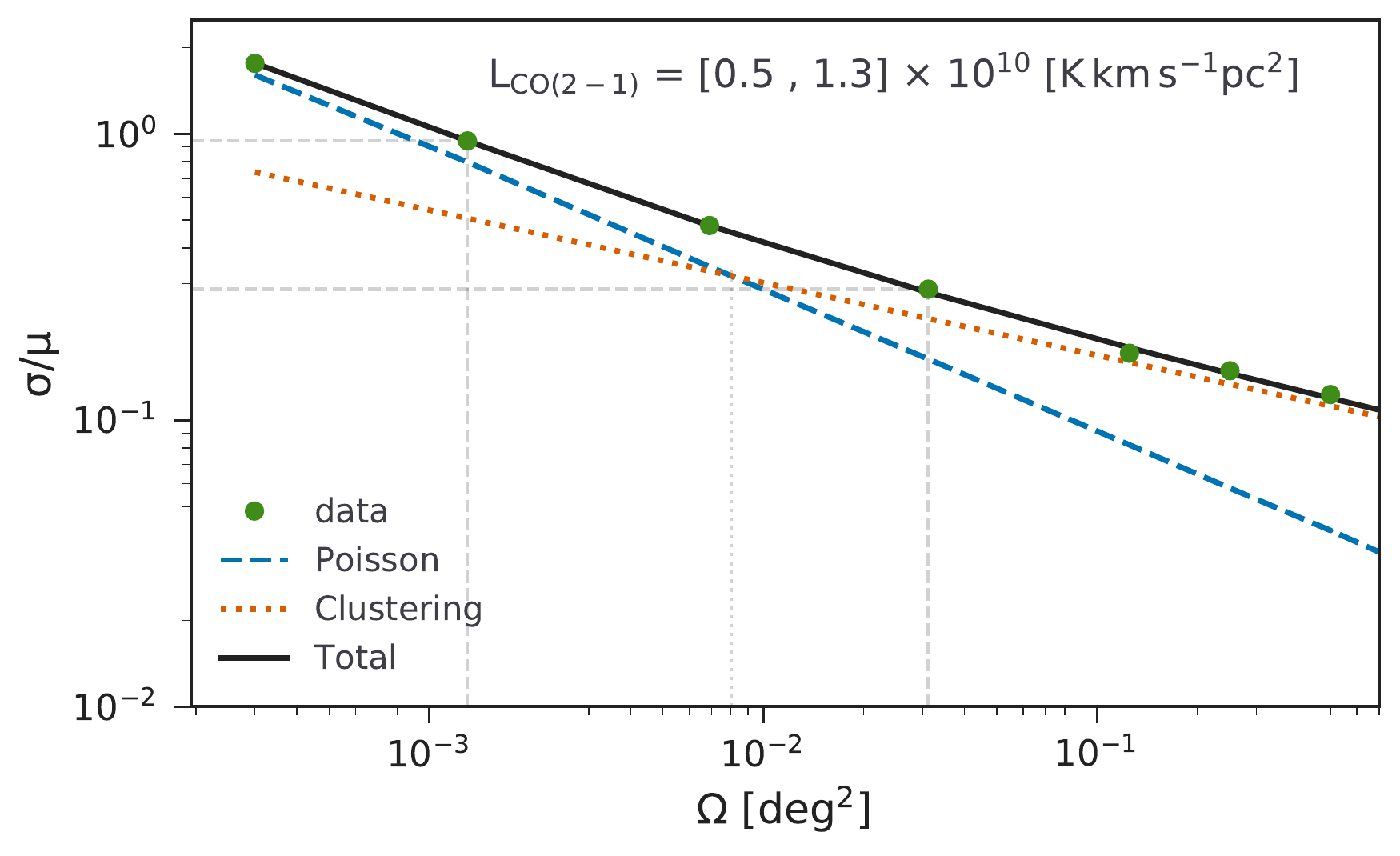}
\caption{Total relative variance and its decomposition into the Poisson and clustering components for the luminosity bin $\rm L_{CO(2-1)} = [0.5, 3.6] \times 10^{10} \, [K \, km \, s^{-1} pc^2]$.}
\label{fig:LF_size_std_one_lbin}
\end{center}
\end{figure}

We show as an example in Fig.\,\ref{fig:LF_size_std_one_lbin} the total variance of the CO(2-1) LF for the luminosity bin $\rm L_{CO(2-1)} = [0.5, 1.3] \times 10^{10} \, [K \, km \, m^{-1}pc^2]$ as a function of survey size at $z$ = 1.2 - 1.6. As expected, the total relative uncertainty increases when the survey size decreases, by a factor of 3 from 0.0312\,$\rm deg^2$ (112\,$\rm arcmin^2$) to 0.0013\,$\rm deg^2$ (4.6\,$\rm arcmin^2$), while we would have expected a factor of 5 for pure Poisson behavior. We also show the decomposition of the total variance in the Poisson and the clustering component. The larger the survey size the more significant the contribution of the clustering component becomes. Equality is reached for a field size of 0.008\,deg$^2$ (28.8\,arcmin$^2$). Beyond this size, the clustering of the galaxies is responsible for larger uncertainties than what is expected from the Poisson law.

\begin{figure}[h]
\begin{center}
\includegraphics[width=\linewidth]{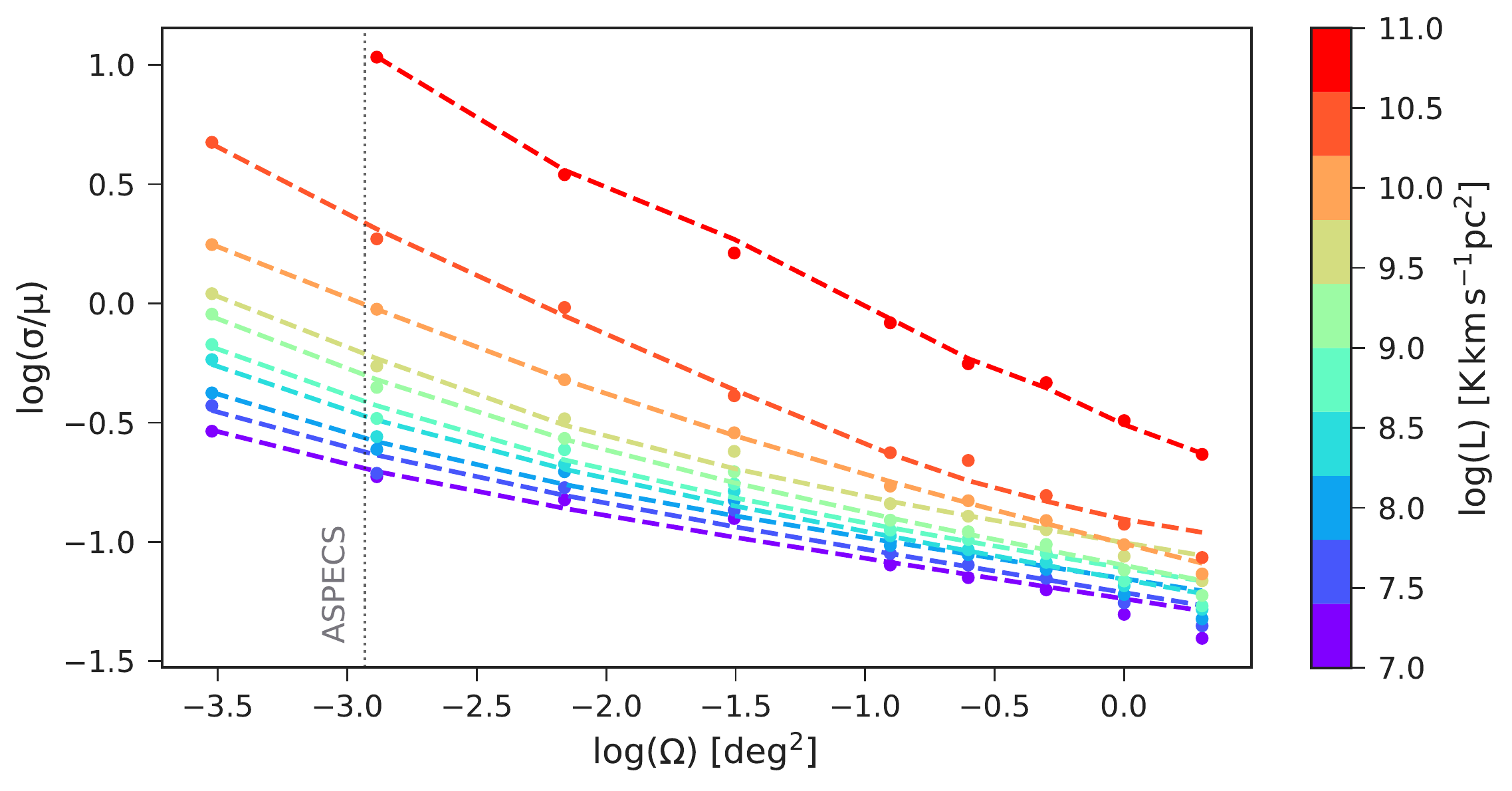}
\caption{Dependence of the total relative variance of the CO(2-1) LF on the survey size. In different colors we show the different luminosity bins of 0.4\,dex size. The points are computed from the SIDES simulations and the dashed lines correspond to the model described by Eq. \ref{eq:lf_variance_final}.}
\label{fig:LF_size_std}
\end{center}
\end{figure}

The relative contribution of the two components varies depending on the choice of the luminosity bin, since it affects both the clustering and the source density. This effect is shown in Appendix\,\ref{ap:LF_variance_appendix} (Fig.\,\ref{fig:LF_var_Lbins}), where we show an equivalent of Fig.\,\ref{fig:LF_size_std_one_lbin} for various luminosity bins. In Fig.\,\ref{fig:LF_size_std} we show the total relative uncertainty as computed directly from the simulation and modeled using Eq.\,\ref{eq:lf_variance_final} as a function of the survey size, for different luminosity bins which are color coded accordingly.

We can finally remark that the relative uncertainty of the low luminosity bins decreases slower with survey size than in the high ones. Even though the constraints in these bins are better because of the larger number of sources, the precision of the constraints is less tightly dependant on the field size than the bright luminosity bins. Therefore, in order to measure the faint-end of the LF it would be more efficient to use multiple independent small fields rather than one large field.

\subsection{Variance of the molecular gas mass density}
\label{subsec:cv_gas_mass}

One of the main factors regulating the evolution of galaxies is their star formation and its evolution with time. As the molecular gas constitutes the fuel for the star formation, measuring the abundance of the molecular gas through cosmic time is receiving considerable interest \citep[e.g.,][]{bisigello2022}. The molecular gas density ($\rho_{\rm H_2}$) is usually computed using the first moment of the observed LF ($L^{\prime}_{\rm CO(1-0)}$) after assuming a conversion factor ($\alpha_{\rm CO}$), 
\begin{equation}
    \rho_{\rm H_2} = \alpha_{\rm CO} \times \int_{0}^{\infty} L^{\prime}_{\rm CO} \Phi(L^{\prime}_{\rm CO}) \, d log(L^{\prime}_{\rm CO}) \,.
    \label{eq:rho_mol}
\end{equation}
The significant intrinsic variance in the LFs (Sect. \ref{subsec:cv_LFs}) is thus propagated to the molecular gas mass density estimation. Usually, a fixed value is assumed for the $\alpha_{\rm CO}$ conversion factor and the conversion from the luminosity of a given transition to  $L^{\prime}_{\rm CO(1-0)}$. We use  $\alpha_{\rm CO} = 3.6\,M_{\odot}$. In this paper, we ignore the impact of the conversion factor on the uncertainties and we focus instead on the impact of the field-to-field variance on the precision of the molecular gas density.

\begin{figure*}[h]
\begin{center}
\includegraphics[width=0.49\linewidth]{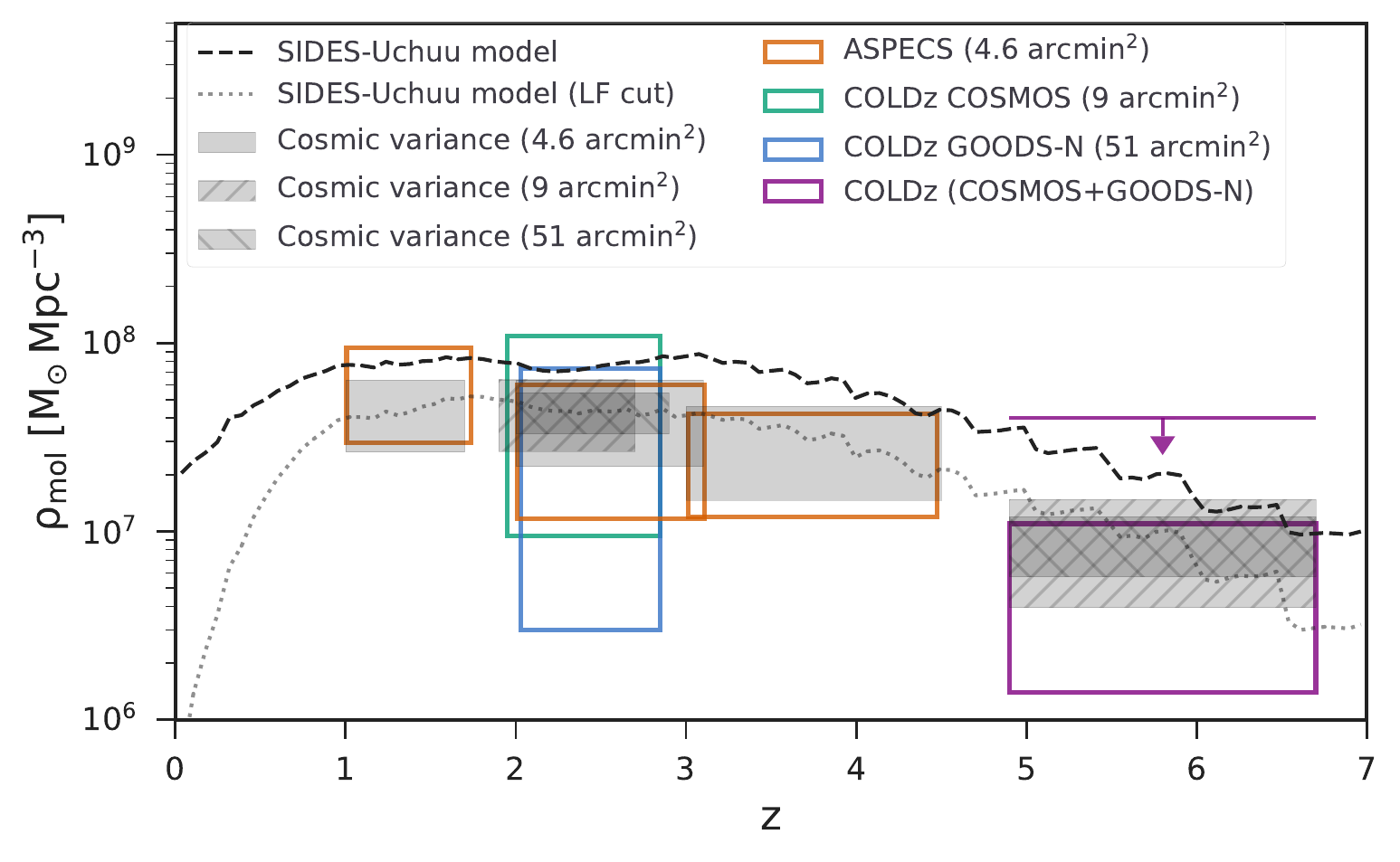}
\includegraphics[width=0.49\linewidth]{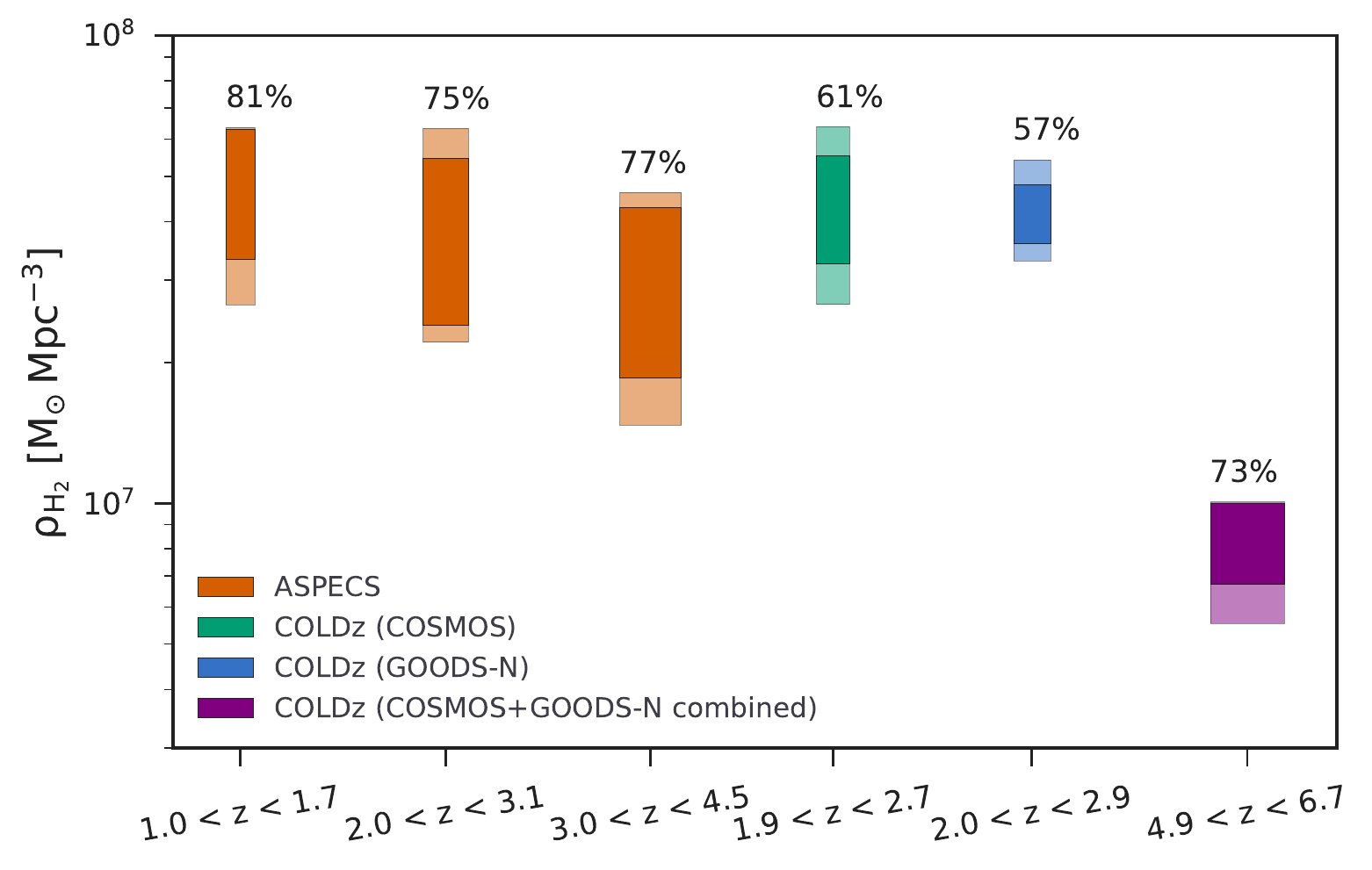}
\caption{Field-to-field variance of the molecular gas mass density, $\rm \rho_{H_2}$, at several redshift slices where observational data are available. Left: Evolution of $\rm \rho_{H_2}$ with redshift. The empty boxes are the observational data from ASPECS \citep{decarli2020} and COLDz \citep{riechers2019}. We also include the upper limit offered by COLDz assuming that all the galaxies could be CO(2-1) emitters at high z. The black dashed line is the SIDES model without a luminosity cut and the black dotted line is the SIDES model with a luminosity cut at $\rm \sim 5 \times 10^{9} \, K\, km \, s^{-1} pc^2$. The steps in both lines are caused by the selected redshift grid that the cosmological simulation snapshots were taken. The shaded gray areas correspond to the $\rm \rho_{H_2}$ field-to-field variance which has been computed for the same redshift ranges and sizes as the corresponding surveys. Right: Field-to-field variance in the molecular gas density in different redshift bins. The light-colored areas show the total field-to-field variance introduced in the molecular gas density estimates, while the dark-colored areas show the Poisson-only variance. The contribution of the latter to the total variance is given as a percentage on top of each rectangle. The different colors indicate the different surveys, hence different size and redshift slice.}
\label{fig:rho_mol_vs_redshift}
\end{center}
\end{figure*}

We first show in Fig.\,\ref{fig:rho_mol_vs_redshift} on the left the evolution of $\rho_{\rm H_2}$ with redshift as computed with SIDES when integrating both the full range of the LFs and only down to a luminosity cut. We additionally compare with measurements from ASPECS \citep{decarli2020} and COLDz \citep{riechers2019}. Both surveys use a luminosity cut at $\rm \sim 5 \times 10^{9} \, K\, km \, s^{-1} pc^2$. The SIDES model without the luminosity cut is systematically higher than the measurements. In contrast, the model that includes the luminosity cut agrees very well with the observations. Integrating the LFs down to $\rm \sim 5 \times 10^{9} \, K\, km \, s^{-1} pc^2$ leads to an underestimation of the molecular gas density. For instance, at z=0.5, 2 and 5 the discrepancy factor is 8.2, 1.9, and 1.7, respectively.

We note that adopting smaller ($\leq$117 deg$^2$) Uchuu fields introduces significant uncertainties. We estimated this variance for several redshift ranges by cutting the simulated area into multiple subfields with size equal to the corresponding survey that we compared our simulation with. However, for the combined COLDz field (COSMOS and GOODS-N, purple square in Fig.\,\ref{fig:rho_mol_vs_redshift}) on the left we selected a 9\,arcmin$^2$-sized SIDES subfield (COLDz COSMOS size) and a 51\,arcmin$^2$-sized subfield (COLDz GOODS-N size) without them overlapping. We created 117 realizations and computed the LF for each one of them. We obtained the molecular gas density value from the CO LF of each subfield using Eq. \ref{eq:rho_mol}. We then computed the 5th and 95th percentile confidence levels of $\rm \rho_{H_2}$ at various redshifts, which are presented in Fig.\,\ref{fig:rho_mol_vs_redshift} on the left as gray shaded rectangular areas. Overall, there is an excellent agreement between the interval predicted by SIDES for various realizations of the cosmic variance and the observational data.

All the CO surveys account for the field-to-field variance using the Poissonian uncertainties. However, they neglect the additional variance due to the clustering of the galaxies. In our study we differentiated between the two variance components. We thus show in Fig.\,\ref{fig:rho_mol_vs_redshift} on the right the total field-to-field variance with the fractional contribution of the Poisson variance indicated for each redshift slice. The difference between the total and Poisson variances reveals how the clustering acts as an extra source of variance. It is significant ($>$20\,\%) especially at z$>$2 and it should be taken into account by observational surveys.

\begin{figure}[h]
\begin{center}
\includegraphics[width=\linewidth]{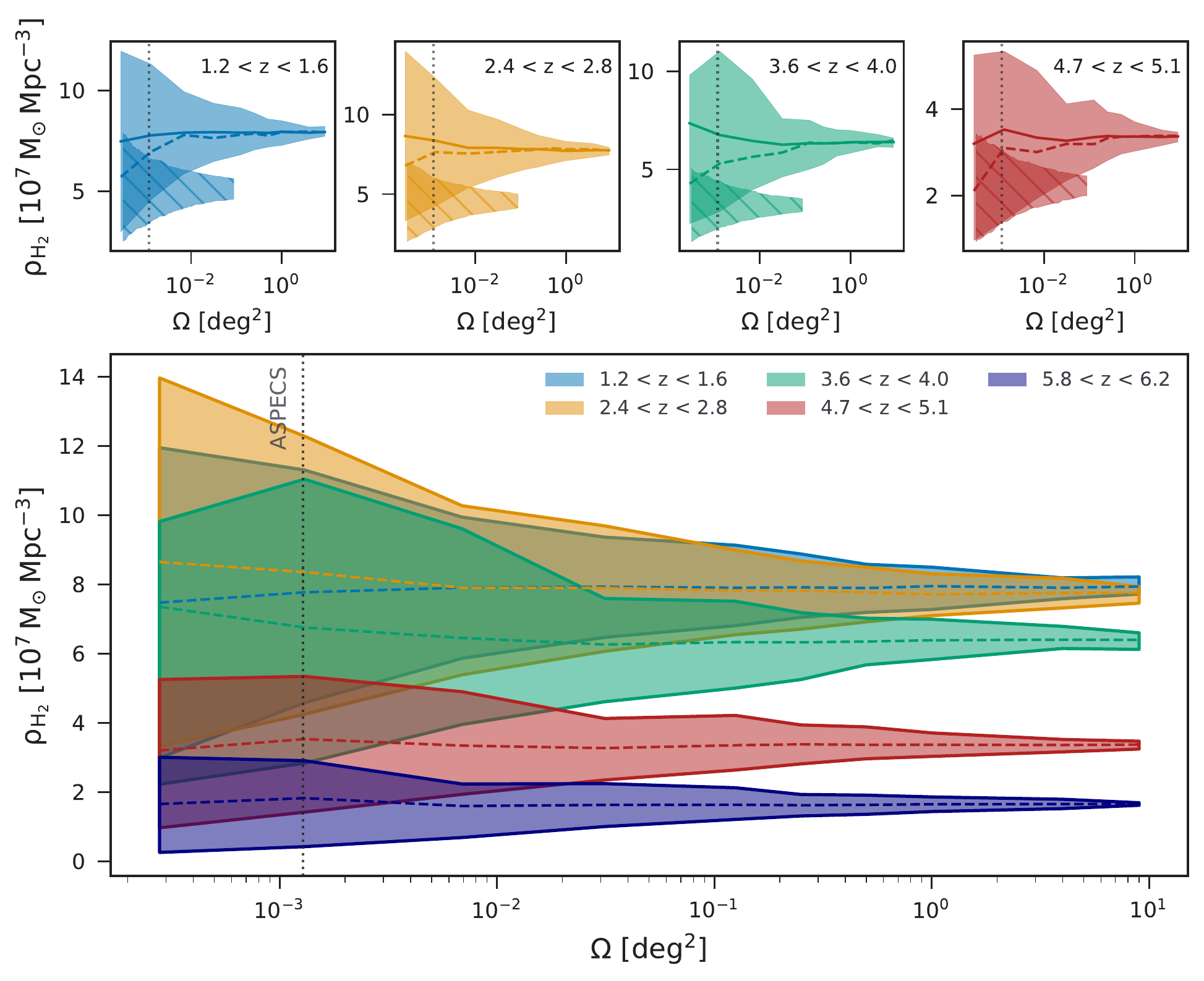}
\caption{Top: Cosmic variance in the $\rm \rho_{H_2}$ values as a function of the survey size for different redshift slices. The solid and dashed lines are the mean and median values, respectively. We have also included the corresponding results from \cite{keenan2020} (the shaded and hatched areas). Bottom: Evolution with redshift of the variance in the molecular gas mass density as a function of survey sizes, where it is easier to spot what is the proper survey size that will allow to probe the evolution of the molecular gas density.}
\label{fig:rho_mol_variance}
\end{center}
\end{figure}

The variance strongly depends on the survey size, as discussed in Sect.\,\ref{subsec:cv_LFs}. Therefore, we would like to address the question of what is the minimum survey size that will allow us to distinguish the evolution of the molecular gas from its intrinsic variance. For that purpose we cut the total simulated catalog in smaller ones of different sizes, varying from 0.00028 $\rm deg^2$ (1 $\rm arcmin^2$) to 9 $\rm deg^2$ and computed the molecular gas mass density (Eq. \ref{eq:rho_mol}) at a fixed redshift bin. We performed this analysis for various redshift bins centered at 1.4, 2.6, 3.8, 4.9, and 6 (taking $\rm \delta z = 0.4$). The results are presented in Fig.\,\ref{fig:rho_mol_variance}, where we also compare our work with the results from \cite{keenan2020}.

There is a shift between the mean $\rm \rho_{H_2}$ of our model and the one from \cite{keenan2020}. This could be explained by their use of a cut off when computing the integral of the first moment of the LF ($\rm L^{\prime}_{\rm CO} = 10^9\,K\,km\,s^{-1}pc^2$) to obtain the molecular gas density. On the contrary, we keep the whole luminosity range given in our simulated catalogs, leading to systematically higher $\rm \rho_{H_2}$ values. This difference can already introduce a discrepancy of the level of $\sim$ 20\%. Furthermore, the two approaches adopt different cosmological simulations, scaling relations to assign CO luminosities to galaxies, and level of scatter on the scaling relations. In spite of this divergence there is an agreement between the two models at the level of 1\,$\rm \sigma$ for survey sizes below $\rm \sim 40 \, arcmin^2$ ($\rm \sim 10^{-1} \, deg^2$).

The bottom panel of Fig.\,\ref{fig:rho_mol_variance} shows how the constraints in the various redshift bins improve with the field size. We need at least 1 $\rm deg^2$ surveys to avoid having an overlap between the $2.4<z<2.8$ and the $3.6<z<4.0$ 1\,$\sigma$ confidence regions and thus start to have hints of an evolution. This is about an order of magnitude larger field than what \cite{keenan2020} suggested, indicating that they may have underestimated the level of the total variance at these redshift ranges. In order to be able to probe the evolution between $3.6<z<4$ and $4.7<z<5.1$, we need at least $\rm 2\times 10^{-2} \, deg^2$ surveys. This value agrees with \cite{keenan2020}.

\subsection{Correlation among the luminosity function bins}
\label{subsec:LF_bins_correlation}

\begin{figure}[h]
\begin{center}
\includegraphics[width=\linewidth]{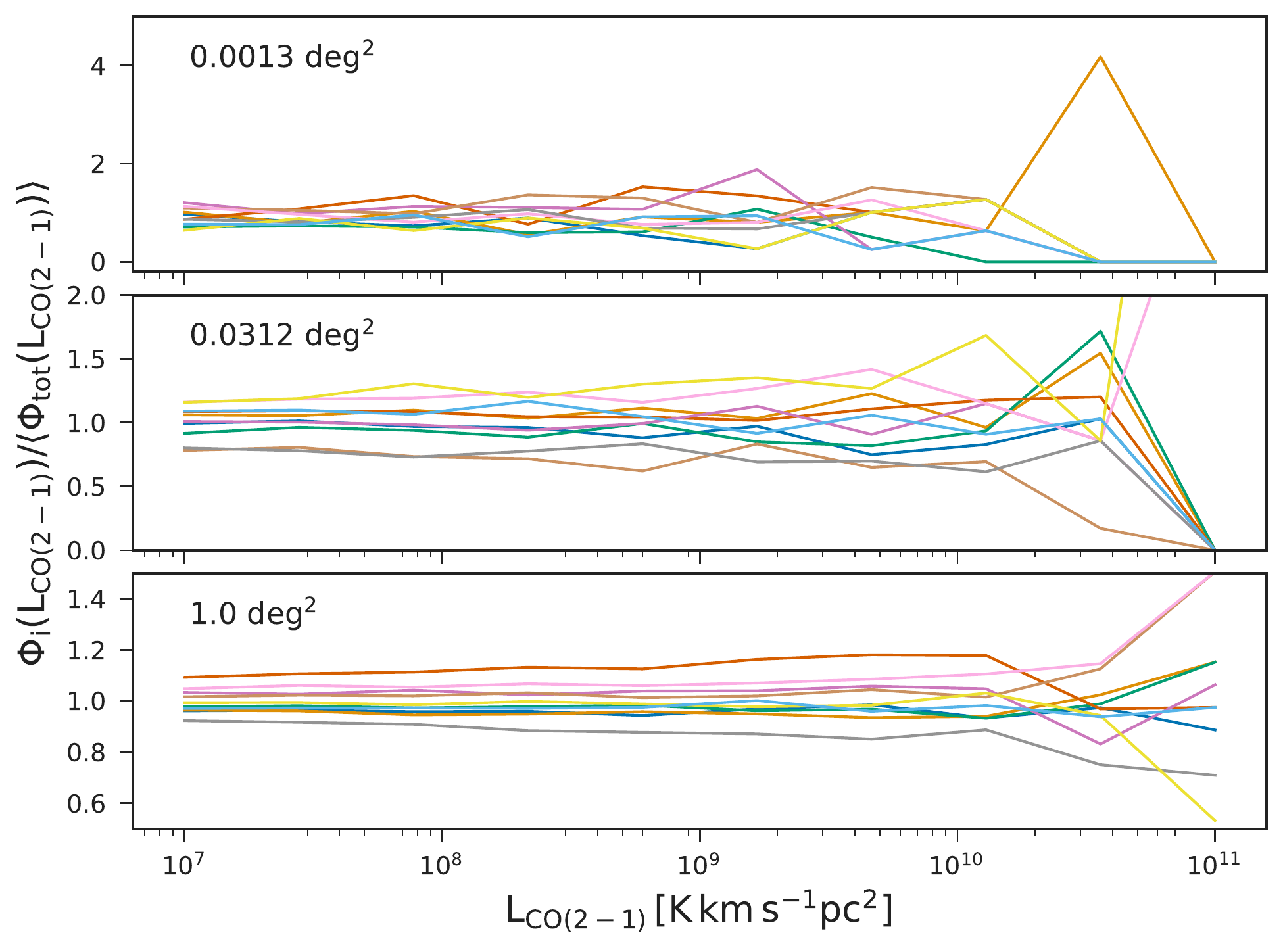}
\caption{Ten randomly drawn LFs for three different survey sizes, of 0.0013 $\rm deg^2$ (4.6 $\rm arcmin^2$), 0.0312 $\rm deg^2$ (112 $\rm arcmin^2$), and 1 $\rm deg^2$ from top to bottom, respectively. Each LF is normalized by dividing by the total Uchuu LF.}
\label{fig:randomly_drawn_LFs}
\end{center}
\end{figure}

The independence of the luminosity bins is a common assumption for the fit of the LF. We used the Uchuu simulation to test its validity. In Fig.\,\ref{fig:randomly_drawn_LFs} we show 10 randomly drawn LFs for three different survey sizes. In the case of the smaller survey size, we can see that the values of the low luminosity bins do not affect the brighter bins. They are rather randomly varying and the luminosity bins can be considered independent. This behavior is different for larger survey sizes and this is more clear for the $\rm 1 \, deg^2$ case, especially for luminosity bins below $10^{10}\,\rm K\,km\,s^{-1}pc^2$. When one luminosity bin is high all the other bins tend to be high as well and vice versa. This reveals a high level of correlation between the different luminosity bins.

\begin{figure*}[h]
\begin{center}
\includegraphics[width=0.95\textwidth]{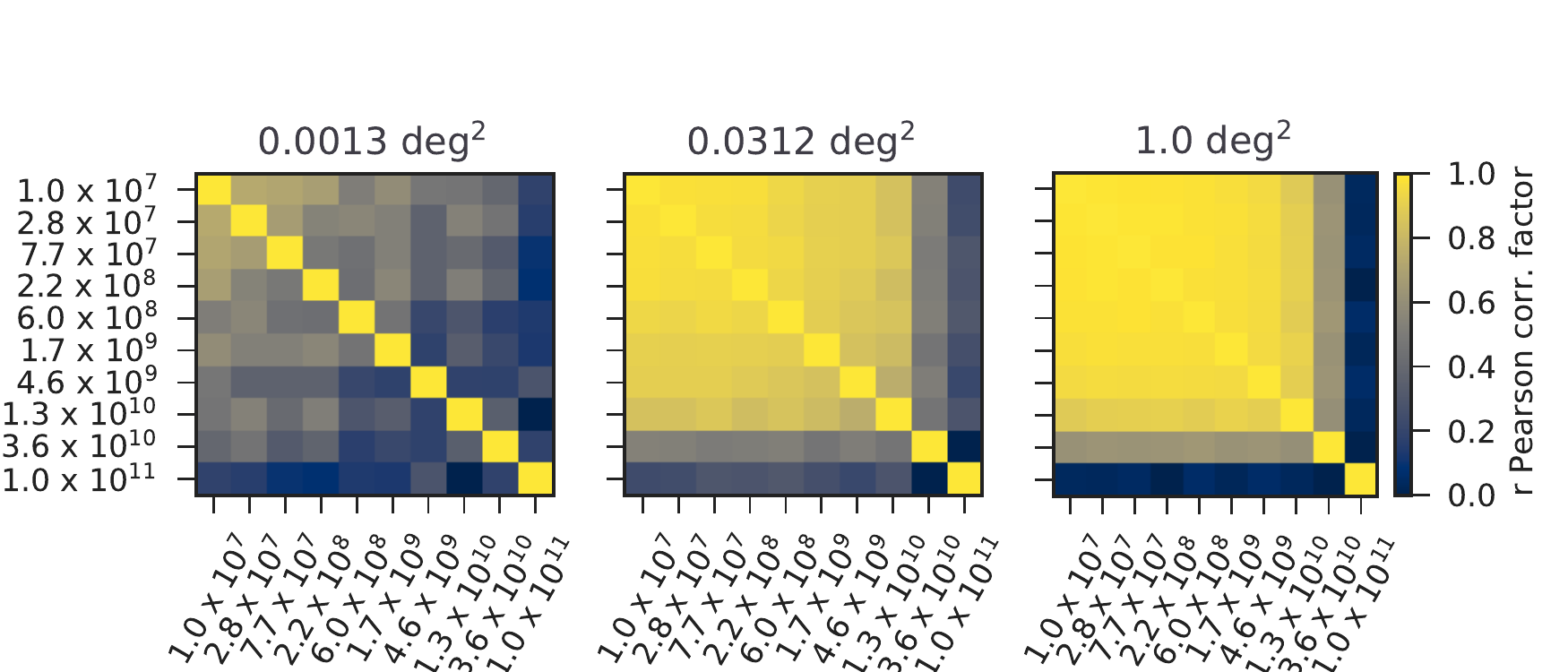}
\caption{Pearson correlation matrix among the luminosity bins used to construct the LFs. Each panel corresponds to different sizes, from left to right: 0.0013 $\rm deg^2$ (4.6 $\rm arcmin^2$), 0.0312 $\rm deg^2$, and 1 $\rm deg^2$. The units of the luminosity bins are $\rm log (L / [K^{-1} \, km \, s^{-1} pc^2])$.}
\label{fig:pearson_cor_matrices_LFs}
\end{center}
\end{figure*}

In order to better visualize the correlation level we computed the Pearson correlation matrix between luminosity bins and show it as a 2D map in Fig.\,\ref{fig:pearson_cor_matrices_LFs}. The larger the survey size the more correlated the luminosity bins. On top of that the faint luminosity bins are more correlated compared to the bright bins, at all survey sizes. For a given size, the faint sources are more abundant compared to the very bright ones, the number of faint sources will thus be high if there is an overdensity or low if there is an underdensity. Therefore, the amplitude of the LF at all the adjacent low luminosity bins is expected to similarly vary from field to field, this is why they appear more correlated. However, the bright sources are rare and their number density could differently vary from one field to another resulting to less correlated bins.

The confirmation of this expected effect is important for all the surveys that use the observed LFs to obtain other physical quantities like the molecular gas density. Bigger survey sizes can offer smaller error bars but the luminosity bins are more correlated with each other. Therefore, one should take into account that the luminosity bins are not independent when fitting the LF.

\subsection{A tool to estimate the variance of line luminosity functions and molecular gas densities}
\label{subsec:script_description_lf_rho_mol}

To help the user investigate the SIDES-Uchuu catalogs \footnote{\url{https://cesamsi.lam.fr/instance/sides/home}}, we offer several simple scripts gathered in a jupyter notebook to derive the most useful quantities. A first set of scripts computes the LF of the requested transition in a specific redshift slice and for a given survey size. The second part of the output is the expected field-to-field variance for the same survey parameters. This part of the script assumes that the fields are squares. Finally, the code can compute the Pearson correlation matrix between luminosity bins.

The second set of scripts compute the  field-to-field variance on the mass density of the molecular gas ($\rm \rho_{H_2}$). The input parameters are similar to the previous set of scripts and it returns $\rm \rho_{H_2}$ as well as the 5\%, 16\%, 84\%, and 95\% confidence levels. The different options available for the user are documented in detail in the notebook.
 
\section{Power spectra variance}
\label{sec:pk_variance}

The total signal of an intensity map can be decomposed into fluctuations at different scales. 
The power spectrum of an intensity map describes how much the fluctuations of each scale contributes to the final map. By studying the CO and [CII] power spectra we can obtain important physical information such as, the SFRD at high redshifts and the clustering of galaxies. The power spectrum consists of the clustering and the shot noise component. The former describes the correlation of the sources at large scales and the latter is the flat component caused by the randomly distributed sources at all scales.

The power spectra obtained by line intensity mapping experiments can significantly vary from one field to another. This could be caused either by several bright sources in the field of the survey, or by the selection of the field itself. Observing an over or underdensity of sources could affect the level of the power spectrum at all scales. This effect has not been studied before, but thanks to the Uchuu simulation we could study and quantify the intrinsic field-to-field variance of such power spectra.

\begin{figure}[h]
  \centering
  \includegraphics[width=\linewidth]{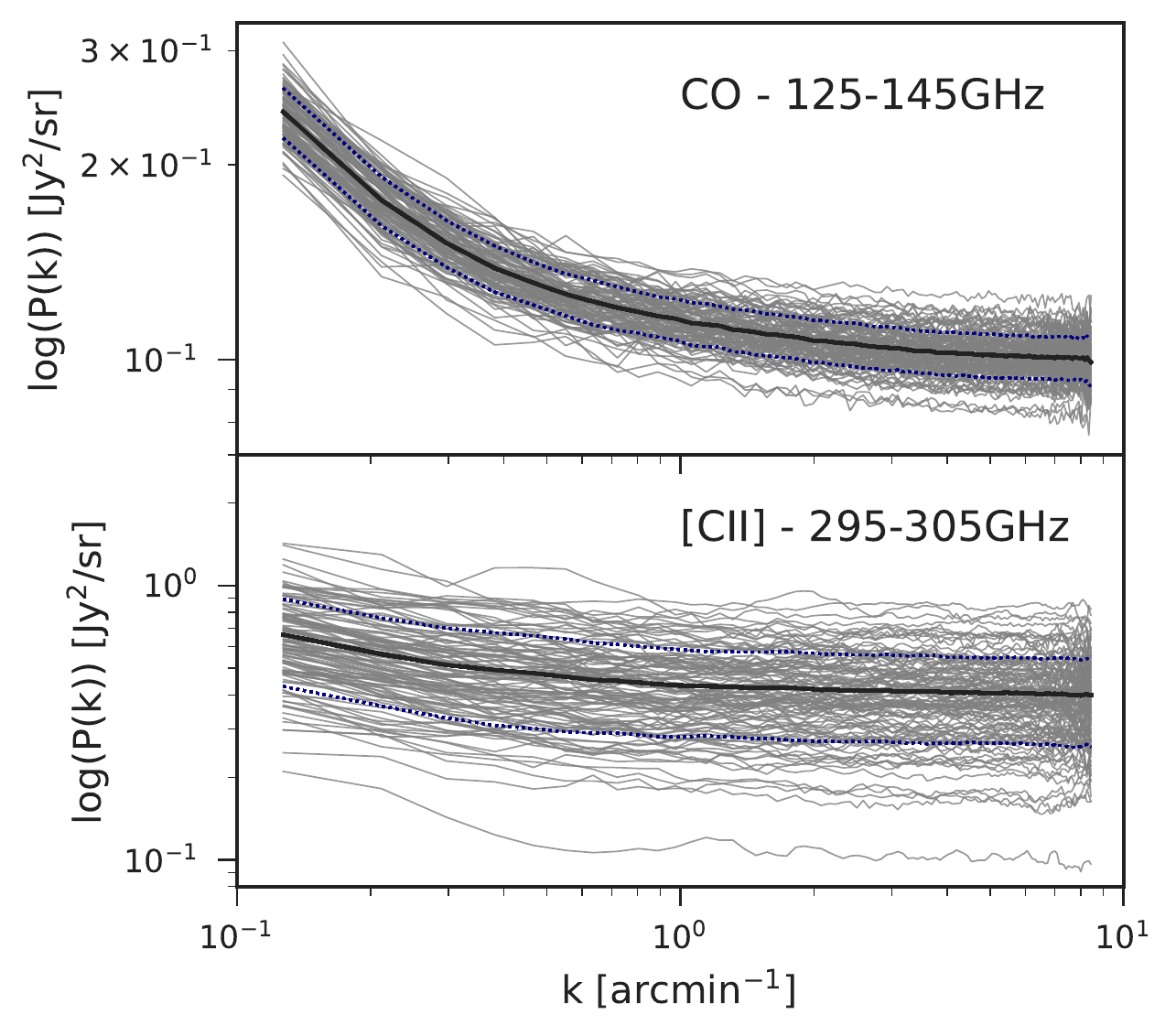}
\caption{Power spectra of 117 1\,deg$^2$-sized Uchuu subfields for the CO and [CII] lines, averaged over the frequency range shown in each plot. Top: CO power spectra. Bottom: [CII] power spectra. In both panels, the solid black line is the mean power spectrum at all k, while the dotted blue lines denote the 1$\, \sigma$ range.}
\label{fig:co_cii_pk_variance}
\end{figure}

Similarly to our study of the LF variance (Sect.\,\ref{subsec:cv_LFs}), we generated cubes corresponding to 117 different 1 $\rm deg^2$ subfields, and computed the power spectrum for each. In Fig.\,\ref{fig:co_cii_pk_variance} we show the resulting power spectra for both the CO and [CII] lines, as well as the corresponding subfields that were used to obtain them. The level of the power spectra can vary by a factor of 2-10 depending on the line, the observing frequency and the scale. Therefore, trying to fit a highly uncertain power spectrum could strongly affect the inference of physical quantities like the clustering of the galaxies and the SFRD.

\subsection{Contribution to the shot noise}
\label{subsec:n_counts}

\begin{figure}[h]
    \begin{subfigure}{\columnwidth}
    \centering
    \includegraphics[width=\linewidth]{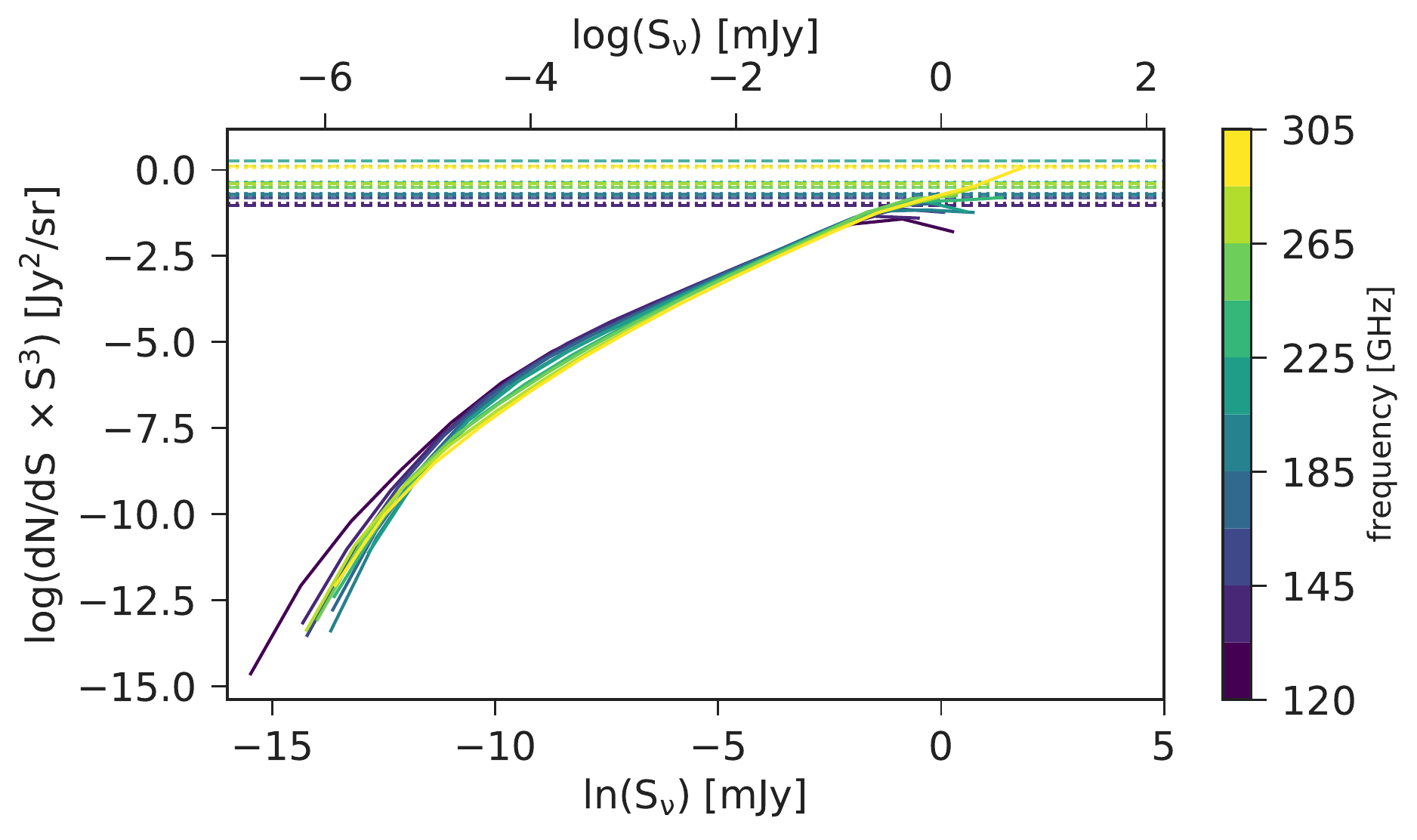}
    \label{fig:CO_number_counts}
    \end{subfigure}
    \begin{subfigure}{\columnwidth}
    \centering
    \includegraphics[width=\linewidth]{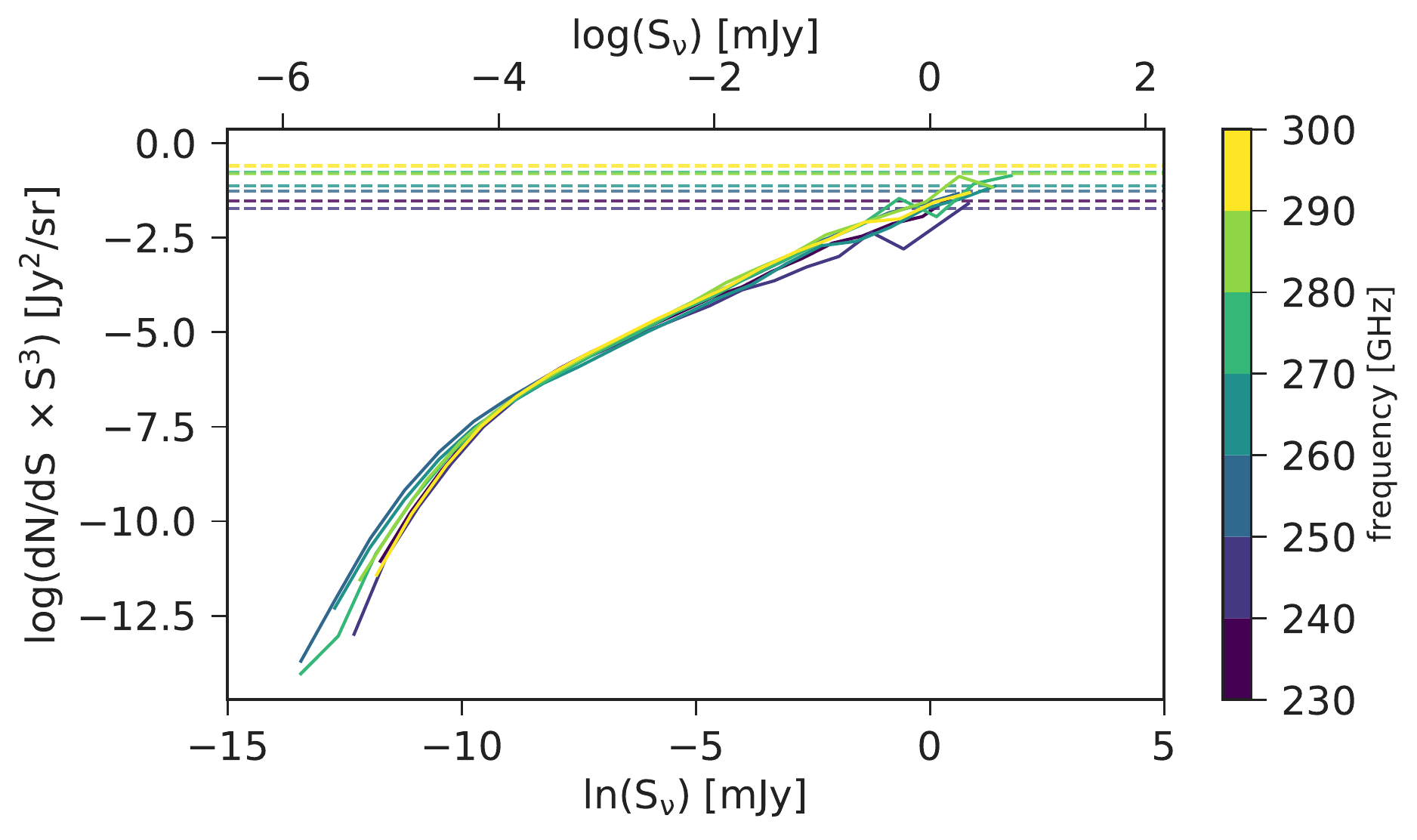}
    \label{fig:CII_number_counts}
    \end{subfigure}
\caption{Source flux contribution to the CO (top) and [CII] (bottom) shot noise power spectrum (see Eq.\,\ref{eq:shot_noise}). The different color lines correspond to different frequency slices with the CONCERTO frequency resolution ($\rm1.5 \, GHz$) within the whole CONCERTO observing frequency range. The dashed lines are the expected shot noise level as computed by Eq.\,\ref{eq:shot_noise}.}
\label{fig:number_counts}
\end{figure}

The shot noise component of the power spectrum is defined as 
 \begin{center}
\begin{equation}
    P_{\rm shot} = \int_{0}^{S_o} S^2 \frac{dN}{dS} \,dS  = \int_{0}^{S_o} S^3 \frac{dN}{dS} \,d{\rm ln}(S) ,
\label{eq:shot_noise}
\end{equation}
\end{center}
where $S_{\rm o}$ is the flux above which the sources are masked or removed in a given survey. In order to investigate which sources contribute the most to the Poisson noise, we focused on the integrand in Eq.\,\ref{eq:shot_noise}. In Fig.\,\ref{fig:number_counts} we show the quantity $\rm S^3 \frac{dN}{dS} \,dln(S)$ as a function of flux. We also show in the same plot, as reference, the level of the shot noise as computed from Eq.\,\ref{eq:shot_noise}.
    
We see from Fig.\,\ref{fig:number_counts} that for both CO and [CII] the brightest sources are the ones that contribute the most to the power of the shot noise and this trend is independent of the observing frequency. We also see that all the curves converge to the level of the shot noise as computed from the integral, which is consistent with the fact that all the shot noise is produced by the highest dlnS bins. The behavior shown in Fig.\,\ref{fig:number_counts} is the opposite with respect to the CIB case, where the faint sources are the most contributing ones \citep{lagache1999}. This means for CO and [CII] that even if we apply flux cuts (to remove the brightest sources) the most contributing sources to the shot noise level will always be the ones that exist right below the flux cut.

 It will be challenging for the first-generation intensity mapping surveys to detect individual bright sources. But even if they detect some, since the shot noise level is very sensitive to the flux cut, applying an incomplete masking could lead to an incorrect interpretation of the shot noise. Finally, the interpretation of the shot noise will offer information about the bright sources just below the detection threshold. LIM experiments will offer statistical information about these sources, but classical surveys are more suitable for studying them \citep[e.g.,][]{yue2019}.
 
\subsection{Power spectrum variance model}
\label{subsec:my_model}

The variance of the power spectra (Fig.\,\ref{fig:co_cii_pk_variance}) depends on the observed line, the survey size ($\rm \Omega$), the observing frequency ($\rm \nu$), and the number of frequency channels ($\rm \Delta \nu$) averaged to obtain the final power spectrum of each subfield. For instance, CONCERTO observes from 130 to 310\,GHz with a spectral resolution of up to 1.3\,GHz, so if we average the power spectra of the first 15 channels, meaning $\rm \Delta \nu = 19.5 \, GHz$, the resulting power spectrum will be at the center of 130 - 149.5\,GHz.

\begin{figure*}[h]
  \centering
  \includegraphics[width=\linewidth]{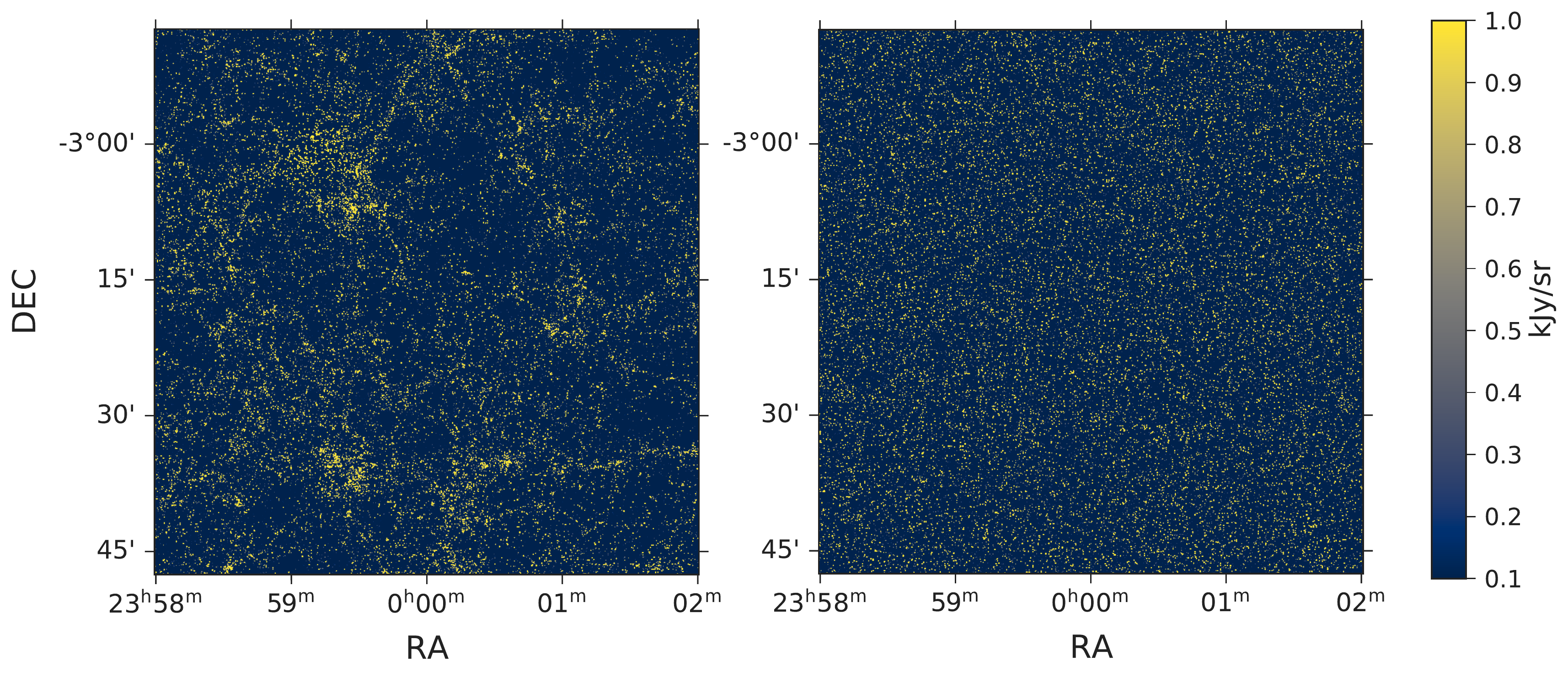}
\caption{Visualization of the 275\,GHz slice of an original CO cube on the left and the shuffled cube on the right. The shuffled cube contains the same sources (same redshift and luminosity) as the one on the left but is randomly distributed.}
\label{fig:original_vs_shuffled_cube}
\end{figure*}

We aimed to measure the variance as a function of these various parameters. However, the variance on the power spectra could be very different for the shot-noise and the clustering components. We thus needed to compute each component separately in each subfield. We created a new set of cubes with the same size and sources as the original ones but with their (RA, DEC) coordinates randomly shuffled, erasing in this way any clustering information embedded in the original cubes (defined as shuffled cubes hereafter). The followed procedure is visualized in Fig.\,\ref{fig:original_vs_shuffled_cube}. This allowed us to measure the shot-noise level. We then subtracted this quantity from the total power spectrum to obtain the clustering component and then computed the variance for each $k$. We found that the variance is not scale dependent. We thus averaged the variance at the various scales to obtain the final variance of the clustering component.

\begin{figure*}[h]
  \centering
  \includegraphics[width=0.8\linewidth]{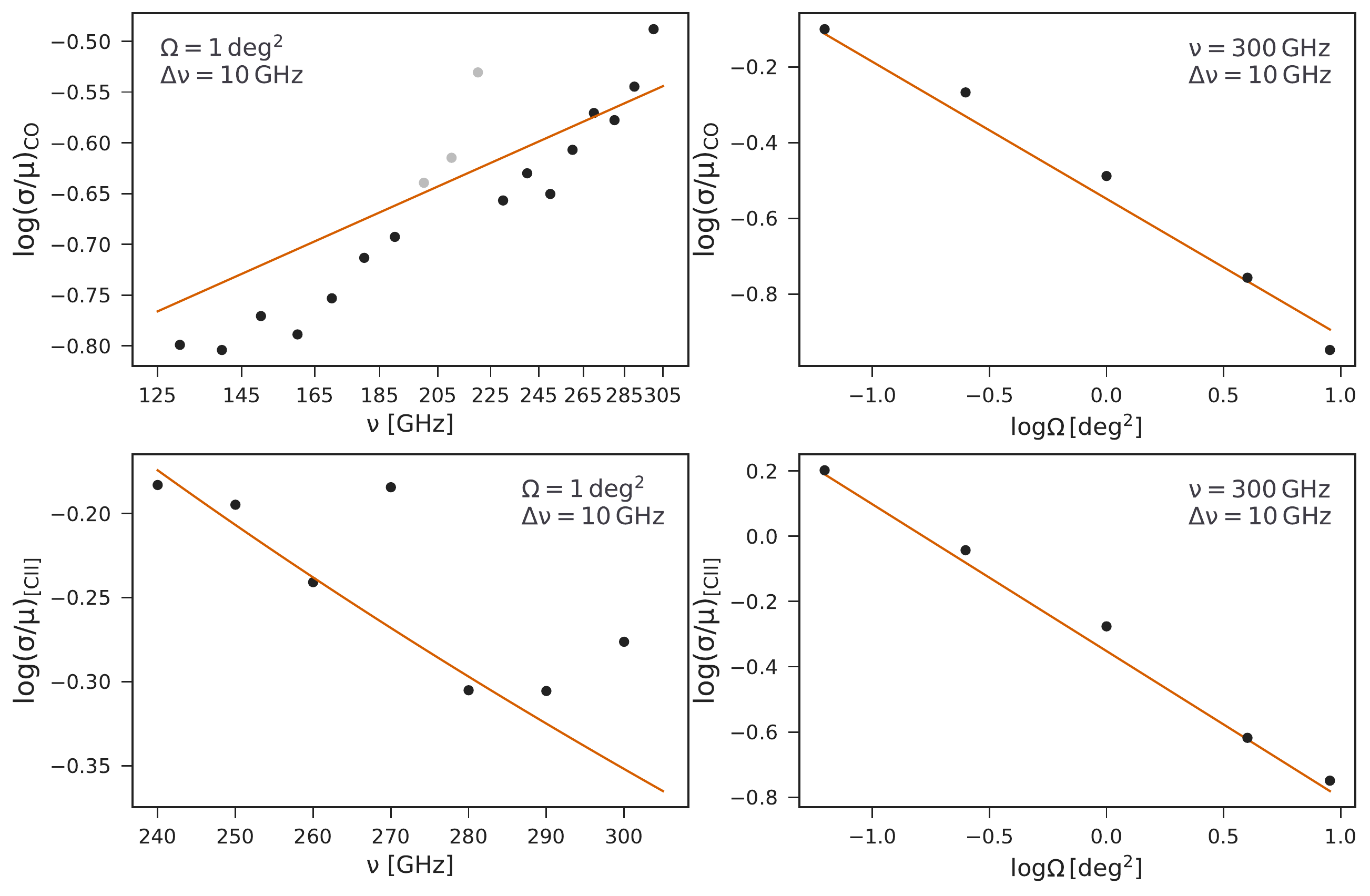}
\caption{Dependence of the clustering power spectrum variance (0.13 < k < 1\,arcmin$^{-1}$) on the observing frequency ($\nu$) and the survey size ($\Omega$). Top row: the black points are the CO data while the high transparency points are the ones excluded due to the CO(2-1) contamination as explained in Sect\,\ref{subsec:my_model}. The solid orange line is the best-fit model. Bottom row: the black points are the [CII] data and the orange solid line is the best-fit model.}
\label{fig:data_fitted-model}
\end{figure*}

We subsequently aimed to obtain a model that would enclose all the information of field-to-field variance dependence. We first created a set of simulated data that our model would be able to fit. For this purpose, we created five data sets corresponding to five different field sizes (0.0625\,deg$^2$, 0.25\,deg$^2$, 1\,deg$^2$, 4\,deg$^2$, and 9\,deg$^2$). In each case we cut the simulation in subfields of the corresponding size and computed the power spectra for the whole range of frequencies (125-305\,GHz) and frequency step ($\Delta \nu$ = 5, 10, 20\,GHz). A subset of the simulated data are shown as black points in Fig.\,\ref{fig:data_fitted-model}. We finally fitted all the simulated data for a given line and a given component (shot-noise or clustering) by the following parametric form:

\begin{center}
    \begin{equation}
        \frac{\sigma}{\mu} = c \left(\frac{\nu}{\nu_o} \right)^{\alpha} \left(\frac{\Delta \nu}{\Delta \nu_o} \right)^{\beta} \left(\frac{\Omega}{\Omega_o} \right)^{\gamma}\,,
    \label{eq:variance_formula_co_cii}    
    \end{equation}
\end{center}
where $\rm c, \alpha, \beta$, and $\rm \gamma$ are parameters determining the scaling, while we set the normalization to $\rm \nu_o = 100 \, GHz$, $\rm \Delta \nu_o = 5 \, GHz $, and $\rm \Omega_o = 1 \, \rm deg^2$.

We observe an excess of variance at around 230\,GHz for CO. This is caused by low-redshift bright sources seen in CO(2-1) as discussed in B22. This excess of the CO power spectra in this frequency range complicates the process of searching for a universal model that could fit all the given data. We therefore chose to exclude the CO data points between 210 and 230 GHz (light-colored points in Fig.\,\ref{fig:data_fitted-model}).

We provide two sets of best-fit parameters for each line: one for the clustering component and one for the shot noise component. All the parameters are summarized in Table \ref{table:variance_models_params}, while in Fig.\,\ref{fig:data_fitted-model} we show the best-fit model and how the relative uncertainty $\left ( \frac{\sigma}{\mu} \right )$ depends on the survey size and frequency. We can thus use it to estimate the expected relative variance of the observed power spectra.

\begin{table}[h]
\begin{center}
    \caption{Best-fit parameters of Eq. \ref{eq:variance_formula_co_cii} for the CO and [CII] lines.}
    \label{table:variance_models_params}
    {\centering
    \begin{tabular}{c | c c | c c}
    \hline\hline 
     & \multicolumn{2}{c|}{CO} & \multicolumn{2}{|c}{[CII]} \\
     & Poisson & Clustering & Poisson & Clustering \\
    \hline 
    c & 0.1   & 0.2 &  5.43  & 6.22  \\
    $\alpha$ & 0.71 & 0.58  & -2.1 & -2.4 \\
    $\beta$ & -0.52  & -0.39 & -0.62  & -0.61 \\ %CII values without bootstrapping
    $\gamma$ & -0.31  & -0.37 & -0.45 & -0.46 \\ %CII values without bootstrapping
    \hline
    \end{tabular}}
\end{center}
\end{table}
    
 The $\rm \alpha$ slope is positive for CO and negative for [CII] revealing the opposite dependence on frequency. In the case of CO the higher the observing frequency the higher the relative variance, while in the case of the [CII] line the trend is the exact opposite. The [CII] trend is not surprising, since the lower frequency probes very high redshifts, where the number of objects are very small and star formation tends to appear in the most overdense regions \citep[e.g.,][]{behroozi2013,bethermin2013}. The CO intensity fluctuations are the result of multiple rotational lines and are more difficult to interpret. 
    
The $\rm \beta$ slope for the CO Poisson component is -0.52. It is very close to the -0.5 expected if the shot noise in the various frequency slices was independent (B22, Appendix D). The values obtained for [CII] and for the CO clustered component are slightly different, which suggests that there may be significant correlations between the power spectra of the various frequency slices.

The $\gamma$ parameter for both components of CO and [CII] is systematically higher than the expected value in a pure Poisson case (-0.5). The variance thus decreases with the field size slower than what is expected in the absence of any clustering. This is similar to what has been found for the LFs (Sect.\,\ref{subsec:cv_LFs}). It is not surprising for the clustering component, but is less intuitive for the Poisson one. However, these results make sense if we consider that the variance of the LFs, from which the Poisson component derives, exceeds the Poisson term.
    
\subsection{Correlation between the Poisson and the clustering components}
\label{subsec:Pclust_Pshot_correlation}
    
The dependence of the variance of the clustering power spectrum with the field size is expected to deviate from the expected behavior of a purely Poisson case. We see a similar trend for the variance of the Poisson component. This could mean that both components are similarly dependent on the local over or underdensity. Of course, we do not expect a perfect correlation, since the shot noise is mainly caused by the bright sources while the sources around the knee of the LFs are those that maximally contribute to the clustered component (e.g., B22). However, as we find in Sect.\,\ref{subsec:LF_bins_correlation}, the various luminosity bins are correlated and we can thus expect a significant effect. In order to quantify it, we measured the level of correlation between the clustering and shot noise component among the various Uchuu subfields.

\begin{figure*}[h]
\begin{center}
\includegraphics[width=0.8\textwidth]{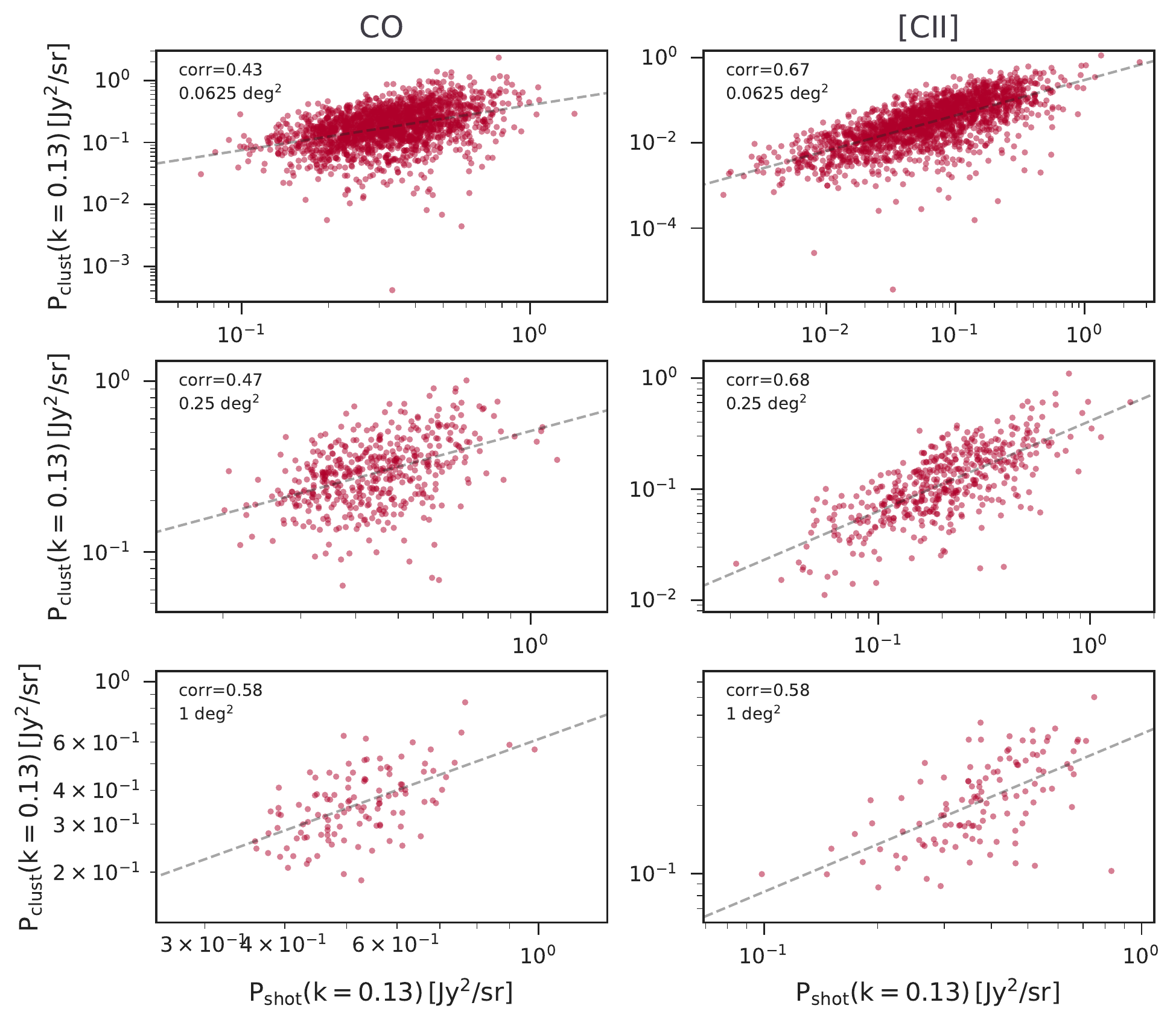}
\caption{Correlation between the power spectra of the clustering and the shot noise from various subfields for the frequency range 295-305\,GHz. The left column is for the CO power spectra for sizes from top to the bottom: 0.0625, 0.25, and 1\,deg$^2$. The right column is for the [CII] power spectra for the same sizes as the CO. On the top left of each panel we show the correlation value, while the dashed line shows the corresponding linear line.}
\label{fig:Pclust_Pshot_correlation}
\end{center}
\end{figure*}

In Fig.\,\ref{fig:Pclust_Pshot_correlation} we show how the amplitude of the clustering and the shot noise are correlated. When the clustering is high in a given subfield, the shot noise level tends to be high. The Pearson correlation coefficient varies between 0.41 and 0.67. As a result, the total variance of the power spectrum is neither the quadratic sum of the two components nor their simple sum. Therefore one should be cautious when combining these two sources of uncertainty.

\subsection{Consequences for line intensity mapping experiments}
\label{subsec:consequencies_for_LIM_experiments}

Using our model we forecast the expected uncertainties on $\rm P(k)_{clust}$ associated with field-to-field variance for CONCERTO and other current or upcoming experiments. TIME will cover a rectangular 0.01\,deg$^2$-sized field and will observe in the 183-326\,GHz frequency range. The South Pole Telescope Summertime Line Intensity Mapper \citep[SPT-SLIM, ][]{karkare2022} survey will cover 1\,deg$^2$ at 120-180\,GHz. Finally, the CO Mapping Array Project \citep[COMAP, ][]{cleary2021} and  the Prime-Cam on FYST \citep{collaboration2021} will survey 4\,deg$^2$ patches at 26-34\,GHz and 210-420\,GHz, respectively. The properties of the LIM experiments we consider are summarized in Table\,\ref{table:IM_experiments}.

\begin{figure}[ht]
\begin{subfigure}{\linewidth}
\includegraphics[width=\linewidth]{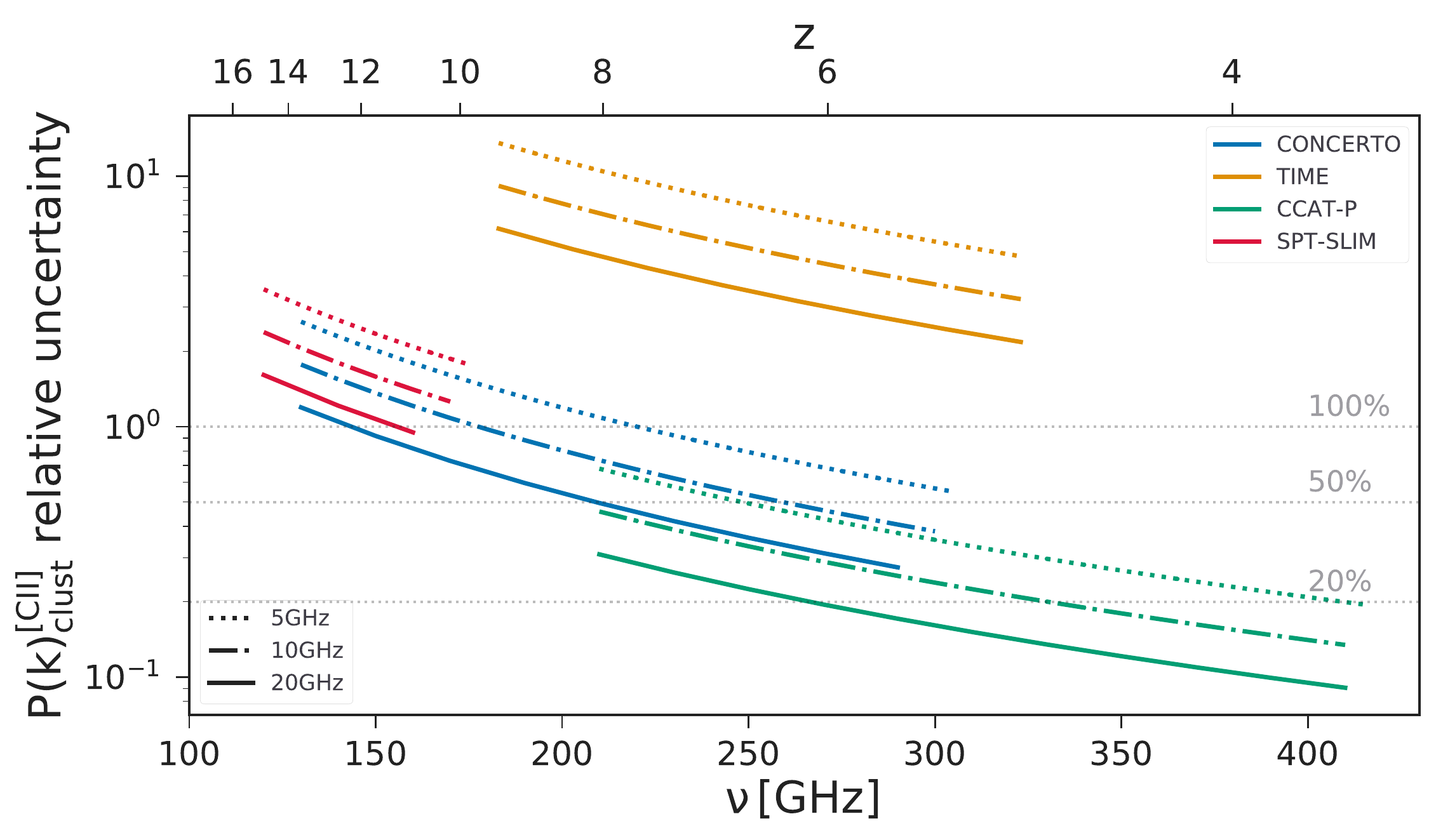}
\end{subfigure}

\begin{subfigure}{\linewidth}
\includegraphics[width=\linewidth]{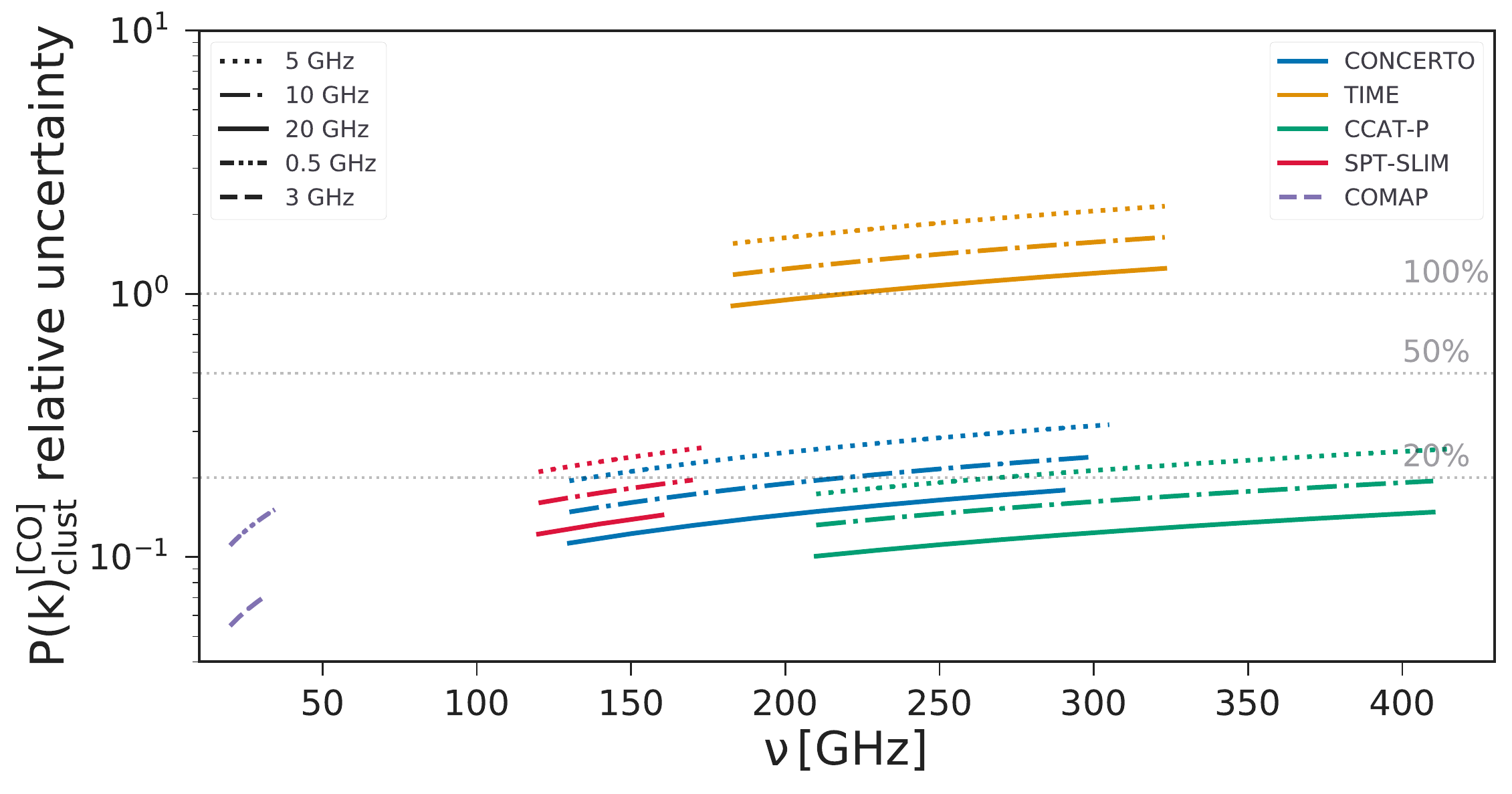}
\end{subfigure}
\caption{Level of relative uncertainty of the clustering component (the standard deviation of $\rm P(k)_{clust}$ divided by the average) of the [CII] power spectra (top panel) and the CO (bottom panel) for various LIM experiments (see Table \ref{table:IM_experiments}). The different line styles correspond to different number of averaged channels ($\Delta \nu$).}
\label{fig:CII_CO_uncertainty_LIMexperiments}
\end{figure}

Given the field size and the frequency range of each experiment we obtained the relative uncertainty of both the CO and [CII] power spectra caused by the field-to-field variance. The results are shown in Fig.\,\ref{fig:CII_CO_uncertainty_LIMexperiments}. We used different line styles for the different number of averaged channels ($\Delta \nu$). Averaging more frequency channels results in lower uncertainties. For instance, the $\Delta \nu = 20\, \rm GHz$ curves are below the $\Delta \nu = 5\, \rm GHz$ ones.

For frequencies below 280\,GHz (z $\gtrsim$ 6 for [CII]) CONCERTO will not robustly constrain the [CII] clustering power spectrum since the field-to-field uncertainties exceed 50\%. However, at higher frequencies (z $\lesssim$ 6) these uncertainties are smaller and it should be possible to provide meaningful constraints if the signal is detected. On the other hand, CONCERTO will be able to get a good estimate of the CO clustering power spectrum below $\sim$250\,GHz where the field-to-field uncertainties are smaller than 20\,\%. However, the interpretation of the shot noise for both [CII] and CO will be difficult due to the high field-to-field variance, the correlation with the clustering component, and because it is mainly produced by rare bright sources (Sect.\,\ref{subsec:n_counts}).

Regarding our forecast for the other [CII] LIM experiments, for TIME it is quite pessimistic with an uncertainty higher than 100\% at all frequencies. However, our model was designed for square field areas which is very different than the TIME strategy (1.3 $\times$ 0.007\,deg$^2$). It is thus possible that we have overestimated its field-to-field uncertainty. The wide field area of the FYST/Prime-Cam experiment could offer the most promising results with a field-to-field uncertainty below 20\% even up to $z$ = 6. Even though at high redshift the uncertainties increase, this still remains the most accurate amongst the experiments. Finally, SPT-LIM will provide good constraints on CO ($\sim$20\% field-to-field variance), which is the main goal of the experiment. However, because of its low frequency coverage, the [CII] power spectrum will be highly uncertain ($>$ 100\%). But [CII] is not a science goal of this experiment.

Regarding the CO power spectrum, the COMAP (Pathfinder and EoR) experiment will be negligibly affected by the field-to-field variance with field-to-field uncertainties well below 20\,\% for its whole frequency range. However, these results come from an extrapolation of our model, which could include some extra uncertainties we have not considered. SPT-LIM and FYST/Prime-Cam will also recover the CO power spectra with small uncertainty ($\lesssim$20\%).

According to the power spectrum formalism, the clustering component is proportional to the square of both the total emission of the galaxies and their linear clustering bias \citep{kovetz2017}. Therefore the uncertainty we have obtained here should be reduced by a factor of $\sim$2 when propagated to the integrated line emission and subsequently to the SFRD or the molecular gas density history. However, this assumes that we will be able to break efficiently the degeneracies between emission and clustering. Considering the high uncertainty when measuring the molecular gas density evolution from current surveys (left Fig.\,\ref{fig:rho_mol_vs_redshift}), the current generation of LIM experiments will still offer more accurate estimates. Future experiments with wider fields and higher sensitivities will offer even more accurate constraints.

\begin{table*}[h]
\begin{center}
    \caption{Characteristics of current and future line intensity mapping experiments.}
    \label{table:IM_experiments}
    {\centering
    \begin{tabular}{c | c | c | c | c | c }
    \hline\hline 
    & TIME & SPT-LIM & CONCERTO & COMAP & CCAT-P \\
    \hline 
    Field size [deg$^2$] & 0.0091 & 1 & 1.4 & 4 & 4 \\
    Targeted line(s) & [CII], CO & CO & [CII], CO & CO(1-0), CO(2-1) & [CII] \\
    Frequency range [GHz] & 183-326 & 120-180 & 130-310 & 26-34 & 210-420\\
    \end{tabular}}
\end{center}
\end{table*}

\section{Conclusion}
\label{sec:conclusion}

Combining the SIDES model with the Uchuu dark matter simulation, which offers a large simulated volume and high mass resolution at the same time, allows us to test the SIDES model with a much higher accuracy than in B22. We compared the SIDES model with the power spectra of the CIB fluctuations in various \textit{Planck} and \textit{Herschel}/SPIRE bands. The overall good agreement confirmed the ability of SIDES to realistically reproduce the clustering of galaxies at large scales. As an additional validation test, we compared the line LFs predicted by SIDES with observational data from surveys like ASPECS and ALPINE. All the SIDES LFs from multiple different subfields lay inside the $1 \, \sigma$ confidence level of the observations confirming the validity of the newly added recipes for the lines (CO and [CII], B22).

Regarding the line LFs we find that for low luminosity bins and large survey sizes, the field-to-field variance is larger than the Poisson uncertainty. This excess is caused by the galaxy clustering. We modeled the contribution of the Poisson and the clustering variance to determine precisely for each field size the luminosity range where the clustering is comparable to the Poisson term and needs to be taken into account. Finally, we show that the luminosity bins tend to be correlated with each other, especially for large fields (e.g., 0.0312\,deg$^2$), which are dominated by the clustering rather than the Poisson variance.

We also studied the effect of the field-to-field variance on the cosmic molecular gas density measurements as a consequence of the CO LF uncertainties. For small survey sizes (e.g., ASPECS) the field-to-field uncertainty is large ($\sim$\,50\%). Comparing our findings with other similar studies, we find that the total variance at certain sizes (e.g., 1\,deg$^2$) could be up to an order of magnitude higher. We show that in order to differentiate between the $\rm \rho_{H_2}$ variation due to the field-to-field variance and the actual evolution at z$<$4 we need survey sizes on the order of 1\,deg$^2$, while at z$>$4 the survey size should be above $2\times 10^{-2} \, \rm deg^2$. We see this steep change because the $\rm \rho_{H_2}$ evolution at z$\gtrsim$4 is faster than at z$\lesssim$4. However, in order for spectral scans to reduce the field-to-field variance, it is not only the covered area on the sky that matters but also the observing frequency range (redshift range) they use, since the goal is to succeed as large comoving volumes as possible.

Both the clustering and the shot noise components of the LIM power spectra significantly vary from one field to another. This can strongly affect the inference of physical quantities, like the line LF and subsequently the SFRD. Moreover, we show that the brightest nonmasked sources, which are usually just below the detection limit, are the ones that contribute the most to the shot noise level. This means that: 1) the shot noise level is sensitive to the flux cut one chooses when masking the brightest sources, 2) the study of the shot noise level will target the subdetection threshold sources, which can be more efficiently studied by classical surveys.

Additionally, we show that there is a significant level of correlation between the clustering and the shot noise power, which increases with the field size and is higher for the [CII] than CO line. This could affect the interpretation of the clustering power leading to over or underestimate of the physical quantities deriving from it. Additionally, because of this correlation, there is no analytical formula that describes how the clustering and Poisson variance should be combined to get the total variance. 

The field-to-field variance of the power spectra (both clustering and shot noise) depends on the survey size, the observed frequency as well as on the frequency slice used to compute the 2D power spectra. We provided two models, one for CO and one for [CII] (separately for the clustering and the shot noise part), that estimate the level of the expected variance given the above mentioned quantities. This is a useful tool for CONCERTO and any other intensity mapping experiment that aims to optimize its configuration.

By using our variance estimation model we forecast how the field-to-field variance will limit the precision of the constraints provided by some current and near-future LIM experiments. The significant field-to-field variance of both the clustering and shot-noise components of the power spectra along with the correlation between them provides a quite pessimistic forecast for CONCERTO. At low frequencies the [CII] power spectrum exceeds the level of 50\% uncertainty, while for higher frequencies it is slightly smaller but still well above 20\%. On the other hand, COMAP and FYST/Prime-Cam, designed to cover a bigger field (4\,deg$^2$), seem to be more promising for the CO and [CII] power spectra accurate modeling.

The Uchuu extension of the SIDES code and its products are publicly available \footnote{\url{https://cesamsi.lam.fr/instance/sides/home}}. We also offer some complementary tools to help the community explore the products of our simulation and adapt them to their needs.

\begin{acknowledgements}
This project has received funding from the European Research Council (ERC) under the European Union’s Horizon 2020 research and innovation programme (grant agreement No 788212) and from the Excellence Initiative of Aix-Marseille University-A*Midex, a French “Investissements d’Avenir” programme. The Uchuu simulations were carried out on Aterui II supercomputer at Center for Computational Astrophysics, CfCA, of National Astronomical Observatory of Japan, and the K computer at the RIKEN Advanced Institute for Computational Science. The Uchuu data effort has made use of the skun@IAA\_RedIRIS and skun6@IAA computer facilities managed by the IAA-CSIC in Spain (MICINN EU-Feder grant EQC2018-004366-P). 

T.I. has been supported by IAAR Research Support Program in Chiba University Japan, 
MEXT/JSPS KAKENHI (Grant Number JP19KK0344 and JP21H01122), 
MEXT as ``Program for Promoting Researches on the Supercomputer Fugaku'' (JPMXP1020200109),
and JICFuS.

AF, AP, LV acknowledges support from the ERC Advanced Grant INTERSTELLAR H2020/740120 (PI: Ferrara).

MA acknowledges support from FONDECYT grant 1211951, ‘ANID + PCI + INSTITUTO MAX PLANCK DE ASTRONOMIA MPG 190030’, ‘ANID + PCI + REDES 190194’, and ANID BASAL project FB210003.

SAC acknowledges funding from  {\it Consejo Nacional de
Investigaciones Cient\'ificas y T\'ecnicas} (CONICET, PIP-2876), 
{\it Agencia Nacional de Promoci\'on de la Investigaci\'on, el Desarrollo Tecnol\'ogico y la Innovaci\'on} (Agencia I+D+i, PICT-2018-3743), and {\it Universidad Nacional de La Plata} (G11-150), Argentina.

GM acknowledges support by Swedish Research Council grant 2020-04691.
\end{acknowledgements}

\bibliographystyle{aa}
\bibliography{aa}

\begin{thebibliography}{109}
\expandafter\ifx\csname natexlab\endcsname\relax\def\natexlab#1{#1}\fi

\bibitem[{Amblard {et~al.}(2011)Amblard, Cooray, Serra, Altieri, Arumugam,
  Aussel, Blain, Bock, Boselli, Buat, {Castro-Rodriguez}, Cava, Chanial,
  Chapin, Clements, Conley, Conversi, Dowell, Dwek, Eales, Elbaz, Farrah,
  Franceschini, Gear, Glenn, Griffin, Halpern, Hatziminaoglou, Ibar, Isaak,
  Ivison, Khostovan, Lagache, Levenson, Lu, Madden, Maffei, Mainetti,
  Marchetti, Marsden, {Mitchell-Wynne}, Nguyen, O'Halloran, Oliver, Omont,
  Page, Panuzzo, Papageorgiou, Pearson, {Perez-Fournon}, Pohlen, Rangwala,
  Roseboom, {Rowan-Robinson}, Portal, Schulz, Scott, Seymour, Shupe, Smith,
  Stevens, Symeonidis, Trichas, Tugwell, Vaccari, Valiante, Valtchanov, Vieira,
  Vigroux, Wang, Ward, Wright, Xu, \& Zemcov}]{amblard2011}
Amblard, A., Cooray, A., Serra, P., {et~al.} 2011, Nature, 470, 510

\bibitem[{{Behroozi} {et~al.}(2020){Behroozi}, {Conroy}, {Wechsler}, {Hearin},
  {Williams}, {Moster}, {Yung}, {Somerville}, {Gottl{\"o}ber}, {Yepes}, \&
  {Endsley}}]{behroozi2020}
{Behroozi}, P., {Conroy}, C., {Wechsler}, R.~H., {et~al.} 2020, \mnras, 499,
  5702

\bibitem[{{Behroozi} {et~al.}(2019){Behroozi}, {Wechsler}, {Hearin}, \&
  {Conroy}}]{behroozi2019}
{Behroozi}, P., {Wechsler}, R.~H., {Hearin}, A.~P., \& {Conroy}, C. 2019,
  \mnras, 488, 3143

\bibitem[{Behroozi {et~al.}(2010)Behroozi, Conroy, \& Wechsler}]{behroozi2010}
Behroozi, P.~S., Conroy, C., \& Wechsler, R.~H. 2010, ApJ, 717, 379

\bibitem[{Behroozi {et~al.}(2013{\natexlab{a}})Behroozi, Wechsler, \&
  Conroy}]{behroozi2013}
Behroozi, P.~S., Wechsler, R.~H., \& Conroy, C. 2013{\natexlab{a}}, ApJ, 770,
  57

\bibitem[{Behroozi {et~al.}(2013{\natexlab{b}})Behroozi, Wechsler, \&
  Wu}]{behroozi2013b}
Behroozi, P.~S., Wechsler, R.~H., \& Wu, H.-Y. 2013{\natexlab{b}}, ApJ, 762,
  109

\bibitem[{Behroozi {et~al.}(2013{\natexlab{c}})Behroozi, Wechsler, Wu, Busha,
  Klypin, \& Primack}]{behroozi2013a}
Behroozi, P.~S., Wechsler, R.~H., Wu, H.-Y., {et~al.} 2013{\natexlab{c}}, ApJ,
  763, 18

\bibitem[{Bertincourt {et~al.}(2016)Bertincourt, Lagache, Martin, Schulz,
  Conversi, Dassas, Maurin, Abergel, Beelen, Bernard, Crill, Dole, Eales,
  Gudmundsson, Lellouch, Moreno, \& Perdereau}]{bertincourt2016}
Bertincourt, B., Lagache, G., Martin, P.~G., {et~al.} 2016, A\&A, 588, A107

\bibitem[{B{\'e}thermin {et~al.}(2015)B{\'e}thermin, Daddi, Magdis, Lagos,
  Sargent, Albrecht, Aussel, Bertoldi, Buat, Galametz, Heinis, Ilbert, Karim,
  Koekemoer, Lacey, Le~Floc'h, Navarrete, Pannella, Schreiber, Smol{\v
  c}i{\'c}, Symeonidis, \& Viero}]{bethermin2015}
B{\'e}thermin, M., Daddi, E., Magdis, G., {et~al.} 2015, A\&A, 573, A113

\bibitem[{B{\'e}thermin {et~al.}(2012)B{\'e}thermin, Daddi, Magdis, Sargent,
  Hezaveh, Elbaz, Le~Borgne, Mullaney, Pannella, Buat, Charmandaris, Lagache,
  \& Scott}]{bethermin2012}
B{\'e}thermin, M., Daddi, E., Magdis, G., {et~al.} 2012, ApJ, 757, L23

\bibitem[{B{\'e}thermin {et~al.}(2010)B{\'e}thermin, Dole, Beelen, \&
  Aussel}]{bethermin2010}
B{\'e}thermin, M., Dole, H., Beelen, A., \& Aussel, H. 2010, A\&A, 512, A78

\bibitem[{B{\'e}thermin {et~al.}(2020)B{\'e}thermin, Fudamoto, Ginolfi,
  Loiacono, Khusanova, Capak, Cassata, Faisst, Le~F{\`e}vre, Schaerer,
  Silverman, Yan, Amorin, Bardelli, Boquien, Cimatti, Davidzon,
  {Dessauges-Zavadsky}, Fujimoto, Gruppioni, Hathi, Ibar, Jones, Koekemoer,
  Lagache, Lemaux, Moreau, Oesch, Pozzi, Riechers, Talia, Toft, Vallini,
  Vergani, Zamorani, \& Zucca}]{bethermin2020}
B{\'e}thermin, M., Fudamoto, Y., Ginolfi, M., {et~al.} 2020, A\&A, 643, A2

\bibitem[{{B{\'e}thermin} {et~al.}(2022){B{\'e}thermin}, {Gkogkou}, {Van
  Cuyck}, {Lagache}, {Beelen}, {Aravena}, {Benoit}, {Bounmy}, {Calvo},
  {Catalano}, {de Batz de Trenquelleon}, {De Breuck}, {Fasano}, {Ferrara},
  {Goupy}, {Hoarau}, {Horellou}, {Hu}, {Julia}, {Knudsen}, {Lambert},
  {Macias-Perez}, {Marpaud}, {Monfardini}, {Pallottini}, {Ponthieu}, {Roehlly},
  {Vallini}, {Walter}, \& {Weiss}}]{bethermin2022}
{B{\'e}thermin}, M., {Gkogkou}, A., {Van Cuyck}, M., {et~al.} 2022, \aap, 667,
  A156

\bibitem[{{B{\'e}thermin} {et~al.}(2013){B{\'e}thermin}, {Wang}, {Dor{\'e}},
  {Lagache}, {Sargent}, {Daddi}, {Cousin}, \& {Aussel}}]{bethermin2013}
{B{\'e}thermin}, M., {Wang}, L., {Dor{\'e}}, O., {et~al.} 2013, \aap, 557, A66

\bibitem[{B{\'e}thermin {et~al.}(2017)B{\'e}thermin, Wu, Lagache, Davidzon,
  Ponthieu, Cousin, Wang, Dore, Daddi, \& Lapi}]{bethermin2017}
B{\'e}thermin, M., Wu, H.-Y., Lagache, G., {et~al.} 2017, A\&A, 607, A89

\bibitem[{{Bisigello} {et~al.}(2022){Bisigello}, {Vallini}, {Gruppioni},
  {Esposito}, {Calura}, {Delvecchio}, {Feltre}, {Pozzi}, \&
  {Rodighiero}}]{bisigello2022}
{Bisigello}, L., {Vallini}, L., {Gruppioni}, C., {et~al.} 2022, \aap, 666, A193

\bibitem[{Blake \& Wall(2002)}]{blake2002}
Blake, C. \& Wall, J. 2002, Monthly Notices of the Royal Astronomical Society,
  337, 993

\bibitem[{Bournaud {et~al.}(2015)Bournaud, Daddi, Weiss, Renaud, Mastropietro,
  \& Teyssier}]{bournaud2015}
Bournaud, F., Daddi, E., Weiss, A., {et~al.} 2015, A\&A, 575, A56

\bibitem[{Bouwens {et~al.}(2015)Bouwens, Illingworth, Oesch, Trenti, Labb{\'e},
  Bradley, Carollo, {van Dokkum}, Gonzalez, Holwerda, Franx, Spitler, Smit, \&
  Magee}]{bouwens2015}
Bouwens, R.~J., Illingworth, G.~D., Oesch, P.~A., {et~al.} 2015, ApJ, 803, 34

\bibitem[{Capak {et~al.}(2015)Capak, Carilli, Jones, Casey, Riechers, Sheth,
  Carollo, Ilbert, Karim, LeFevre, Lilly, Scoville, Smolcic, \&
  Yan}]{capak2015}
Capak, P.~L., Carilli, C., Jones, G., {et~al.} 2015, Nature, 522, 455

\bibitem[{Carilli \& Walter(2013)}]{carilli2013}
Carilli, C. \& Walter, F. 2013, Annu. Rev. Astron. Astrophys., 51, 105

\bibitem[{Carlson \& White(2010)}]{carlson2010}
Carlson, J. \& White, M. 2010, ApJS, 190, 311

\bibitem[{{Carniani} {et~al.}(2020){Carniani}, {Ferrara}, {Maiolino},
  {Castellano}, {Gallerani}, {Fontana}, {Kohandel}, {Lupi}, {Pallottini},
  {Pentericci}, {Vallini}, \& {Vanzella}}]{carniani2020}
{Carniani}, S., {Ferrara}, A., {Maiolino}, R., {et~al.} 2020, \mnras, 499, 5136

\bibitem[{{CCAT-Prime collaboration} {et~al.}(2021){CCAT-Prime collaboration},
  Aravena, Austermann, Basu, Battaglia, Beringue, Bertoldi, Bigiel, Bond,
  Breysse, Broughton, Bustos, Chapman, Charmetant, Choi, Chung, Clark, Cothard,
  Crites, Dev, Douglas, Duell, Ebina, Erler, Fich, Fissel, Foreman, Gao,
  Garc{\'i}a, Giovanelli, Haynes, Hensley, Herter, Higgins, Huber, Hubmayr,
  Johnstone, Karoumpis, Keating, Komatsu, Li, Magnelli, Matthews, Meerburg,
  Meyers, Muralidhara, Murray, Niemack, Nikola, Okada, Riechers, Rosolowsky,
  Roy, Sadavoy, Schaaf, Schilke, Scott, Simon, Sinclair, Sivakoff, Stacey,
  Stutz, Stutzki, Tahani, Thanjavur, Timmermann, Ullom, {van Engelen},
  Vavagiakis, Vissers, Wheeler, White, Zhu, \& Zou}]{collaboration2021}
{CCAT-Prime collaboration}, C.-P., Aravena, M., Austermann, J.~E., {et~al.}
  2021, {{CCAT-prime Collaboration}}: {{Science Goals}} and {{Forecasts}} with
  {{Prime-Cam}} on the {{Fred Young Submillimeter Telescope}}

\bibitem[{Chabrier(2003)}]{chabrier2003}
Chabrier, G. 2003, PUBL ASTRON SOC PAC, 115, 763

\bibitem[{Cheng {et~al.}(2020)Cheng, Chang, \& Bock}]{cheng2020}
Cheng, Y.-T., Chang, T.-C., \& Bock, J.~J. 2020, ApJ, 901, 142

\bibitem[{Cleary {et~al.}(2021)Cleary, Borowska, Breysse, Catha, Chung, Church,
  Dickinson, Eriksen, Foss, Gundersen, Harper, Harris, Hobbs, H{\aa}vard, Ihle,
  Kim, Kocz, Lamb, Lunde, Padmanabhan, Pearson, Philip, Powell, Rasmussen,
  Readhead, Rennie, Silva, Stutzer, Uzgil, Watts, Wehus, Woody, Basoalto, Bond,
  Dunne, Gaier, Hensley, Keating, Lawrence, Murray, Reeves, Viero, \&
  Wechsler}]{cleary2021}
Cleary, K.~A., Borowska, J., Breysse, P.~C., {et~al.} 2021, {{COMAP Early
  Science}}: {{I}}. {{Overview}}

\bibitem[{Conroy {et~al.}(2006)Conroy, Wechsler, \& Kravtsov}]{conroy2006}
Conroy, C., Wechsler, R.~H., \& Kravtsov, A.~V. 2006, ApJ, 647, 201

\bibitem[{Cooray {et~al.}(2010)Cooray, Amblard, Wang, Altieri, Arumugam, Auld,
  Aussel, Babbedge, Blain, Bock, Boselli, Buat, Burgarella, {Castro-Rodriguez},
  Cava, Chanial, Clements, Conley, Conversi, Dowell, Dwek, Eales, Elbaz,
  Farrah, Fox, Franceschini, Gear, Glenn, Griffin, Halpern, Hatziminaoglou,
  Ibar, Isaak, Ivison, Khostovan, Lagache, Levenson, Lu, Madden, Maffei,
  Mainetti, Marchetti, Marsden, {Mitchell-Wynne}, Mortier, Nguyen, O'Halloran,
  Oliver, Omont, Page, Panuzzo, Papageorgiou, Pearson, {P{\'e}rez-Fournon},
  Pohlen, Rawlings, Raymond, Rigopoulou, Rizzo, Roseboom, {Rowan-Robinson},
  Portal, Schulz, Scott, Serra, Seymour, Shupe, Smith, Stevens, Symeonidis,
  Trichas, Tugwell, Vaccari, Valtchanov, Vieira, Vigroux, Ward, Wright, Xu, \&
  Zemcov}]{cooray2010}
Cooray, A., Amblard, A., Wang, L., {et~al.} 2010, A\&A, 518, L22

\bibitem[{Crites {et~al.}(2014)Crites, Bock, Bradford, Chang, Cooray, Duband,
  Gong, {Hailey-Dunsheath}, Hunacek, Koch, Li, O'Brient, Prouve, Shirokoff,
  Silva, Staniszewski, Uzgil, \& Zemcov}]{crites2014}
Crites, A.~T., Bock, J.~J., Bradford, C.~M., {et~al.} 2014, in {{SPIE
  Astronomical Telescopes}} + {{Instrumentation}}, ed. W.~S. Holland \&
  J.~Zmuidzinas, {Montr\'eal, Quebec, Canada}, 91531W

\bibitem[{Crocce {et~al.}(2006)Crocce, Pueblas, \& Scoccimarro}]{crocce2006}
Crocce, M., Pueblas, S., \& Scoccimarro, R. 2006, Monthly Notices of the Royal
  Astronomical Society, 373, 369

\bibitem[{Daddi {et~al.}(2015)Daddi, Dannerbauer, Liu, Aravena, Bournaud,
  Walter, Riechers, Magdis, Sargent, B{\'e}thermin, Carilli, Cibinel,
  Dickinson, Elbaz, Gao, Gobat, Hodge, \& Krips}]{daddi2015}
Daddi, E., Dannerbauer, H., Liu, D., {et~al.} 2015, A\&A, 577, A46

\bibitem[{Daddi {et~al.}(2007)Daddi, Dickinson, Morrison, Chary, Cimatti,
  Elbaz, Frayer, Renzini, Pope, Alexander, Bauer, Giavalisco, Huynh, Kurk, \&
  Mignoli}]{daddi2007}
Daddi, E., Dickinson, M., Morrison, G., {et~al.} 2007, ApJ, 670, 156

\bibitem[{Davidzon {et~al.}(2018)Davidzon, Ilbert, Faisst, Sparre, \&
  Capak}]{davidzon2018}
Davidzon, I., Ilbert, O., Faisst, A.~L., Sparre, M., \& Capak, P.~L. 2018, ApJ,
  852, 107

\bibitem[{De~Looze {et~al.}(2014)De~Looze, Cormier, Lebouteiller, Madden, Baes,
  Bendo, Boquien, Boselli, Clements, Cortese, Cooray, Galametz, Galliano,
  {Gracia-Carpio}, Isaak, Karczewski, Parkin, Pellegrini, {Remy-Ruyer},
  Spinoglio, Smith, \& Sturm}]{delooze2014}
De~Looze, I., Cormier, D., Lebouteiller, V., {et~al.} 2014, A\&A, 568, A62

\bibitem[{Decarli {et~al.}(2020)Decarli, Aravena, Boogaard, Carilli,
  {Gonz{\'a}lez-L{\'o}pez}, Walter, Cortes, Cox, {da Cunha}, Daddi,
  {D{\'i}az-Santos}, Hodge, Inami, Neeleman, Novak, Oesch, Popping, Riechers,
  Smail, Uzgil, {van der Werf}, Wagg, \& Weiss}]{decarli2020}
Decarli, R., Aravena, M., Boogaard, L., {et~al.} 2020, ApJ, 902, 110

\bibitem[{{Decarli} {et~al.}(2022){Decarli}, {Pensabene}, {Venemans}, {Walter},
  {Ba{\~n}ados}, {Bertoldi}, {Carilli}, {Cox}, {Fan}, {Farina}, {Ferkinhoff},
  {Groves}, {Li}, {Mazzucchelli}, {Neri}, {Riechers}, {Uzgil}, {Wang}, {Wang},
  {Weiss}, {Winters}, \& {Yang}}]{decarli2022}
{Decarli}, R., {Pensabene}, A., {Venemans}, B., {et~al.} 2022, \aap, 662, A60

\bibitem[{Decarli {et~al.}(2019)Decarli, Walter, {G{\'o}nzalez-L{\'o}pez},
  Aravena, Boogaard, Carilli, Cox, Daddi, Popping, Riechers, Uzgil, Weiss,
  Assef, Bacon, Bauer, Bertoldi, Bouwens, Contini, Cortes, da~Cunha,
  {D{\'i}az-Santos}, Elbaz, Inami, Hodge, Ivison, F{\`e}vre, Magnelli, Novak,
  Oesch, Rix, Sargent, Smail, Swinbank, Somerville, van~der Werf, Wagg, \&
  Wisotzki}]{decarli2019}
Decarli, R., Walter, F., {G{\'o}nzalez-L{\'o}pez}, J., {et~al.} 2019, ApJ, 882,
  138

\bibitem[{{Despali} {et~al.}(2016){Despali}, {Giocoli}, {Angulo}, {Tormen},
  {Sheth}, {Baso}, \& {Moscardini}}]{despali2016}
{Despali}, G., {Giocoli}, C., {Angulo}, R.~E., {et~al.} 2016, \mnras, 456, 2486

\bibitem[{Driver \& Robotham(2010)}]{driver2010}
Driver, S.~P. \& Robotham, A. S.~G. 2010, 10

\bibitem[{Faisst {et~al.}(2020)Faisst, Schaerer, Lemaux, Oesch, Fudamoto,
  Cassata, B{\'e}thermin, Capak, Le~F{\`e}vre, Silverman, Yan, Ginolfi,
  Koekemoer, Morselli, Amor{\'i}n, Bardelli, Boquien, Brammer, Cimatti,
  {Dessauges-Zavadsky}, Fujimoto, Gruppioni, Hathi, Hemmati, Ibar, Jones,
  Khusanova, Loiacono, Pozzi, Talia, Tasca, Riechers, Rodighiero, Romano,
  Scoville, Toft, Vallini, Vergani, Zamorani, \& Zucca}]{faisst2020}
Faisst, A.~L., Schaerer, D., Lemaux, B.~C., {et~al.} 2020, ApJS, 247, 61

\bibitem[{{Ferrara} {et~al.}(2019){Ferrara}, {Vallini}, {Pallottini},
  {Gallerani}, {Carniani}, {Kohandel}, {Decataldo}, \& {Behrens}}]{ferrara2019}
{Ferrara}, A., {Vallini}, L., {Pallottini}, A., {et~al.} 2019, \mnras, 489, 1

\bibitem[{Fudamoto {et~al.}(2021)Fudamoto, Oesch, Schouws, Stefanon, Smit,
  Bouwens, Bowler, Endsley, Gonzalez, Inami, Labbe, Stark, Aravena, Barrufet,
  {da Cunha}, Dayal, Ferrara, Graziani, Hodge, Hutter, Li, De~Looze,
  Nanayakkara, Pallottini, Riechers, Schneider, Ucci, {van der Werf}, \&
  White}]{fudamoto2021}
Fudamoto, Y., Oesch, P.~A., Schouws, S., {et~al.} 2021, Nature, 597, 489

\bibitem[{Gruppioni {et~al.}(2020)Gruppioni, B{\'e}thermin, Loiacono,
  Le~F{\`e}vre, Capak, Cassata, Faisst, Schaerer, Silverman, Yan, Bardelli,
  Boquien, Carraro, Cimatti, {Dessauges-Zavadsky}, Ginolfi, Fujimoto, Hathi,
  Jones, Khusanova, Koekemoer, Lagache, Lemaux, Oesch, Pozzi, Riechers,
  Rodighiero, Romano, Talia, Vallini, Vergani, Zamorani, \&
  Zucca}]{gruppioni2020}
Gruppioni, C., B{\'e}thermin, M., Loiacono, F., {et~al.} 2020, A\&A, 643, A8

\bibitem[{Hall {et~al.}(2010)Hall, Keisler, Knox, Reichardt, Ade, Aird, Benson,
  Bleem, Carlstrom, Chang, Cho, Crawford, Crites, {de Haan}, Dobbs, George,
  Halverson, Holder, Holzapfel, Hrubes, Joy, Lee, Leitch, Lueker, McMahon,
  Mehl, Meyer, Mohr, Montroy, Padin, Plagge, Pryke, Ruhl, Schaffer, Shaw,
  Shirokoff, Spieler, Stalder, Staniszewski, Stark, Switzer, Vanderlinde,
  Vieira, Williamson, \& Zahn}]{hall2010}
Hall, N.~R., Keisler, R., Knox, L., {et~al.} 2010, ApJ, 718, 632

\bibitem[{Hezaveh \& Holder(2011)}]{hezaveh2011}
Hezaveh, Y.~D. \& Holder, G.~P. 2011, ApJ, 734, 52

\bibitem[{Hilbert {et~al.}(2007)Hilbert, White, Hartlap, \&
  Schneider}]{hilbert2007}
Hilbert, S., White, S. D.~M., Hartlap, J., \& Schneider, P. 2007, Monthly
  Notices of the Royal Astronomical Society, 382, 121

\bibitem[{{Hopkins} {et~al.}(2018){Hopkins}, {Wetzel}, {Kere{\v{s}}},
  {Faucher-Gigu{\`e}re}, {Quataert}, {Boylan-Kolchin}, {Murray}, {Hayward},
  {Garrison-Kimmel}, {Hummels}, {Feldmann}, {Torrey}, {Ma},
  {Angl{\'e}s-Alc{\'a}zar}, {Su}, {Orr}, {Schmitz}, {Escala}, {Sanderson},
  {Grudi{\'c}}, {Hafen}, {Kim}, {Fitts}, {Bullock}, {Wheeler}, {Chan},
  {Elbert}, \& {Narayanan}}]{hopkins2018}
{Hopkins}, P.~F., {Wetzel}, A., {Kere{\v{s}}}, D., {et~al.} 2018, \mnras, 480,
  800

\bibitem[{Ishigaki {et~al.}(2018)Ishigaki, Kawamata, Ouchi, Oguri, Shimasaku,
  \& Ono}]{ishigaki2018}
Ishigaki, M., Kawamata, R., Ouchi, M., {et~al.} 2018, ApJ, 854, 73

\bibitem[{Ishiyama {et~al.}(2009)Ishiyama, Fukushige, \& Makino}]{ishiyama2009}
Ishiyama, T., Fukushige, T., \& Makino, J. 2009, Publ Astron Soc Jpn, 61, 1319

\bibitem[{{Ishiyama} {et~al.}(2012){Ishiyama}, {Nitadori}, \&
  {Makino}}]{ishiyama2012}
{Ishiyama}, T., {Nitadori}, K., \& {Makino}, J. 2012, arXiv e-prints,
  arXiv:1211.4406

\bibitem[{Ishiyama {et~al.}(2021)Ishiyama, Prada, Klypin, Sinha, Metcalf,
  Jullo, Altieri, Cora, Croton, {de la Torre}, {Mill{\'a}n-Calero}, Oogi,
  Ruedas, \& {Vega-Mart{\'i}nez}}]{ishiyama2021}
Ishiyama, T., Prada, F., Klypin, A.~A., {et~al.} 2021, Monthly Notices of the
  Royal Astronomical Society, 506, 4210

\bibitem[{Karkare {et~al.}(2022)Karkare, Anderson, Barry, Benson, Carlstrom,
  Cecil, Chang, Dobbs, Hollister, Keating, Marrone, McMahon, Montgomery, Pan,
  Robson, Rouble, Shirokoff, \& Smecher}]{karkare2022}
Karkare, K.~S., Anderson, A.~J., Barry, P.~S., {et~al.} 2022, J Low Temp Phys
  [\eprint[arXiv]{2111.04631}]

\bibitem[{Keenan {et~al.}(2020)Keenan, Marrone, \& Keating}]{keenan2020}
Keenan, R.~P., Marrone, D.~P., \& Keating, G.~K. 2020, ApJ, 904, 127

\bibitem[{Kennicutt(1998)}]{kennicutt1998}
Kennicutt, R.~C. 1998, Annual Review of Astronomy and Astrophysics, 36, 189

\bibitem[{Khusanova {et~al.}(2021)Khusanova, Bethermin, Le~F{\`e}vre, Capak,
  Faisst, Schaerer, Silverman, Cassata, Yan, Ginolfi, Fudamoto, Loiacono,
  Amorin, Bardelli, Boquien, Cimatti, {Dessauges-Zavadsky}, Gruppioni, Hathi,
  Jones, Koekemoer, Lagache, Maiolino, Lemaux, Oesch, Pozzi, Riechers, Romano,
  Talia, Toft, Vergani, Zamorani, \& Zucca}]{khusanova2021}
Khusanova, Y., Bethermin, M., Le~F{\`e}vre, O., {et~al.} 2021, A\&A, 649, A152

\bibitem[{Klypin {et~al.}(2011)Klypin, {Trujillo-Gomez}, \&
  Primack}]{klypin2011}
Klypin, A.~A., {Trujillo-Gomez}, S., \& Primack, J. 2011, ApJ, 740, 102

\bibitem[{Kovetz {et~al.}(2017)Kovetz, Viero, Lidz, Newburgh, Rahman, Switzer,
  Kamionkowski, Aguirre, Alvarez, Bock, Bond, Bower, Bradford, Breysse, Bull,
  Chang, Cheng, Chung, Cleary, Corray, Crites, Croft, Dor{\'e}, Eastwood,
  Ferrara, Fonseca, Jacobs, Keating, Lagache, Lakhlani, Liu, Moodley, Murray,
  P{\'e}nin, Popping, Pullen, Reichers, Saito, Saliwanchik, Santos, Somerville,
  Stacey, Stein, {Villaescusa-Navarro}, Visbal, Weltman, Wolz, \&
  Zemcov}]{kovetz2017}
Kovetz, E.~D., Viero, M.~P., Lidz, A., {et~al.} 2017, arXiv:1709.09066
  [astro-ph] [\eprint[arXiv]{1709.09066}]

\bibitem[{Kravtsov {et~al.}(2004)Kravtsov, Berlind, Wechsler, Klypin,
  Gottlober, Allgood, \& Primack}]{kravtsov2004}
Kravtsov, A.~V., Berlind, A.~A., Wechsler, R.~H., {et~al.} 2004, ApJ, 609, 35

\bibitem[{Lagache {et~al.}(2007)Lagache, Bavouzet, {Fernandez-Conde}, Ponthieu,
  Rodet, Dole, {Miville-Desch{\^e}nes}, \& Puget}]{lagache2007}
Lagache, G., Bavouzet, N., {Fernandez-Conde}, N., {et~al.} 2007, ApJ, 665, L89

\bibitem[{Lagache {et~al.}(2020)Lagache, Bethermin, Montier, Serra, \&
  Tucci}]{lagache2020}
Lagache, G., Bethermin, M., Montier, L., Serra, P., \& Tucci, M. 2020, A\&A,
  642, A232

\bibitem[{Lagache {et~al.}(2018)Lagache, Cousin, \& Chatzikos}]{lagache2018}
Lagache, G., Cousin, M., \& Chatzikos, M. 2018, A\&A, 609, A130

\bibitem[{{Lagache} {et~al.}(1999){Lagache}, {Puget}, \&
  {Gispert}}]{lagache1999}
{Lagache}, G., {Puget}, J.-L., \& {Gispert}, R. 1999, \apss, 269, 263

\bibitem[{Le~F{\`e}vre {et~al.}(2020)Le~F{\`e}vre, B{\'e}thermin, Faisst,
  Jones, Capak, Cassata, Silverman, Schaerer, Yan, Amorin, Bardelli, Boquien,
  Cimatti, {Dessauges-Zavadsky}, Giavalisco, Hathi, Fudamoto, Fujimoto,
  Ginolfi, Gruppioni, Hemmati, Ibar, Koekemoer, Khusanova, Lagache, Lemaux,
  Loiacono, Maiolino, Mancini, Narayanan, Morselli, {M{\'e}ndez-Hern{\`a}ndez},
  Oesch, Pozzi, Romano, Riechers, Scoville, Talia, Tasca, Thomas, Toft,
  Vallini, Vergani, Walter, Zamorani, \& Zucca}]{lefevre2020}
Le~F{\`e}vre, O., B{\'e}thermin, M., Faisst, A., {et~al.} 2020, A\&A, 643, A1

\bibitem[{Loiacono {et~al.}(2021)Loiacono, Decarli, Gruppioni, Talia, Cimatti,
  Zamorani, Pozzi, Yan, Lemaux, Riechers, Le~F{\`e}vre, B{\`e}thermin, Capak,
  Cassata, Faisst, Schaerer, Silverman, Bardelli, Boquien, Burkutean,
  {Dessauges-Zavadsky}, Fudamoto, Fujimoto, Ginolfi, Hathi, Jones, Khusanova,
  Koekemoer, Lagache, Lubin, Massardi, Oesch, Romano, Vallini, Vergani, \&
  Zucca}]{loiacono2021}
Loiacono, F., Decarli, R., Gruppioni, C., {et~al.} 2021, A\&A, 646, A76

\bibitem[{Madau \& Dickinson(2014)}]{madau2014}
Madau, P. \& Dickinson, M. 2014, Annu. Rev. Astron. Astrophys., 52, 415

\bibitem[{Maddox {et~al.}(2010)Maddox, Dunne, Rigby, Eales, Cooray, Scott,
  Peacock, Negrello, Smith, Benford, Amblard, Auld, Baes, Bonfield, Burgarella,
  Buttiglione, Cava, Clements, Dariush, {de Zotti}, Dye, Frayer, Fritz,
  {Gonzalez-Nuevo}, Herranz, Ibar, Ivison, Jarvis, Lagache, Leeuw,
  {Lopez-Caniego}, Pascale, Pohlen, Rodighiero, Samui, Serjeant, Temi,
  Thompson, \& Verma}]{maddox2010}
Maddox, S.~J., Dunne, L., Rigby, E., {et~al.} 2010, A\&A, 518, L11

\bibitem[{Magdis {et~al.}(2012)Magdis, Daddi, Bethermin, Sargent, Elbaz,
  Pannella, Dickinson, Dannerbauer, Da~Cunha, Walter, Rigopoulou, Charmandaris,
  Hwang, \& Kartaltepe}]{magdis2012}
Magdis, G.~E., Daddi, E., Bethermin, M., {et~al.} 2012, ApJ, 760, 6

\bibitem[{Maniyar {et~al.}(2018)Maniyar, B{\'e}thermin, \&
  Lagache}]{maniyar2018}
Maniyar, A., B{\'e}thermin, M., \& Lagache, G. 2018, A\&A, 614, A39

\bibitem[{Maniyar {et~al.}(2021)Maniyar, B{\'e}thermin, \&
  Lagache}]{maniyar2021}
Maniyar, A., B{\'e}thermin, M., \& Lagache, G. 2021, A\&A, 645, A40

\bibitem[{Moster {et~al.}(2013)Moster, Naab, \& White}]{moster2013}
Moster, B.~P., Naab, T., \& White, S. D.~M. 2013, Monthly Notices of the Royal
  Astronomical Society, 428, 3121

\bibitem[{Moster {et~al.}(2011)Moster, Somerville, Newman, \& Rix}]{moster2011}
Moster, B.~P., Somerville, R.~S., Newman, J.~A., \& Rix, H.-W. 2011, The
  Astrophysical Journal, 8

\bibitem[{Niemiec {et~al.}(2019)Niemiec, Jullo, Giocoli, Limousin, \&
  Jauzac}]{niemiec2019}
Niemiec, A., Jullo, E., Giocoli, C., Limousin, M., \& Jauzac, M. 2019, Monthly
  Notices of the Royal Astronomical Society, 487, 653

\bibitem[{{Pallottini} {et~al.}(2017){Pallottini}, {Ferrara}, {Bovino},
  {Vallini}, {Gallerani}, {Maiolino}, \& {Salvadori}}]{pallottini2017}
{Pallottini}, A., {Ferrara}, A., {Bovino}, S., {et~al.} 2017, \mnras, 471, 4128

\bibitem[{{Pallottini} {et~al.}(2022){Pallottini}, {Ferrara}, {Gallerani},
  {Behrens}, {Kohandel}, {Carniani}, {Vallini}, {Salvadori}, {Gelli},
  {Sommovigo}, {D'Odorico}, {Di Mascia}, \& {Pizzati}}]{pallottini2022}
{Pallottini}, A., {Ferrara}, A., {Gallerani}, S., {et~al.} 2022, \mnras, 513,
  5621

\bibitem[{{Planck Collaboration} {et~al.}(2014){Planck Collaboration}, Ade,
  Aghanim, {Armitage-Caplan}, Arnaud, Ashdown, {Atrio-Barandela}, Aumont,
  Baccigalupi, Banday, Barreiro, Bartlett, Battaner, Benabed, Beno{\^i}t,
  {Benoit-L{\'e}vy}, Bernard, Bersanelli, Bethermin, Bielewicz, Blagrave,
  Bobin, Bock, Bonaldi, Bond, Borrill, Bouchet, Boulanger, Bridges, Bucher,
  Burigana, Butler, Cardoso, Catalano, Challinor, Chamballu, Chen, Chiang,
  Chiang, Christensen, Church, Clements, Colombi, Colombo, Couchot, Coulais,
  Crill, Curto, Cuttaia, Danese, Davies, Davis, {de Bernardis}, {de Rosa}, {de
  Zotti}, Delabrouille, Delouis, D{\'e}sert, Dickinson, Diego, Dole, Donzelli,
  Dor{\'e}, Douspis, Dupac, Efstathiou, En{\ss}lin, Eriksen, Finelli, Forni,
  Frailis, Franceschi, Galeotta, Ganga, Ghosh, Giard, {Giraud-H{\'e}raud},
  {Gonz{\'a}lez-Nuevo}, G{\'o}rski, Gratton, Gregorio, Gruppuso, Hansen,
  Hanson, Harrison, Helou, {Henrot-Versill{\'e}}, {Hern{\'a}ndez-Monteagudo},
  Herranz, Hildebrandt, Hivon, Hobson, Holmes, Hornstrup, Hovest, Huffenberger,
  Jaffe, Jaffe, Jones, Juvela, Kalberla, Keih{\"a}nen, Kerp, Keskitalo, Kisner,
  Kneissl, Knoche, Knox, Kunz, {Kurki-Suonio}, Lacasa, Lagache,
  L{\"a}hteenm{\"a}ki, Lamarre, Langer, Lasenby, Laureijs, Lawrence, Leonardi,
  {Le{\'o}n-Tavares}, Lesgourgues, Liguori, Lilje, {Linden-V{\o}rnle},
  {L{\'o}pez-Caniego}, Lubin, {Mac{\'i}as-P{\'e}rez}, Maffei, Maino, Mandolesi,
  Maris, Marshall, Martin, {Mart{\'i}nez-Gonz{\'a}lez}, Masi, Matarrese,
  Matthai, Mazzotta, Melchiorri, Mendes, Mennella, Migliaccio, Mitra,
  {Miville-Desch{\^e}nes}, Moneti, Montier, Morgante, Mortlock, Munshi, Murphy,
  Naselsky, Nati, Natoli, Netterfield, {N{\o}rgaard-Nielsen}, Noviello,
  Novikov, Novikov, Osborne, Oxborrow, Paci, Pagano, Pajot, Paladini, Paoletti,
  Partridge, Pasian, Patanchon, Perdereau, Perotto, Perrotta, Piacentini, Piat,
  Pierpaoli, Pietrobon, Plaszczynski, Pointecouteau, Polenta, Ponthieu, Popa,
  Poutanen, Pratt, Pr{\'e}zeau, Prunet, Puget, Rachen, Reach, Rebolo, Reinecke,
  Remazeilles, Renault, Ricciardi, Riller, Ristorcelli, Rocha, Rosset, Roudier,
  {Rowan-Robinson}, {Rubi{\~n}o-Mart{\'i}n}, Rusholme, Sandri, Santos, Savini,
  Scott, Seiffert, Serra, Shellard, Spencer, Starck, Stolyarov, Stompor,
  Sudiwala, Sunyaev, Sureau, Sutton, {Suur-Uski}, Sygnet, Tauber, Tavagnacco,
  Terenzi, Toffolatti, Tomasi, Tristram, Tucci, Tuovinen, T{\"u}rler,
  Valenziano, Valiviita, Van~Tent, Vielva, Villa, Vittorio, Wade, Wandelt,
  Welikala, White, White, Winkel, Yvon, Zacchei, \&
  Zonca}]{planckcollaboration2014}
{Planck Collaboration}, Ade, P. A.~R., Aghanim, N., {et~al.} 2014, A\&A, 571,
  A30

\bibitem[{{Planck Collaboration} {et~al.}(2011){Planck Collaboration}, Ade,
  Aghanim, Arnaud, Ashdown, Aumont, Baccigalupi, Balbi, Banday, Barreiro,
  Bartlett, Battaner, Benabed, Beno{\^i}t, Bernard, Bersanelli, Bhatia,
  Blagrave, Bock, Bonaldi, Bonavera, Bond, Borrill, Bouchet, Bucher, Burigana,
  Cabella, Cardoso, Catalano, Cay{\'o}n, Challinor, Chamballu, Chiang, Chiang,
  Christensen, Clements, Colombi, Couchot, Coulais, Crill, Cuttaia, Danese,
  Davies, Davis, {de Bernardis}, {de Gasperis}, {de Rosa}, {de Zotti},
  Delabrouille, Delouis, D{\'e}sert, Dole, Donzelli, Dor{\'e}, D{\"o}rl,
  Douspis, Dupac, Efstathiou, En{\ss}lin, Eriksen, Finelli, Forni, Fosalba,
  Frailis, Franceschi, Galeotta, Ganga, Giard, Giardino, {Giraud-H{\'e}raud},
  {Gonz{\'a}lez-Nuevo}, G{\'o}rski, Grain, Gratton, Gregorio, Gruppuso, Hansen,
  Harrison, Helou, {Henrot-Versill{\'e}}, Herranz, Hildebrandt, Hivon, Hobson,
  Holmes, Hovest, Hoyland, Huffenberger, Jaffe, Jones, Juvela, Keih{\"a}nen,
  Keskitalo, Kisner, Kneissl, Knox, {Kurki-Suonio}, Lagache, Lamarre, Lasenby,
  Laureijs, Lawrence, Leach, Leonardi, Leroy, Lilje, {Linden-V{\o}rnle},
  Lockman, {L{\'o}pez-Caniego}, Lubin, {Mac{\'i}as-P{\'e}rez}, MacTavish,
  Maffei, Maino, Mandolesi, Mann, Maris, Martin, {Mart{\'i}nez-Gonz{\'a}lez},
  Masi, Matarrese, Matthai, Mazzotta, Melchiorri, Mendes, Mennella, Mitra,
  {Miville-Desch{\^e}nes}, Moneti, Montier, Morgante, Mortlock, Munshi, Murphy,
  Naselsky, Natoli, Netterfield, {N{\o}rgaard-Nielsen}, Novikov, Novikov,
  O'Dwyer, Oliver, Osborne, Pajot, Pasian, Patanchon, Perdereau, Perotto,
  Perrotta, Piacentini, Piat, Pinheiro~Gon{\c c}alves, Plaszczynski,
  Pointecouteau, Polenta, Ponthieu, Poutanen, Pr{\'e}zeau, Prunet, Puget,
  Rachen, Reach, Reinecke, Remazeilles, Renault, Ricciardi, Riller,
  Ristorcelli, Rocha, Rosset, {Rowan-Robinson}, {Rubi{\~n}o-Mart{\'i}n},
  Rusholme, Sandri, Santos, Savini, Scott, Seiffert, Shellard, Smoot, Starck,
  Stivoli, Stolyarov, Stompor, Sudiwala, Sunyaev, Sygnet, Tauber, Terenzi,
  Toffolatti, Tomasi, Torre, Tristram, Tuovinen, Umana, Valenziano, Vielva,
  Villa, Vittorio, Wade, Wandelt, White, Yvon, Zacchei, \&
  Zonca}]{planckcollaboration2011}
{Planck Collaboration}, Ade, P. A.~R., Aghanim, N., {et~al.} 2011, A\&A, 536,
  A18

\bibitem[{{Planck Collaboration} {et~al.}(2016{\natexlab{a}}){Planck
  Collaboration}, Ade, Aghanim, Arnaud, Ashdown, Aumont, Baccigalupi, Banday,
  Barreiro, Bartlett, Bartolo, Battaner, Battye, Benabed, Benoit,
  {Benoit-Levy}, Bernard, Bersanelli, Bielewicz, Bock, Bonaldi, Bonavera, Bond,
  Borrill, Bouchet, Boulanger, Bucher, Burigana, Butler, Calabrese, Cardoso,
  Catalano, Challinor, Chamballu, Chary, Chiang, Chluba, Christensen, Church,
  Clements, Colombi, Colombo, Combet, Coulais, Crill, Curto, Cuttaia, Danese,
  Davies, Davis, {de Bernardis}, {de Rosa}, {de Zotti}, Delabrouille, Desert,
  Di~Valentino, Dickinson, Diego, Dolag, Dole, Donzelli, Dore, Douspis, Ducout,
  Dunkley, Dupac, Efstathiou, Elsner, Ensslin, Eriksen, Farhang, Fergusson,
  Finelli, Forni, Frailis, Fraisse, Franceschi, Frejsel, Galeotta, Galli,
  Ganga, Gauthier, Gerbino, Ghosh, Giard, {Giraud-Heraud}, Giusarma, Gjerlow,
  {Gonzalez-Nuevo}, Gorski, Gratton, Gregorio, Gruppuso, Gudmundsson, Hamann,
  Hansen, Hanson, Harrison, Helou, {Henrot-Versille}, {Hernandez-Monteagudo},
  Herranz, Hildebrandt, Hivon, Hobson, Holmes, Hornstrup, Hovest, Huang,
  Huffenberger, Hurier, Jaffe, Jaffe, Jones, Juvela, Keihanen, Keskitalo,
  Kisner, Kneissl, Knoche, Knox, Kunz, {Kurki-Suonio}, Lagache, Lahteenmaki,
  Lamarre, Lasenby, Lattanzi, Lawrence, Leahy, Leonardi, Lesgourgues, Levrier,
  Lewis, Liguori, Lilje, {Linden-Vornle}, {Lopez-Caniego}, Lubin,
  {Macias-Perez}, Maggio, Maino, Mandolesi, Mangilli, Marchini, Martin,
  Martinelli, {Martinez-Gonzalez}, Masi, Matarrese, Mazzotta, McGehee,
  Meinhold, Melchiorri, Melin, Mendes, Mennella, Migliaccio, Millea, Mitra,
  {Miville-Deschenes}, Moneti, Montier, Morgante, Mortlock, Moss, Munshi,
  Murphy, Naselsky, Nati, Natoli, Netterfield, {Norgaard-Nielsen}, Noviello,
  Novikov, Novikov, Oxborrow, Paci, Pagano, Pajot, Paladini, Paoletti,
  Partridge, Pasian, Patanchon, Pearson, Perdereau, Perotto, Perrotta,
  Pettorino, Piacentini, Piat, Pierpaoli, Pietrobon, Plaszczynski,
  Pointecouteau, Polenta, Popa, Pratt, Prezeau, Prunet, Puget, Rachen, Reach,
  Rebolo, Reinecke, Remazeilles, Renault, Renzi, Ristorcelli, Rocha, Rosset,
  Rossetti, Roudier, {d'Orfeuil}, {Rowan-Robinson}, {Rubino-Martin}, Rusholme,
  Said, Salvatelli, Salvati, Sandri, Santos, Savelainen, Savini, Scott,
  Seiffert, Serra, Shellard, Spencer, Spinelli, Stolyarov, Stompor, Sudiwala,
  Sunyaev, Sutton, {Suur-Uski}, Sygnet, Tauber, Terenzi, Toffolatti, Tomasi,
  Tristram, Trombetti, Tucci, Tuovinen, Turler, Umana, Valenziano, Valiviita,
  Van~Tent, Vielva, Villa, Wade, Wandelt, Wehus, White, White, Wilkinson, Yvon,
  Zacchei, \& Zonca}]{planckcollaboration2016c}
{Planck Collaboration}, Ade, P. A.~R., Aghanim, N., {et~al.}
  2016{\natexlab{a}}, A\&A, 594, A13

\bibitem[{{Planck Collaboration} {et~al.}(2020){Planck Collaboration}, Aghanim,
  Akrami, Ashdown, Aumont, Baccigalupi, Ballardini, Banday, Barreiro, Bartolo,
  Basak, Battye, Benabed, Bernard, Bersanelli, Bielewicz, Bock, Bond, Borrill,
  Bouchet, Boulanger, Bucher, Burigana, Butler, Calabrese, Cardoso, Carron,
  Challinor, Chiang, Chluba, Colombo, Combet, Contreras, Crill, Cuttaia, {de
  Bernardis}, {de Zotti}, Delabrouille, Delouis, Di~Valentino, Diego, Dor{\'e},
  Douspis, Ducout, Dupac, Dusini, Efstathiou, Elsner, En{\ss}lin, Eriksen,
  Fantaye, Farhang, Fergusson, {Fernandez-Cobos}, Finelli, Forastieri, Frailis,
  Fraisse, Franceschi, Frolov, Galeotta, Galli, Ganga, {G{\'e}nova-Santos},
  Gerbino, Ghosh, {Gonz{\'a}lez-Nuevo}, G{\'o}rski, Gratton, Gruppuso,
  Gudmundsson, Hamann, Handley, Hansen, Herranz, Hildebrandt, Hivon, Huang,
  Jaffe, Jones, Karakci, Keih{\"a}nen, Keskitalo, Kiiveri, Kim, Kisner, Knox,
  Krachmalnicoff, Kunz, {Kurki-Suonio}, Lagache, Lamarre, Lasenby, Lattanzi,
  Lawrence, Le~Jeune, Lemos, Lesgourgues, Levrier, Lewis, Liguori, Lilje,
  Lilley, Lindholm, {L{\'o}pez-Caniego}, Lubin, Ma, {Mac{\'i}as-P{\'e}rez},
  Maggio, Maino, Mandolesi, Mangilli, {Marcos-Caballero}, Maris, Martin,
  Martinelli, {Mart{\'i}nez-Gonz{\'a}lez}, Matarrese, Mauri, McEwen, Meinhold,
  Melchiorri, Mennella, Migliaccio, Millea, Mitra, {Miville-Desch{\^e}nes},
  Molinari, Montier, Morgante, Moss, Natoli, {N{\o}rgaard-Nielsen}, Pagano,
  Paoletti, Partridge, Patanchon, Peiris, Perrotta, Pettorino, Piacentini,
  Polastri, Polenta, Puget, Rachen, Reinecke, Remazeilles, Renzi, Rocha,
  Rosset, Roudier, {Rubi{\~n}o-Mart{\'i}n}, {Ruiz-Granados}, Salvati, Sandri,
  Savelainen, Scott, Shellard, Sirignano, Sirri, Spencer, Sunyaev, {Suur-Uski},
  Tauber, Tavagnacco, Tenti, Toffolatti, Tomasi, Trombetti, Valenziano,
  Valiviita, Van~Tent, Vibert, Vielva, Villa, Vittorio, Wandelt, Wehus, White,
  White, Zacchei, \& Zonca}]{planckcollaboration2020}
{Planck Collaboration}, Aghanim, N., Akrami, Y., {et~al.} 2020, A\&A, 641, A6

\bibitem[{{Planck Collaboration} {et~al.}(2016{\natexlab{b}}){Planck
  Collaboration}, Aghanim, Ashdown, Aumont, Baccigalupi, Ballardini, Banday,
  Barreiro, Bartolo, Basak, Benabed, Bernard, Bersanelli, Bielewicz, Bonavera,
  Bond, Borrill, Bouchet, Boulanger, Burigana, Calabrese, Cardoso, Carron,
  Chiang, Colombo, Comis, Couchot, Coulais, Crill, Curto, Cuttaia, {de
  Bernardis}, {de Zotti}, Delabrouille, Di~Valentino, Dickinson, Diego,
  Dor{\'e}, Douspis, Ducout, Dupac, Dusini, Elsner, En{\ss}lin, Eriksen,
  Falgarone, Fantaye, Finelli, Forastieri, Frailis, Fraisse, Franceschi,
  Frolov, Galeotta, Galli, Ganga, {G{\'e}nova-Santos}, Gerbino, Ghosh,
  {Giraud-H{\'e}raud}, {Gonz{\'a}lez-Nuevo}, G{\'o}rski, Gruppuso, Gudmundsson,
  Hansen, Helou, {Henrot-Versill{\'e}}, Herranz, Hivon, Huang, Jaffe, Jones,
  Keih{\"a}nen, Keskitalo, Kiiveri, Kisner, Krachmalnicoff, Kunz,
  {Kurki-Suonio}, Lamarre, Langer, Lasenby, Lattanzi, Lawrence, Jeune, Levrier,
  Lilje, Lilley, Lindholm, {L{\'o}pez-Caniego}, Ma, {Mac{\'i}as-P{\'e}rez},
  Maggio, Maino, Mandolesi, Mangilli, Maris, Martin,
  {Mart{\'i}nez-Gonz{\'a}lez}, Matarrese, Mauri, McEwen, Melchiorri, Mennella,
  Migliaccio, {Miville-Desch{\^e}nes}, Molinari, Moneti, Montier, Morgante,
  Moss, Natoli, Oxborrow, Pagano, Paoletti, Patanchon, Perdereau, Perotto,
  Pettorino, Piacentini, Plaszczynski, Polastri, Polenta, Puget, Rachen,
  Racine, Reinecke, Remazeilles, Renzi, Rocha, Rosset, Rossetti, Roudier,
  {Rubi{\~n}o-Mart{\'i}n}, {Ruiz-Granados}, Salvati, Sandri, Savelainen, Scott,
  Sirignano, Sirri, Soler, Spencer, {Suur-Uski}, Tauber, Tavagnacco, Tenti,
  Toffolatti, Tomasi, Tristram, Trombetti, Valiviita, Van~Tent, Vielva, Villa,
  Vittorio, Wandelt, Wehus, Zacchei, \& Zonca}]{planckcollaboration2016}
{Planck Collaboration}, Aghanim, N., Ashdown, M., {et~al.} 2016{\natexlab{b}},
  A\&A, 596, A109

\bibitem[{Potter {et~al.}(2017)Potter, Stadel, \& Teyssier}]{potter2017}
Potter, D., Stadel, J., \& Teyssier, R. 2017, Comput. Astrophys., 4, 2

\bibitem[{Reddick {et~al.}(2013)Reddick, Wechsler, Tinker, \&
  Behroozi}]{reddick2013}
Reddick, R.~M., Wechsler, R.~H., Tinker, J.~L., \& Behroozi, P.~S. 2013, ApJ,
  771, 30

\bibitem[{Riechers {et~al.}(2019)Riechers, Pavesi, Sharon, Hodge, Decarli,
  Walter, Carilli, Aravena, {da Cunha}, Daddi, Dickinson, Smail, Capak, Ivison,
  Sargent, Scoville, \& Wagg}]{riechers2019}
Riechers, D.~A., Pavesi, R., Sharon, C.~E., {et~al.} 2019, ApJ, 872, 7

\bibitem[{{Rodr{\'i}guez-Puebla} {et~al.}(2016){Rodr{\'i}guez-Puebla},
  Behroozi, Primack, Klypin, Lee, \& Hellinger}]{rodriguez-puebla2016}
{Rodr{\'i}guez-Puebla}, A., Behroozi, P., Primack, J., {et~al.} 2016, Mon. Not.
  R. Astron. Soc., 462, 893

\bibitem[{Sargent {et~al.}(2012)Sargent, B{\'e}thermin, Daddi, \&
  Elbaz}]{sargent2012}
Sargent, M.~T., B{\'e}thermin, M., Daddi, E., \& Elbaz, D. 2012, ApJ, 747, L31

\bibitem[{Sargent {et~al.}(2014)Sargent, Daddi, B{\'e}thermin, Aussel, Magdis,
  Hwang, Juneau, Elbaz, \& {da Cunha}}]{sargent2014}
Sargent, M.~T., Daddi, E., B{\'e}thermin, M., {et~al.} 2014, ApJ, 793, 19

\bibitem[{Schaerer {et~al.}(2020)Schaerer, Ginolfi, Bethermin, Fudamoto, Oesch,
  Fevre, Faisst, Capak, Cassata, Silverman, Yan, Jones, Amorin, Bardelli,
  Boquien, Cimatti, {Dessauges-Zavadsky}, Giavalisco, Hathi, Fujimoto, Ibar,
  Koekemoer, Lagache, Lemaux, Loiacono, Maiolino, Narayanan, Morselli,
  {Mendez-Hernandez}, Pozzi, Riechers, Talia, Toft, Vallini, Vergani, Zamorani,
  \& Zucca}]{schaerer2020}
Schaerer, D., Ginolfi, M., Bethermin, M., {et~al.} 2020, A\&A, 643, A3

\bibitem[{Schreiber {et~al.}(2015)Schreiber, Pannella, Elbaz, B{\'e}thermin,
  Inami, Dickinson, Magnelli, Wang, Aussel, Daddi, Juneau, Shu, Sargent, Buat,
  Faber, Ferguson, Giavalisco, Koekemoer, Magdis, Morrison, Papovich, Santini,
  \& Scott}]{schreiber2015}
Schreiber, C., Pannella, M., Elbaz, D., {et~al.} 2015, A\&A, 575, A74

\bibitem[{Shankar {et~al.}(2006)Shankar, Lapi, Salucci, De~Zotti, \&
  Danese}]{shankar2006}
Shankar, F., Lapi, A., Salucci, P., De~Zotti, G., \& Danese, L. 2006, ApJ, 643,
  14

\bibitem[{Skillman {et~al.}(2014)Skillman, Warren, Turk, Wechsler, Holz, \&
  Sutter}]{skillman2014}
Skillman, S.~W., Warren, M.~S., Turk, M.~J., {et~al.} 2014, arXiv:1407.2600
  [astro-ph] [\eprint[arXiv]{1407.2600}]

\bibitem[{Stacey {et~al.}(2018)Stacey, Aravena, Basu, Battaglia, Beringue,
  Bertoldi, Bond, Breysse, Bustos, Chapman, Chung, Cothard, Erler, Fich,
  Foreman, Gallardo, Giovanelli, Graf, Haynes, Herrera-Camus, Herter, Hložek,
  Johnstone, Keating, Magnelli, Meerburg, Meyers, Murray, Niemack, Nikola,
  Nolta, Parshley, Riechers, Schilke, Scott, Stein, Stevens, Stutzki,
  Vavagiakis, \& Viero}]{stacey2018}
Stacey, G.~J., Aravena, M., Basu, K., {et~al.} 2018, in Ground-based and
  Airborne Telescopes VII, ed. H.~K. Marshall \& J.~Spyromilio, Vol. 10700,
  International Society for Optics and Photonics (SPIE), 107001M

\bibitem[{{Tacconi} {et~al.}(2020){Tacconi}, {Genzel}, \&
  {Sternberg}}]{tacconi2020}
{Tacconi}, L.~J., {Genzel}, R., \& {Sternberg}, A. 2020, \araa, 58, 157

\bibitem[{Thacker {et~al.}(2013)Thacker, Cooray, Smidt, De~Bernardis,
  {Mitchell-Wynne}, Amblard, Auld, Baes, Clements, Dariush, De~Zotti, Dunne,
  Eales, Hopwood, Hoyos, Ibar, Jarvis, Maddox, Micha{\l}owski, Pascale, Scott,
  Serjeant, Smith, Valiante, \& {van der Werf}}]{thacker2013}
Thacker, C., Cooray, A., Smidt, J., {et~al.} 2013, ApJ, 768, 58

\bibitem[{{The CONCERTO collaboration} {et~al.}(2020){The CONCERTO
  collaboration}, Ade, Aravena, Barria, Beelen, Benoit, B{\'e}thermin, Bounmy,
  Bourrion, Bres, De~Breuck, Calvo, Cao, Catalano, D{\'e}sert, Dur{\'a}n,
  Fasano, Fenouillet, Garcia, Garde, Goupy, Groppi, Hoarau, Lagache, Lambert,
  Leggeri, {Levy-Bertrand}, {Macias-Perez}, Mani, Marpaud, Mauskopf,
  Monfardini, Pisano, Ponthieu, Prieur, Roni, Roudier, Tourres, \&
  Tucker}]{theconcertocollaboration2020}
{The CONCERTO collaboration}, Ade, P., Aravena, M., {et~al.} 2020, A\&A, 642,
  A60

\bibitem[{Vale \& Ostriker(2004)}]{vale2004}
Vale, A. \& Ostriker, J.~P. 2004, Monthly Notices of the Royal Astronomical
  Society, 353, 189

\bibitem[{{Vallini} {et~al.}(2018){Vallini}, {Pallottini}, {Ferrara},
  {Gallerani}, {Sobacchi}, \& {Behrens}}]{vallini2018}
{Vallini}, L., {Pallottini}, A., {Ferrara}, A., {et~al.} 2018, \mnras, 473, 271

\bibitem[{{Venemans} {et~al.}(2017){Venemans}, {Walter}, {Decarli},
  {Ferkinhoff}, {Wei{\ss}}, {Findlay}, {McMahon}, {Sutherland}, \&
  {Meijerink}}]{venemans2017}
{Venemans}, B.~P., {Walter}, F., {Decarli}, R., {et~al.} 2017, \apj, 845, 154

\bibitem[{Viero {et~al.}(2009)Viero, Ade, Bock, Chapin, Devlin, Griffin,
  Gundersen, Halpern, Hargrave, Hughes, Klein, MacTavish, Marsden, Martin,
  Mauskopf, Moncelsi, Negrello, Netterfield, Olmi, Pascale, Patanchon, Rex,
  Scott, Semisch, Thomas, Truch, Tucker, Tucker, \& Wiebe}]{viero2009}
Viero, M.~P., Ade, P. A.~R., Bock, J.~J., {et~al.} 2009, ApJ, 707, 1766

\bibitem[{Viero {et~al.}(2019)Viero, Reichardt, Benson, Bleem, Bock, Carlstrom,
  Chang, Cho, Crawford, Crites, {de Haan}, Dobbs, Everett, George, Halverson,
  Harrington, Holder, Holzapfel, Hou, Hrubes, Knox, Lee, {Luong-Van}, Marrone,
  McMahon, Meyer, Millea, Mocanu, Mohr, Moncelsi, Padin, Pryke, Ruhl, Schaffer,
  Serra, Shirokoff, Staniszewski, Stark, Story, Vanderlinde, Vieira,
  Williamson, \& Zemcov}]{viero2019}
Viero, M.~P., Reichardt, C.~L., Benson, B.~A., {et~al.} 2019, ApJ, 881, 96

\bibitem[{Viero {et~al.}(2013)Viero, Wang, Zemcov, Addison, Amblard, Arumugam,
  Aussel, B{\'e}thermin, Bock, Boselli, Buat, Burgarella, Casey, Clements,
  Conley, Conversi, Cooray, De~Zotti, Dowell, Farrah, Franceschini, Glenn,
  Griffin, Hatziminaoglou, Heinis, Ibar, Ivison, Lagache, Levenson, Marchetti,
  Marsden, Nguyen, O'Halloran, Oliver, Omont, Page, Papageorgiou, Pearson,
  {P{\'e}rez-Fournon}, Pohlen, Rigopoulou, Roseboom, {Rowan-Robinson}, Schulz,
  Scott, Seymour, Shupe, Smith, Symeonidis, Vaccari, Valtchanov, Vieira,
  Wardlow, \& Xu}]{viero2013}
Viero, M.~P., Wang, L., Zemcov, M., {et~al.} 2013, ApJ, 772, 77

\bibitem[{{Vizgan} {et~al.}(2022){Vizgan}, {Greve}, {Olsen}, {Zanella},
  {Narayanan}, {Dav{\`e}}, {Magdis}, {Popping}, {Valentino}, \&
  {Heintz}}]{vizgan2022}
{Vizgan}, D., {Greve}, T.~R., {Olsen}, K.~P., {et~al.} 2022, \apj, 929, 92

\bibitem[{{Walter} {et~al.}(2020){Walter}, {Carilli}, {Neeleman}, {Decarli},
  {Popping}, {Somerville}, {Aravena}, {Bertoldi}, {Boogaard}, {Cox}, {da
  Cunha}, {Magnelli}, {Obreschkow}, {Riechers}, {Rix}, {Smail}, {Weiss},
  {Assef}, {Bauer}, {Bouwens}, {Contini}, {Cortes}, {Daddi}, {Diaz-Santos},
  {Gonz{\'a}lez-L{\'o}pez}, {Hennawi}, {Hodge}, {Inami}, {Ivison}, {Oesch},
  {Sargent}, {van der Werf}, {Wagg}, \& {Yung}}]{walter2020}
{Walter}, F., {Carilli}, C., {Neeleman}, M., {et~al.} 2020, \apj, 902, 111

\bibitem[{Walter {et~al.}(2014)Walter, Decarli, Sargent, Carilli, Dickinson,
  Riechers, Ellis, Stark, Weiner, Aravena, Bell, Bertoldi, Cox, Da~Cunha,
  Daddi, Downes, Lentati, Maiolino, Menten, Neri, Rix, \& Weiss}]{walter2014}
Walter, F., Decarli, R., Sargent, M., {et~al.} 2014, ApJ, 782, 79

\bibitem[{Wang {et~al.}(2021)Wang, Gao, Best, Duncan, Hardcastle, Kondapally,
  Malek, McCheyne, Sabater, Shimwell, Tasse, Bonato, Bondi, Cochrane, Farrah,
  Gurkan, Haskell, Pearson, Prandoni, Rottgering, Smith, Vaccari, \&
  Williams}]{wang2021}
Wang, L., Gao, F., Best, P.~N., {et~al.} 2021, A\&A, 648, A8

\bibitem[{Whitaker {et~al.}(2021)Whitaker, Williams, Mowla, Spilker, Toft,
  Narayanan, Pope, Magdis, {van Dokkum}, Akhshik, Bezanson, Brammer, Leja, Man,
  Nelson, Richard, Pacifici, Sharon, \& Valentino}]{whitaker2021}
Whitaker, K.~E., Williams, C.~C., Mowla, L., {et~al.} 2021, Nature, 597, 485

\bibitem[{Yan {et~al.}(2020)Yan, Sajina, Loiacono, Lagache, B{\`e}thermin,
  Faisst, Ginolfi, F{\`e}vre, Gruppioni, Capak, Cassata, Schaerer, Silverman,
  Bardelli, {Dessauges-Zavadsky}, Cimatti, Hathi, Lemaux, Ibar, Jones,
  Koekemoer, Oesch, Talia, Pozzi, Riechers, Tasca, Toft, Vallini, Vergani,
  Zamorani, \& Zucca}]{yan2020}
Yan, L., Sajina, A., Loiacono, F., {et~al.} 2020, ApJ, 905, 147

\bibitem[{{Yue} \& {Ferrara}(2019)}]{yue2019}
{Yue}, B. \& {Ferrara}, A. 2019, \mnras, 490, 1928

\bibitem[{Zanella {et~al.}(2018)Zanella, Daddi, Magdis, Diaz~Santos, Cormier,
  Liu, Cibinel, Gobat, Dickinson, Sargent, Popping, Madden, Bethermin, Hughes,
  Valentino, Rujopakarn, Pannella, Bournaud, Walter, Wang, Elbaz, \&
  Coogan}]{zanella2018}
Zanella, A., Daddi, E., Magdis, G., {et~al.} 2018, Monthly Notices of the Royal
  Astronomical Society, 481, 1976

\bibitem[{Zavala {et~al.}(2021)Zavala, Casey, Manning, Aravena, Bethermin,
  Caputi, Clements, da~Cunha, Drew, Finkelstein, Fujimoto, Hayward, Hodge,
  Kartaltepe, Knudsen, Koekemoer, Long, Magdis, Man, Popping, Sanders,
  Scoville, Sheth, Staguhn, Toft, Treister, Vieira, \& Yun}]{zavala2021}
Zavala, J.~A., Casey, C.~M., Manning, S.~M., {et~al.} 2021, ApJ, 909, 165

\end{thebibliography}

\begin{appendix} 

\section{Comparing SIDES with SPT and SPTxSPIRE power spectra}
\label{ap:spt_and_spt_x_spire}

\begin{figure*}[h]
  \centering
  \includegraphics[width=\linewidth]{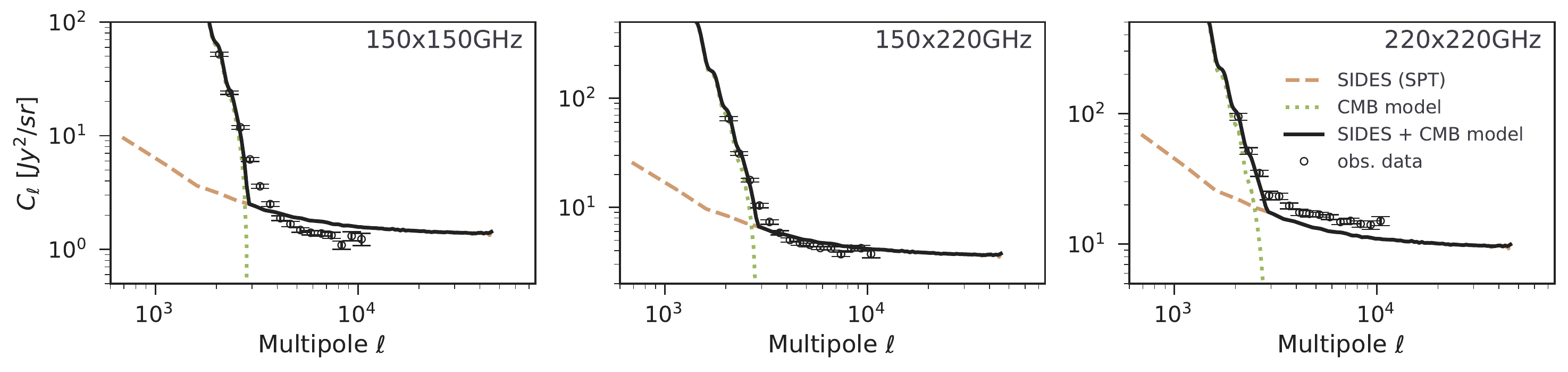}
  \caption{Comparison of the SPT auto and cross power spectra of the 150 and 220\,GHz bands (black points, \citealt{viero2019}) and the SIDES model (brown dashed line). The green dotted line model is the contribution of the CMB in the data. The sum of the SIDES and CMB model is shown with the black solid line.}
  \label{fig:spt_pk_maps}
\end{figure*}

We generated simulated maps for the South Pole Telescope (SPT) frequency bands at 150 and 220\,GHz \citep{hall2010} and compared the SIDES power spectra with both the SPT and SPTxSPIRE power spectra. The SPT data are contaminated with the cosmic microwave background (CMB) emission, especially at low frequency. In order to account for this we add a CMB model\footnote{\url{http://pla.esac.esa.int/pla/#home}} on top of the SIDES model when comparing with the data (Fig.\,\ref{fig:spt_pk_maps}). The masking procedure we follow for the SPIRE bands is the one explained in Sect.\,\ref{subsec:maps_pk_estimation_color_corrections}, while for SPT we mask all the sources with flux above 6.4\,mJy at 150\,GHz, as done in \cite{viero2019}. The applied color corrections are summarized in Table\,\ref{table:cc}.

The sum of the SIDES and CMB model can reproduce the SPT measurements below $\rm \ell = 3 \times 10^3$. However, at higher multipoles the SIDES model agrees with the data only for the SPT150$\times$220 cross power spectrum. It fails to reproduce the SPT150x150 and SPT220X220 data. The discrepancy between SPT220$\times$220 data and SIDES is on the order of $\sim$20\%, while SIDES can reproduce the \textit{Planck}217$\times$217 data at $\sim$10\%, although at much larger scales (Sect.\,\ref{subsec:cib_pk_model_vs_data}).

\begin{figure*}[h]
  \centering
  \includegraphics[width=\linewidth]{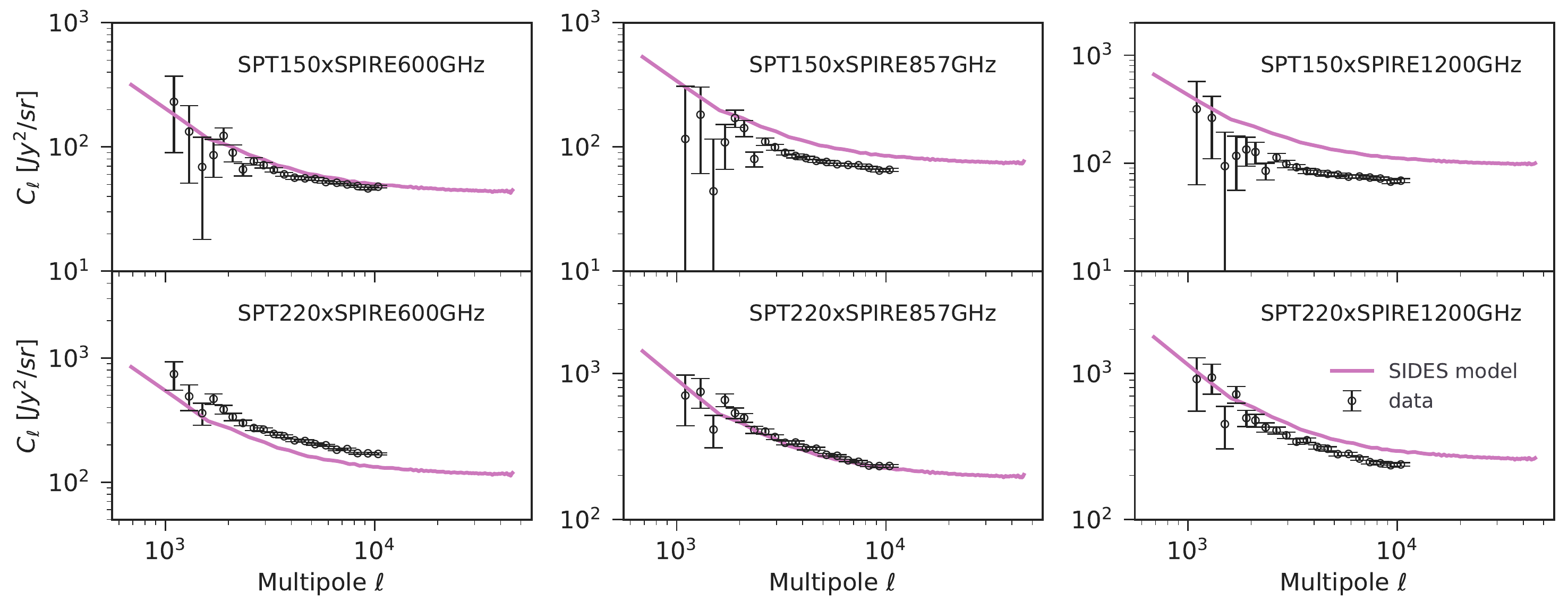}
  \caption{Comparison of the SPT$\times$SPIRE data (black points, \citealt{viero2019}) with the SIDES model (purple solid line).}
  \label{fig:spt_x_spire_pk_maps}
\end{figure*}

Additionally, we see significant discrepancies between SPT$\times$SPIRE measurements and our model in Fig. \ref{fig:spt_x_spire_pk_maps}. SIDES is systematically above the data, except for SPT220$\times$SPIRE600 where it is below by $\sim \,$20\%, and for SPT220$\times$SPIRE857 where there is a good agreement between the two, despite the fact that both SPT220x220 (Fig.\,\ref{fig:spt_pk_maps}) and SPIRE857$\times$857 (Fig.\,\ref{fig:spire_pk_maps}) measurements are below the model.
Even though the SPIRE data can be overall reproduced by the SIDES model as shown in Sect.\,\ref{subsec:cib_pk_model_vs_data}, the discrepancies in the SPT$\times$SPIRE case can even exceed 40\%. We thus suspect that the inconsistencies between the measurements and the model may come from these specific CIB measurements published in \cite{viero2019}.

\section{Luminosity function variance with survey size: Poisson and clustering contribution}
\label{ap:LF_variance_appendix}

\begin{figure*}[h]
  \centering
  \includegraphics[width=\linewidth]{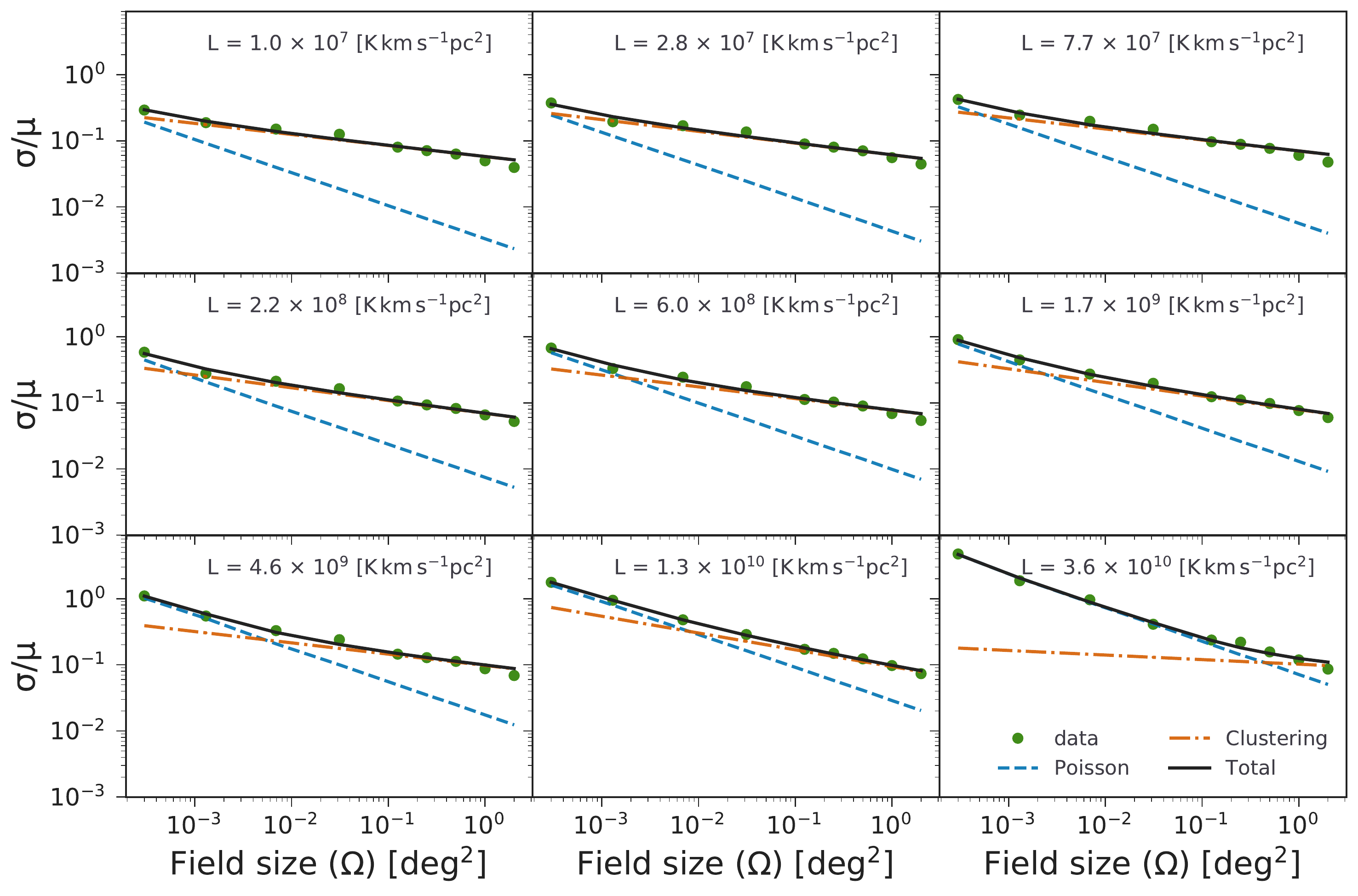}
  \caption{Decomposition of the total variance of the CO LFs into the Poisson and clustering components. The variance is shown as a function of the survey size for different luminosity bins in order to visualize at which luminosities and sizes the clustering component excessively contributes to the total variance.}
  \label{fig:LF_var_Lbins}
\end{figure*}

In Fig.\,\ref{fig:LF_var_Lbins} we show the decomposition of the total variance in the CO LFs into the Poisson and clustering components as a function of the survey size, for different luminosity bins. At the highest luminosities ($3.6 \times 10^{10}$\,L$_\odot$), the Poisson component is the dominant source of variance up to 1\,deg$^2$, after which the clustering starts to dominate. For lower luminosities, this transition appears at lower field sizes. For very low luminosities ($<10^{8}$\,L$_\odot$), even 0.001\,deg$^2$ fields are dominated by the clustering component.

\end{appendix}

\end{document}